\documentclass[aps,prc,showkeys,nofootinbib]{revtex4}
\usepackage{graphicx}% Include figure files

\usepackage{isotope}% isotope editing
\usepackage{braket}% Macros for Dirac bra–ket notation and sets {|}, Donald Arseneau: https://ctan.org/pkg/braket
\usepackage{physics}% physics packet, Sergio C. de la Barrera, 2012 (from ctan.org)

\usepackage{feyn}

\newcommand{\sigmabf}{\mbox{\boldmath $\sigma$}}
\newcommand{\xibf}{\mbox{\boldmath $\xi$}}

\begin{document}
%-----------------------------------------------------------------------------------------------------------------------

%-----------------------------------------------------------------------------------------------------------------------
\title{% (1)
Incoherent processes in dileptons production in proton-nucleus scattering at high energies
% \vspace{2mm}
% \newline
% (2) Dilepton production in proton-nucleus scattering (heavy-ions collisions) at high energies
% \vspace{2mm}
% \newline
% (2) Enhancement (Усиление) of incoherent $\gamma$-production in proton-nucleus scattering in the $\Delta$-resonance region
% \vspace{2mm}
% \newline
% (3) Anomalous magnetic moments of nucleons in nuclear matter and phenomenon of suppression of incoherent bremsstrahlung
}
% \author{Peng-Ming~Zhang$^{1}$}\email{zhangpm5@mail.sysu.edu.cn}%
% \author{Li-Ping~Zou$^{1}$}\email{zoulp@impcas.ac.cn} %
% \author{\ldots}%
\author{Sergei~P.~Maydanyuk$^{1,2}$}\email{sergei.maydanyuk@wigner.hu}%
\author{Gyorgy~Wolf$^{(1)}$}\email{wolf.gyorgy@wigner.hu}%
\affiliation{$^{(1)}$Wigner Research Centre for Physics, Budapest, 1121, Hungary}
\affiliation{$^{(2)}$Institute for Nuclear Research, National Academy of Sciences of Ukraine, Kyiv, 03680, Ukraine}
% \affiliation{$^{(1)}$School of Physics and Astronomy, Sun Yat-sen University, Zhuhai, China}
% \affiliation{$^{(2)}$Institute of Modern Physics, Chinese Academy of Sciences, Lanzhou, 730000, China}

\date{\small\today}
%-----------------------------------------------------------------------------------------------------------------------

%-----------------------------------------------------------------------------------------------------------------------
\begin{abstract}
\noindent
\textbf{Purpose:}
Incoherent processes in production of lepton pairs in the scattering of protons off nuclei are investigated. % in the paper at intermediate energy region.
\textbf{Methods:}
New quantum mechanical model is constructed
which uses generalization of the nuclear model of emission of photons in the proton-nucleus reactions
in region from low up to high energies
with inclusion of formalism of production of lepton pairs.
% A focus is directed on question of how much the bremsstrahlung spectrum is changed after transition of one nucleon in nucleus to $\Delta$-resonance.
% including properties o $\Delta$-resonance in the nucleus-target.
%
\textbf{Results:}
% On the basis of such a model
% We find the following.
%
% \begin{itemize}
% \item
% \noindent
(1) Model with the coherent matrix elements is tested for the scattering of protons on the \isotope[]{Be} nuclei at energy of proton beam $E_{\rm p}$ of 2.1~GeV.
The calculated cross section of production of lepton pairs is in good agreement with experimental data obtained by DLS Collaboration.
% This is good test of the model and algorithms, that is important for further analysis.
% we test formalism and calculations for the \isotope[197]{Au} nucleus at $E_{\rm p}=190$~MeV on the experimental data~\cite{Goethem.2002.PRL}
% (ratio between such contributions is about $10^{6}$--$10^{7}$ for $p + \isotope[197]{Au}$ at $E_{\rm p}=190$~MeV,
% This confirms that inclusion of incoherent processes in study of $\Delta$-resonance in the proton-nucleus scattering is important.
% This aspect has never been studied yet.
% we test formalism and calculations for the \isotope[197]{Au} nucleus at $E_{\rm p}=190$~MeV on the experimental data~\cite{Goethem.2002.PRL}
% (ratio between such contributions is about $10^{6}$--$10^{7}$ for $p + \isotope[197]{Au}$ at $E_{\rm p}=190$~MeV,
% this result is in agreement with results in Refs.~\cite{Maydanyuk_Zhang.2015.PRC,Maydanyuk_Zhang_Zou.2016.PRC,Liu_Maydanyuk_Zhang_Liu.2019.PRC.hypernuclei}).
% This incoherent contribution is dependent on magnetic moments of nucleons of nucleus-target
% This confirms that inclusion of incoherent processes in study of $\Delta$-resonance in the proton-nucleus scattering is important.
% This aspect has never been studied yet.
% *******************************************************************************************************************
%
% *******************************************************************************************************************
% \item
% \noindent
(2) We analyzed dilepton production for
% in region from low up to large masses of
nuclei-targets
\isotope[9]{Be}, \isotope[12]{C}, \isotope[16]{O}, \isotope[24]{Mg}, \isotope[44]{Ca}, \isotope[197]{Au}
at $E_{\rm p}=2.1$~GeV.
Coherent cross sections of dilepton production are monotonously decreased with increasing of nucleus-target mass.
This new effect is explained by suppressing of production of dileptons (by nuclear matter of nucleus-target).
% The cross section of dilepton production is smaller for nuclei-target with larger mass.
% *******************************************************************************************************************
%
% *******************************************************************************************************************
% (3) The spectra are not sensitive to choice of isotopes with the same charge number, both for light and heavy nuclei-targets.
% This is clearly explained by dependencies of effective charge $Z_{\rm eff}^{\rm (mon)} (\vb{k}_{\rm ph})$ on
% different masses and charges of two participating nuclei in reaction in formula for matrix elements.
% *******************************************************************************************************************
%
% *******************************************************************************************************************
(3) Production of lepton pairs is more intensive at larger $E_{\rm p}$.
% We analyze dependence of cross section of production of leptons pair on energy of proton beam for the same fixed nucleus-target
% (a) incoherent emission is essentially more intensive than the coherent emission,
% This result confirms the important (and not small) role of incoherent emission in bremsstrahlung.
% *******************************************************************************************************************
%
% *******************************************************************************************************************
(4) For $p + \isotope[9]{Be}$ at $E_{\rm p}=2.1$~GeV
we analyzed role of incoherent processes in production of pairs of leptons. % and find the following.
% (a) incoherent emission is essentially more intensive than the coherent emission,
% (b) role of incoherent processes is increased at increasing of energy of photon,
% This result confirms the important (and not small) role of incoherent emission in bremsstrahlung.
Inclusion of incoherent proton-nucleon
processes to the model improves agreement with experimental data a little.
% (it allows to achieve better description of experimental data at low energies smaller 250~MeV).
% Our predictions in regions of low energies $M <250$~MeV and high energies $M > 1000$~MeV can be recommended for tests in possible experimental study in future.
A new phenomenon of suppression of production of lepton pairs at low energies due to incoherent processes is observed.
This is explained by dominant coherent contribution at very low energies.
% *******************************************************************************************************************
%
% *******************************************************************************************************************
(5) Adding of the longitudinal amplitude
of virtual photon suppresses the cross section of dilepton production a little
(effect is observed for $p + \isotope[9]{Be}$ at $E_{\rm p}=2.1$~GeV).
% We analyzed influence of longitudinal part of virtual photon on the calculated cross sections of production of dilepton pairs.
% From analysis of the scattering of $p + \isotope[9]{Be}$ at $E_{\rm p}=2.1$~GeV.
% factor $f_{\rm incoh}$ which suppresses the intensity of incoherent processes.
% The spectra have similar shapes of monotonous type at different virtualities $\chi$.
% \emph{ratio between incoherent and coherent contributions is about 10--100 inside main region of invariant mass values}.
% \emph{we observe a new phenomenon of suppression of production of lepton pairs at low energy region due to incoherent processes.}
% This is explained by that at very low energies coherent contribution is dominant in comparison with incoherent one.
% *******************************************************************************************************************
%
% *******************************************************************************************************************
% \item
% The incoherent contribution is essentially larger than the coherent one.
% \noindent
(6) The incoherent contribution has a leading role in dilepton production
% than the coherent one
(ratio between incoherent and coherent contributions is 10--100).
% This explains improving agreement between calculations and experimental data after inclusion of incoherent terms.
% For inclusion of transition from proton to $\Delta^{+}$-resonance ($p\,N \to \Delta^{+} N$) in nucleus-target
% % formation of $\Delta$-resonance in nucleus-target
% in the proton-nucleus scattering to the model, we use scheme of the coherent processes in Ref.~\cite{Gil_Oset.1998.PLB.v416}.
% % We estimate change in the bremsstrahlung emission after such an inclusion to the model.
% To find nuclei with maximal reinforcement of bremsstrahlung due to transition $p\,N \to \Delta^{+} N$, we obtain condition~(\ref{eq.resultingformulas.2}) determining ratio between protons and neutrons for the normal nucleus.
% (i.e., there is no enhancement of bremsstrahlung due to transition $p\,N \to \Delta^{+} N$, if incoherent processes are not included to the model).
%
% We also analyzed role of incoherent processes in production of pairs of leptons
% in the scattering of $p + \isotope[93]{Nb}$ at $E_{\rm p}=3.5$~GeV and find the following.
%
Also our model provides the tendencies of the full spectrum for
$p + \isotope[93]{Nb}$ at $E_{\rm p}=3.5$~GeV in good agreement with
experimental data obtained by HADES collaboration, and shows large role of incoherent processes.
%
% Incoherent contribution for this reaction is essentially larger than
% for $p + \isotope[9]{Be}$ at $E_{\rm p}=2.1$~GeV [see Fig.~\ref{fig.6}~(b)].
%
% (3) Incoherent processes are highly increased at increasing atomic mass of the nucleus-target.
%
%
% \vspace{1.0mm}
% \noindent
\textbf{Conclusion:}
Results above confirms importance of incoherent processes in study of dilepton production in this reaction.
% , which have never been studied yet.
%-----------------------------------------------------------------------------------------------------------------------

%-----------------------------------------------------------------------------------------------------------------------
% This property can be used for proposal for future experiments with measurements of photons,
% as tools to distinguish process of formation of $\Delta$-resonance in the nucleus-target.
% Including properties of $\Delta$-resonance in the nucleus-target to the bremsstrahlung model, we find the following.
% (1) Ratio between incoherent emission and coherent emission is about $10^{6}$--$10^{7}$
% for $p + \isotope[197]{Au}$ at energy of proton beam $E_{\rm p}$ of 190~MeV,
% where the calculated full bremsstrahlung spectrum is in good agreement with experimental data.
%-----------------------------------------------------------------------------------------------------------------------
%
%-----------------------------------------------------------------------------------------------------------------------
\end{abstract}
%-----------------------------------------------------------------------------------------------------------------------

%-----------------------------------------------------------------------------------------------------------------------
% \textbf{PACS numbers:}
 % Multipole matrix elements (in electromagnetic transitions)

% 27.80.+w % A is greater than or equal to 190 and is less than or equal to 219 (properties of specific nuclei listed by mass ranges)
% 23.60.+e, % Alpha decay
% 23.70.+j, % Heavy-particle decay
% 25.70.-z, % Low and intermediate energy heavy-ion reactions
% 24.75.+i, % General properties of fission
% 25.85.Ca, % Spontaneous fission
% 25.70.Gh, % Compound nucleus


\maketitle
% *******************************************************************************************************************

% *******************************************************************************************************************
% \newpage
\section{Introduction
\label{sec.introduction}}

Heavy-ion collisions in the energy regime from roughly 20 MeV/nucleon to 2 GeV/nucleon offer a unique
possibility to study the properties of dense and hot nuclear matter
\cite{Csernai.Kapusta.1986.PhysRep,
Oare.Strottmann.1986.PhysRep,
Stock.1986.PhysRep,
Stocker.Greiner.1986.PhysRep,
Schflrmann.1987.PhysRep,
Bertsch.DasGupta.1988.PhysRep}.
%
% Compressed dense matter in heavy ion collisions is hot topic studied last few decades \cite{Wolf.1993.PPNP}.
Many particles are created when two nuclei collide.
Particles produced
% plays important role in such investigations of compressed dense matter,
as mesonic and electromagnetic probes bring important information about
properties of compressive phase of nuclear matter which can be formed during
the collision of heavy ions
\cite{Mosel.1991.AnnRevNuclPartSci}.
$e^{+} e^{-}$ pairs (dileptons),
$\pi$, $\rho$, $\eta$-mesons are produced and they are included to the unified picture of the studied process~\cite{Salabura.2021.PPNP}.
One of
% \textcolor[rgb]{1.00,0.00,0.00}{\textbf{
the possibilities is production of lepton pairs which can be measured in experiments and provide useful information
on high density phase of nuclear matter~\cite{Wolf.1993.PPNP}.

Properties of dilepton production in heavy-ion collisions
at bombarding energies from 1 to 2 GeV/A
were successfully studied by the transport model of the Boltzmann~Uehling-Uhlenbeck (BUU) type
[see Refs.~\cite{Bertsch.DasGupta.1988.PhysRep,Cassing.1990.PhysRep,Wolf.1990.NPA},
also Refs.~\cite{Bleicher_Bratkovskaya.2022.ProgPartNuclPhys,Linnyk_Bratkovskaya_Cassing.2016.ProgPartNuclPhys}
for general aspects of transport models BUU at different energies with classification, reference therein].
Many effects such as
the dynamical evolution of the nucleus-nucleus collision,
nucleon resonances in nuclei,
the electromagnetic form factor of hadrons,
medium properties,
effects of pion and $\Delta$ self-energies on the pion dynamics,
etc.
were well described by such an approach and good agreement with experimental data was obtained.

As the simplest mechanism, process of dilepton production can go through exchange of virtual photon, which is emitted by one nucleon of
the nuclei participated in collision.
The full picture of emission of photons is obtained due to evolution of moving
electric charges and magnetic momentums of nucleons (or nucleon resonances), which belong to the nuclei in reaction.
This leads to dependence of dilepton production on the different mechanisms and aspects of the studied nuclear reaction
(which depends on the chosen way of description of nuclear interactions),
and even properties of nuclei.
From this point of view, the question appears: how important is
the role of such mechanisms in nuclear reaction in the study of production of dileptons?
It could be interesting to clarify, how many-nucleon dynamics influences on production of dileptons in the nuclear collision.

From the previous study it has been known that
physics of many-nucleon dynamics in the nuclear reactions (in study of emission of photons) is related with
incoherent processes which are significantly larger than coherent ones~\cite{Maydanyuk_Zhang.2015.PRC,Maydanyuk_Zhang_Zou.2016.PRC}. %,Maydanyuk.2023.PRC.delta}.
Thus, in frameworks of formalism~\cite{Maydanyuk.2023.PRC.delta} (see reference in that paper for different aspects of that approach),
the full operator of emission of photons in the proton nucleus scattering can be explicitly separated
on two groups of terms~\cite{Maydanyuk_Vasilevsky.2023.fold.arXiv}.
One group (coherent bremsstrahlung) includes terms with momentum defined on relative distance between center-of-mass of the nucleus-target and the scattered proton
(for example, for the $\alpha$-nucleus scattering this is Eq.~(B10) in Ref.~\cite{Liu_Maydanyuk_Zhang_Liu.2019.PRC.hypernuclei}).
The second group (incoherent bremsstrahlung) includes rest of terms without momentum from relative distance between the nucleus-target and the scattered proton.
It includes momenta of relative distances between individual nucleons of the nucleus-target and the scattering proton
(for the $\alpha$-nucleus scattering this is Eq.~(B11) with addition (B12) in Ref.~\cite{Liu_Maydanyuk_Zhang_Liu.2019.PRC.hypernuclei}).
By simple words, incoherent bremsstrahlung has origin from many nucleon dynamics, while coherent bremsstrahlung is related with two body (proton-nucleus) dynamics.
Such incoherent processes have leading role for the proton nucleus reaction at energies up to 1~GeV~\cite{Maydanyuk.2023.PRC.delta}.
In particular, ratio between the incoherent contribution and
coherent one is about $10^{+6}$ -- $2 \cdot 10^{+7}$ times
for the $p + \isotope[197]{Au}$ scattering at energy of proton beam $E_{\rm p}=190$~MeV
\cite{Maydanyuk_Zhang.2015.PRC}
(see Figs.~1~(a), 2 in Ref.~\cite{Maydanyuk.2023.PRC.delta};
also see less accurate calculations
for $p + ^{208}{\rm Pb}$ at $E_{\rm p} = 140$ and 145~MeV,
and $p + ^{12}{\rm C}$, $p + ^{58}{\rm Ni}$, $p + ^{107}{\rm Ag}$ and $p + ^{197}{\rm Au}$ at $E_{\rm p} = 190$~MeV
in Ref.~\cite{Maydanyuk_Zhang.2015.PRC}).

% In particular, it forms incoherent type of emission, which can be understood as emission from relations between individual nucleons from different nuclei.

Another important issue is that the incoherent emission has shape of the spectrum different from the shape of the coherent contribution after its renormalisations.
Inclusion of the incoherent terms to the calculations increases agreement with experimental data for bremsstrahlung emission in the nuclear scattering
~\cite{Maydanyuk_Zhang.2015.PRC,Maydanyuk_Zhang_Zou.2016.PRC}.
%
% (for example, experimental data were used in such analysis and they were obtained with highest accuracy, up today).
%
By the reasons above, we suppose that production of lepton pairs in the nuclear scattering formed via exchange of virtual photon includes both coherent and incoherent processes,
where the incoherent contribution should have the dominant role.
This is motivation of our study to include incoherent processes in study of production of dileptons in heavy ion collisions.
Such a question on the basis of many-nucleon quantum mechanics has not been studied yet.
From technical point of view this problem is complicated.
After inclusion of the incoherent processes we suppose to obtain essential changes of the full cross section of dilepton productions
estimated before on the basis of coherent processes only.
So, in order to perform such a research, in this paper we do not consider relevant channels related with resonances, meson decays, Dalitz decay
(for example, see Refs.~\cite{Zetenyi.Wolf.2005.ActaPHun,Zetenyi_Wolf.2012.PRC,Wolf.1993.PPNP}).
It could be interesting to estimate such processes, where
to generalize the formalism in Ref.~\cite{Maydanyuk.2023.PRC.delta},
which previously was tested well on existed experimental information on bremsstrahlung.
The simplest nuclear reaction which includes incoherent processes in dilepton production is the proton nucleus scattering.
So, we will consider the proton nucleus scattering for this study in this paper.

In the 1--5 GeV/nucleon beam energy range the dileptons production has been studied by
the DiLepton Spectrometer (DLS) at LBL \cite{Wilson.DSL-collab.1998.PRC,Porter.DSL-collab.1997.PRL} and
the High Acceptance Di-Electron Spectrometer (HADES) at
GSI~\cite{Agakishiev.HADEScollab.2012.plb,Weber.HADEScollab.2011.JPConfSer,%
Agakichiev.HADEScollab.2010.PLB,Holzmanna.HADEScollab.2010.NPA,Agakichiev.HADEScollab.2009.EPJA,%
Salabura.HADEScollab.2005.NPA,Salabura.HADEScollab.2004.ActaPhysPol,Salabura.HADEScollab.2004.PPNP}.
%
% \vspace{1.0mm}
% \noindent
For the proton-nucleus scattering at energy of proton beam $E_{\rm p}$ below 10~GeV,
experimental data
% \textcolor[rgb]{1.00,0.00,0.00}{\textbf{
on dilepton productions have been obtained
for \isotope[]{Be} at $E_{\rm p}=1.0$~GeV, 2.1~GeV and 4.9~GeV (DLS, Ref.~\cite{Naudet.1989.PRL}),
for \isotope[]{Be} at $E_{\rm p}=4.9$~GeV (DLS, Ref.~\cite{Roche.1988.PRL}),
for \isotope[]{Nb} at $E_{\rm p}=3.5$~GeV (HADES, Ref.~\cite{Agakishiev.HADEScollab.2012.plb,Weber.HADEScollab.2011.JPConfSer}).
% }}
%
% Experimental measurements for the proton-nucleus scattering were obtained by DLS Collaboration for
% the \isotope[9]{Be} nucleus at proton beam energies of 1.0~GeV, 2.1~GeV and 4~GeV~\cite{Naudet.1989.PRL}.
%
% The calculated cross sections of production of leptons pair (with coherent terms)
% in the scattering of protons off the \isotope[9]{Be} nuclei at energy of proton beam of $E_{\rm p}=2.1$~GeV
% for different virtualities $\chi$ of photon
% in comparison with experimental data~\cite{Naudet.1989.PRL}
%
So, we will use those data in our analysis.
Note that also
% for the proton-nucleus scattering,
experimental data of dilepton productions have been measured
for $p + \isotope[]{C}$ and $p + \isotope[]{Cu}$ at $E_{\rm p}=12.0$~GeV in KEK~\cite{Ozawa.2001.PRL}.
% }}

% \vspace{3.5mm}
The paper is organized in the following way.
In Sec.~\ref{sec.4} the $S$-matrix formalism for production of pair of leptons in different nuclear processes is analyzed,
with a new inclusion on processes with participation of many nucleons in the proton nucleus scattering.
Here, the leptonic and nuclear matrix elements are defined. % for such a process.
In Sec.~\ref{sec.6-pp} explicit formulas for the leptonic tensor and the full matrix element of dilepton production are derived for the proton-nucleon scattering.
In Sec.~\ref{sec.6-nucleus} the similar leptonic and nuclear tensors, and the full matrix element of dilepton production are generalized for the proton-nucleus scattering.
In Sec.~\ref{sec.model.bremprobability} the cross section of dilepton production based on the previous formalism for the proton-nucleus scattering is calculated.
In Sec.~\ref{sec.model.incoh} new formalism needed for study of incoherent processes in production of leptons pairs in the proton-nucleus scattering is developed.
In Sec.~\ref{sec.analysis}
different aspects of dilepton production in the proton-nucleus scattering are studied in comparison with experimental data.
% role of incoherent processes is studied,
% virtulal photon is analyzed.
% The spectra of production of dilepton pair in the scattering $p + \isotope[9]{Be}$ and experimental data \label{sec.analysis.1}.
% estimation of energies for zero-point vibrations and quasibound states, etc..
% In Sec.~\ref{sec.screening} influence of plasma screening on properties of the pycnonuclear reaction is studied on example
% of $\isotope[12]{C} + \isotope[12]{C}$.
Conclusions and perspectives are summarized in Sec.~\ref{sec.conclusions}.
Tensor forms for hadronic and leptonic parts of the matrix elements for dilepton production are reviewed in App.~\ref{sec.app.1.DIS_leptonic}.
Derivation of the matrix elements for dilepton production
in the many-nucleon nuclear scattering are presented in details in App.~\ref{sec.app.2}.
A new formalism of inclusion of longitudinal polarization of virtual photon to the model is presented in App.~\ref{sec.virtual}.
% Cross section of dilepton production in dependence on invariant mass is formulated in App.~\ref{sec.app.model.crosssection.2}.
%-----------------------------------------------------------------------------------------------------------------------

%-----------------------------------------------------------------------------------------------------------------------
% \newpage
% \section{Processes of production of dilepton pair via virtual photon in the many-nucleon systems in scattering
% \label{sec.3}}

\section{Formalism of $S$-matrix
\label{sec.4}}

\subsection{Connection between processes of emission of real photon and
emission of dileptons via virtual photon
\label{sec.3.1}}

Let us compare two processes:
(1) process of emission of real photon by nucleon, and
(2) process of production of pair of leptons by virtual photon which is emitted by nucleon.
We write down below graphs of these two and also add matrix elements of the studied processes as
[for example, see book~\cite{Halzen.book.1987}, p.~149, Eq.~(6.17)]
\begin{equation}
\begin{array}{llclll}
\vspace{2.5mm}
1) & \mbox{\rm Emission of photon:} &

\Diagram{\vertexlabel^{p'} \\
fdV \\
& gA\vertexlabel_{\varepsilon_{\mu}} \\
\vertexlabel_{p} fuA\\
} & \to &

  -\, i\, \mathcal{M} =
  \bar{u} (p')\: (i e \gamma^{\mu})\: u (p)
  \cdot e_{\mu}, \\

\vspace{2.5mm}
2) & \mbox{\rm Emission of dilepton:} &

\Diagram{\vertexlabel^{p'} & & & \vertexlabel^{e^{+}} \\
fdV & & fu\\
& g & \\
\vertexlabel_{p} fuA & & fd \\
  & & & & \vertexlabel^{e^{-}}
}  & \to &

  -\, i\, \mathcal{M} =
  \bar{u} (p')\: (i e \gamma^{\mu})\: u (p) \cdot
  \Bigl( -\, \displaystyle\frac{i\, g_{\mu\nu}}{q^{2}} \Bigr) \cdot
  \bar{u} (k')\: (i e \gamma^{\nu})\: u (k).
\end{array}
\label{eq.3.1.1}
\end{equation}
Here,
$u$ is bispinor for nucleon or electron,
% $\varepsilon_{\mu}^{\lambda}$ are vectors of polarization of photon.
$e_{\mu}$ are vectors of polarization of photon.
In the second line we use virtual photon (instead of real photon in the first line) and production of pair $e^{-} e^{+}$ from this photon.
To draw graph and write down the matrix element for each process, we use Feynman rules for QED
(see Tabl.~6.2 in book~\cite{Halzen.book.1987}, p.~183).

Process indicated in the first line is used in problem of bremsstrahlung emission of (real) photon in the scattering of proton (in beam) on nucleus (in target) consisted from nucleons. There is detailed formalism describing this process.
So, it needs to understand how to construct additional new part which allows to include process indicated in the second line to the problem of scattering of proton on nucleus.

% \vspace{3mm}
% \noindent
% \textcolor[rgb]{1.00,0.00,0.00}{\textbf{Conclusion (from analysis above):}}

% \noindent
% At first, we need to resolve problem with one nucleon:
% To construct formalism of production of pair of leptons [line 2 in Eq.~(\ref{eq.3.1.1})] on the basis of known formalism for emission of real photon by nucleon
% [line 1 in Eq.~(\ref{eq.3.1.1})].
%-----------------------------------------------------------------------------------------------------------------------

%-----------------------------------------------------------------------------------------------------------------------
% \section{Formalism of $S$-matrix
% \label{sec.4}}

\subsection{Matrix elements of production of dileptons via virtual photon
and generalization of scattering of proton on nucleus with number of nucleons
\label{sec.4.1}}

We will construct formalism describing production of dileptons from nucleon-nucleon scattering via emission of virtual photon.
% \textcolor[rgb]{1.00,0.00,0.00}{\textbf{%
We have the process shown in the first line (for emission of real photon) in Eq.~(\ref{eq.3.1.1}), and
we have to develop formalism and describe process shown in the second line (for production of leptons pair) in Eq.~(\ref{eq.3.1.1}).
%
% \begin{equation}
% \begin{array}{cccllllll}
% \vspace{2.5mm}
%
% \Diagram{
% & \displaystyle{ gvA} \vertexlabel_{\varepsilon_{\mu}} \\
% \vertexlabel_{p} fuA & fdA \vertexlabel_{p'}\\
% } & \Longrightarrow &
%
% \Diagram{
% \vertexlabel^{e^{+}} fdA & fuA \vertexlabel^{e^{+}} \\
% & \displaystyle{ gvA} \vertexlabel_{\varepsilon_{\mu}} \\
% \vertexlabel_{p} fuA & fdA \vertexlabel_{p'}\\
% } \\
%
% \vspace{2.5mm}
% \mbox{\rm Emission of photon} & &
%
% \mbox{\rm Production of leptons pair}
% \end{array}
% \label{eq.4.1.1}
% \end{equation}
%
% But, in this our new development we should take the following into account.
In contrast to QED~\cite{Ahiezer.1981}, where wave functions for proton are proportional to plane waves,
we will take wave functions for proton from the problem of the proton-nucleus scattering, where nucleus is composed on nucleons, following to formalism of bremsstrahlung emission in nuclear processes~\cite{Maydanyuk.2012.PRC,Maydanyuk_Zhang.2015.PRC,Maydanyuk_Zhang_Zou.2018.PRC}.
By such a way, we will add properties of the proton-nucleus scattering (and structure of nucleus) to the studied process of production of lepton pair.

The simplest processes of interaction of fermions (electrons, nucleons, etc.) and photons are described by S-matrix of the 2-nd order.
In general, there is only 6 topologically different processes of the second order\cite{Ahiezer.1981} [see Fig.~3.3 in that book, p.~147].
% We show them in next figures below, following to book~\cite{Ahiezer.1981} [see Fig.~3.3 in that book, p.~147]:
%
% \begin{equation}
% \begin{array}{cccllllll}
% \vspace{5.5mm}
% 1)\;
% \Diagram{
% & \displaystyle{ gv} \\
% \vertexlabel_{p} fuA & fdA \vertexlabel_{p'}\\
% } \quad
%
% \Diagram{
% & \displaystyle{ gv} \\
% \vertexlabel_{p} fuA & fdA \vertexlabel_{p'}\\
% } & \quad
%
% 2)\;
% \Diagram{
% \vertexlabel^{e^{+}} fdA & fuA \vertexlabel^{e^{+}} \\
% & \displaystyle{ gv} \\
% \vertexlabel_{p} fuA & fdA \vertexlabel_{p'}
% } & \quad
%
% 3)\;
% \Diagram{
%   & \displaystyle{ gv} &    & \displaystyle{ gv}    \\
%                        & fA &                       \\
%   \vertexlabel_{p} fuA &    & & fdA \vertexlabel_{p'}
% } \\
%
% \vspace{5.5mm}
%   (\bar{\psi} \hat{A} \psi)\, (\bar{\psi} \hat{A} \psi) & \quad
%   (\bar{\psi} \hat{A}_{a} \psi)\, (\bar{\psi} \hat{A}_{a} \psi) & \quad
%   (\bar{\psi} \hat{A} \psi_{b})\, (\bar{\psi}_{b} \hat{A} \psi) \\
%
% \vspace{6.5mm}
% 4)\;
% \Diagram{
%   g f0  & fluA flV & f0 g \\
% } & \quad
%
% 5)\;
% \Diagram{
%   fs fA  f0  & fluA gl & f0 fA fs \\
% } & \quad
%
% 6)\;
% \Diagram{
%   g & fluA flV & g \\
% } \\
%
%   (\bar{\psi}_{c} \hat{A} \psi_{b})\, (\bar{\psi}_{b} \hat{A} \psi_{c}) & \quad
%   (\bar{\psi}_{c} \hat{A} \psi_{b})\, (\bar{\psi}_{b} \hat{A} \psi_{c}) & \quad
%   (\bar{\psi}_{c} \hat{A} \psi_{b})\, (\bar{\psi}_{b} \hat{A} \psi_{c})
%
% \mbox{\rm Production of leptons pair}
% \end{array}
% \label{eq.4.1.2}
% \end{equation}
%
% In figures we indicate the symbolic formulas under integral of S-matrix (without symbol $N$ of normal multiplication).
%
According to Eq.~(3.2.15) in Ref.~\cite{Ahiezer.1981},
% [see p.~142, 222 in that book],
all these processes are described by S-matrix of the second order in form:%
\footnote{To be closer to detailed formalism in Ref.~\cite{Ahiezer.1981},
formalism in Sect.~\ref{sec.4} is given in Euclidean metric of spacetime,
while in other Sections it is in Pseudo-Euclidean metric.
See Sect.~1.2.8 in Ref.~\cite{Ahiezer.1981}, p.~30--32 for connection between different formalisms.}
\begin{equation}
\begin{array}{llll}
  S^{(2)} & = &
  \displaystyle\frac{e^{2}}{2}\,
  \displaystyle\int
    {\mathcal T}\, \Bigl\{
      {\mathcal N} \bigl[ \bar{\psi} (x_{1})\, \hat{A} (x_{1})\, \psi(x_{1}) \bigr]_{\rm p}\;
      {\mathcal N} \bigl[ \bar{\psi} (x_{2})\, \hat{A} (x_{2})\, \psi(x_{2}) \bigr]_{\rm e}\,
    \Bigr\}\; d^{4}x_{1}\: d^{4}x_{2}, &
  \hat{A} (x) = \gamma_{\mu}\, A_{\mu} (x),
\end{array}
\label{eq.4.1.3}
\end{equation}
where
$\psi(x_{i})$ is operator wave functions for nucleon or electron ($i = 1,2$),
$\bar{\psi}(x_{i}) = \psi^{+}(x_{i})\, \gamma_{4}$,
$A (x_{j})$ is operator wave function of photon ($j = 1,2$).
Symbol $\mathcal T$ denotes the time-order product (or chronological product) of operators
[$\mathcal T$ is often called the ``time-ordering operator'',
for example, see Eq.~(2.3.20) in Ref.~\cite{Ahiezer.1981}, p.~99] and
symbol $\mathcal N$ denotes the normal-order product (or normal product) of operators
[for example, see Eq.~(2.3.21) in Ref.~\cite{Ahiezer.1981}, p.~99].
In order to simplify this formula, one can apply $2^{\rm nd}$ theorem of Wick
[for example, see p.~142 in Ref.~\cite{Ahiezer.1981}].
Following to it, mixed $\mathcal T$--product of operators of fields equals to sum of different $\mathcal N$--products,
where all operators are connected by all possible contractions, % connections,
with exception of contraction % connections
between operators within any given $\mathcal N$--product.
% Using such a theorem, one can finally obtain scheme of graphs in~(\ref{eq.4.1.2}).

% Production of leptons pair corresponds to graph (2) in (\ref{eq.4.1.2}) above.
According to Eq.~(3.2.15) in Ref.~\cite{Ahiezer.1981} [see p.~142, 222 in that book], we write down for production of leptons pair
\begin{equation}
\begin{array}{llll}
  S^{(2)} & = &
  \displaystyle\frac{e^{2}}{2}\,
  \displaystyle\int
    {\mathcal N} \Bigl\{
        \bigl[ \bar{\psi} (x_{1})\, \hat{A}^{a} (x_{1})\, \psi(x_{1}) \bigr]_{\rm p}\;
        \bigl[ \bar{\psi} (x_{2})\, \hat{A}^{a} (x_{2})\, \psi(x_{2}) \bigr]_{\rm e}
    \Bigr\}\; d^{4}x_{1}\: d^{4}x_{2}.
\end{array}
\label{eq.4.1.4}
\end{equation}
%-----------------------------------------------------------------------------------------------------------------------

%-----------------------------------------------------------------------------------------------------------------------
\subsection{Connection of $S$-matrix formalism (\ref{eq.4.1.4}) with QFT formalism in Sect.~\ref{sec.app.1.DIS_leptonic} for production of dileptons
\label{sec.4.2}}

We will connect $S$-matrix formalism above with Eq.~(\ref{eq.4.1.4}) for production of dileptons
and the second formula in Eqs.~(\ref{eq.3.1.1}).
We take into account
contraction
between operator wave functions of photon [see Ref.~\cite{Ahiezer.1981}, Eqs.~(2.3.24), p.~99],
which after calculations has no operators of creation and annihilation for photon,
i.e. it is $c$-number
[see Ref.~\cite{Ahiezer.1981}, Eqs.~(2.3.25), (2.3.26), p.~100]
\begin{equation}
\begin{array}{llll}
  % \contraction{}{A}{B}{C}
  A_{\mu}^{a} (x)\, A_{\nu}^{a} (x') =
  \displaystyle\frac{1}{2V}\: D_{c} (x - x')\: \delta_{\mu\nu}, & % \\

   D_{c} (x) =
   \displaystyle\frac{1}{2\, (2\pi)^{3}}
   \displaystyle\int
     e^{i\, (\vb{kr} - w_{\rm ph} |t|)}
     \displaystyle\frac{\vb{d^{3}k}}{w_{\rm ph}}, &

   w_{\rm ph} = |\vb{k}|.
\end{array}
\label{eq.4.2.1}
\end{equation}
Then Eq.~(\ref{eq.4.1.4}) is rewritten as
\begin{equation}
\begin{array}{llll}
  S^{(2)} & = &
%   \displaystyle\frac{e^{2}}{2}\,
%   \displaystyle\int
%       N \Bigl(
%           \bigl[ \bar{\psi} (x_{1})\, \gamma_{\mu} A_{\mu}^{a} (x_{1})\, \psi(x_{1}) \bigr]_{\rm n}\;
%           \bigl[ \bar{\psi} (x_{2})\, \gamma_{\nu} A_{\nu}^{a} (x_{2})\, \psi(x_{2}) \bigr]_{\rm e}
%       \Bigr)\,
%     \Bigr\}\; d^{4}x_{1}\: d^{4}x_{2} = \\

%   & = &
%   \displaystyle\frac{e^{2}}{2}\,
%   N
%   \displaystyle\int
%     \Bigl(
%       \bar{\psi}_{\rm n} (x_{1})\, \gamma_{\mu}\, \psi_{\rm n} (x_{1}) \cdot
%       \displaystyle\frac{1}{2V}\: D_{c} (x_{1} - x_{2})\: \delta_{\mu\nu} \cdot
%       \bar{\psi}_{\rm e} (x_{2})\, \gamma_{\nu}\, \psi_{\rm e} (x_{2})
%     \Bigr\}\; d^{4}x_{1}\: d^{4}x_{2} = \\

%   & = &
  \displaystyle\frac{e^{2}}{2}\,
   {\mathcal N}
  \displaystyle\int
    \Bigl(
      \bar{\psi}_{\rm n} (x_{1})\, \gamma_{\mu}\, \psi_{\rm n} (x_{1}) \cdot
      \displaystyle\frac{1}{2V}\: D_{c} (x_{1} - x_{2}) \cdot
      \bar{\psi}_{\rm e} (x_{2})\, \gamma_{\mu}\, \psi_{\rm e} (x_{2})
    \Bigr\}\; d^{4}x_{1}\: d^{4}x_{2}.
\end{array}
\label{eq.4.2.2}
\end{equation}
This formula is operator for production of dileptons given in space representation.
% As a next step, we will calculate matrix element on the basis of such an operator in momentum representation.

In order to obtain matrix element from operator in Eq.~(\ref{eq.4.2.2}), one can change operators for fermion on wave functions of fermion, following to logic
in Ref.~~\cite{Ahiezer.1981} [see p.~222 in that book].
We have the following wave functions for nucleon and electron as
[see Eq.~(1.1.23) in p.~15,
Eq.~(4.2.2) in p.~223, Ref.~\cite{Ahiezer.1981}]
\begin{equation}
\begin{array}{llllllll}
  \psi_{p\, a} (x)         = \displaystyle\frac{1}{\sqrt{2\,V\, \varepsilon}}\; u_{a}\,(+p)\, e^{ipx}, &
  \bar{\psi}_{p\, a} (x) = \displaystyle\frac{1}{\sqrt{2\,V\, \varepsilon}}\; \bar{u}_{a}\,(-p)\, e^{-ipx}, &
  a = 1,2,
%   A_{\mu}^{\lambda} (x)    = \displaystyle\frac{1}{\sqrt{2\, w}}\; \varepsilon_{\mu}^{\lambda}\, e^{ikx},
\end{array}
\label{eq.4.2.3}
\end{equation}
where $u$ is bispinor for nucleon or electron.
In momentum representation (after Fourier transformation) we obtain
[see Ref.~\cite{Ahiezer.1981}, Eqs.~(2.3.30), p.~101]
\begin{equation}
\begin{array}{llll}
  D_{c} (k) & \sim & \displaystyle\frac{1}{k^{2}}\,
\end{array}
\label{eq.4.2.4}
\end{equation}
and matrix element from Eq.~(\ref{eq.4.2.2}) is written as
\begin{equation}
\begin{array}{llll}
  \langle f \bigl|\, S^{(2)} \bigr|\, i \rangle & \sim &
  \bar{u} (p)\, (ie\gamma_{\mu})\, u (p) \cdot
  \displaystyle\frac{g_{\mu\nu}}{k^{2}} \cdot
  \bar{u} (k)\, (ie\gamma_{\nu})\, u (k).
\end{array}
\label{eq.4.2.5}
\end{equation}
One can see that this formula up to proportionality coincides with matrix element
in Eq.~(\ref{eq.3.1.1}) for production of dileptons
[see also Eq.~(\ref{eq.3.2.2})].
%-----------------------------------------------------------------------------------------------------------------------
%
%-----------------------------------------------------------------------------------------------------------------------
% \subsection{Formalism with explicit formulas for fermions and photons
% \label{sec.4.3}}
%
% Let us rewrite formula (\ref{eq.4.1.4}) above via explicit formulas for wave functions for fermion and photon.
We have the following wave function for photon
[see Eq.~(2.1.7) in p.~83, Ref.~\cite{Ahiezer.1981}]
\begin{equation}
\begin{array}{llllllll}
%   \psi_{p\, a} (x)         = \displaystyle\frac{1}{\sqrt{2\,V\, \varepsilon}}\; u_{a}\,(+p)\, e^{ipx}, &
%   \bar{\psi}_{p\, a} (x) = \displaystyle\frac{1}{\sqrt{2\,V\, \varepsilon}}\; \bar{u}_{a}\,(-p)\, e^{-ipx}, &
  A_{\mu}^{\lambda} (x)    = \displaystyle\frac{1}{\sqrt{2\, w\, V}}\; e_{\mu}^{\lambda}\, e^{ikx},
\end{array}
\label{eq.4.3.1}
\end{equation}
where
% $u$ is bispinor for nucleon or electron,
% $\varepsilon_{\mu}^{\lambda}$ are vectors of polarization of photon.
$e_{\mu}^{\lambda}$ are vectors of polarization of photon.
Then Eq.~(\ref{eq.4.1.4}) is rewritten as ($V=1$)
\begin{equation}
\begin{array}{llll}
% \vspace{0.5mm}
  S^{(2)} & = &
%   \displaystyle\frac{e^{2}}{2}\,
%   \displaystyle\int
%     N \Bigl\{
%           \bigl[ \bar{\psi}_{p'} (x_{1})\, \hat{A}_{a} (x_{1})\, \psi_{p} (x_{1}) \bigr]_{\rm n}\;
%           \bigl[ \bar{\psi}_{k'} (x_{2})\, \hat{A}_{a} (x_{2})\, \psi_{k }(x_{2}) \bigr]_{\rm e}\,
%     \Bigr\}\; d^{4}x_{1}\: d^{4}x_{2} = \\

% \vspace{0.5mm}
%   & = &
  \displaystyle\frac{e^{2}}{2}\,
   {\mathcal N} \,
  \displaystyle\int
    \Bigl\{
      \Bigl(
          \bigl[
            \displaystyle\frac{1}{\sqrt{2\, \varepsilon}}\; \bar{u}_{p'}\, e^{ip'x_{1}}
          % \bar{\psi}_{p'} (x_{1})\,
          \gamma_{\mu}\, A_{\mu}^{a} (x_{1})\,
          % \psi_{p} (x_{1})
          \displaystyle\frac{1}{\sqrt{2\, \varepsilon}}\; u_{p}\, e^{ipx_{1}}
          \bigr]_{\rm n}\; % \times \\
% \vspace{0.5mm}
%   & \times &
          \bigl[
            % \bar{\psi}_{p'} (x_{2})\,
            \displaystyle\frac{1}{\sqrt{2\, \varepsilon}}\; \bar{u}_{k'}\, e^{ik'x_{2}}\,
            \gamma_{\nu}\, A_{\nu}^{a} (x_{2})\,
            % \psi_{p}(x_{2})
            \displaystyle\frac{1}{\sqrt{2\, \varepsilon}}\; u_{k}\, e^{ikx_{2}}
          \bigr]_{\rm e}
      \Bigr)\,
    \Bigr\}\; d^{4}x_{1}\: d^{4}x_{2}.
\end{array}
\label{eq.4.3.2}
\end{equation}
%-----------------------------------------------------------------------------------------------------------------------

%-----------------------------------------------------------------------------------------------------------------------
\subsection{Rewritting S-matrix via fluxes for fermion
\label{sec.4.4}}

In QED, S-matrix can be rewritten via fluxes for fermions.
S-matrix can be expanded over components with different orders as
[see Eqs.~(3.2.11), (3.2.12), Ref.~\cite{Ahiezer.1981}, p.~140]
\begin{equation}
\begin{array}{lllllll}
% \vspace{0.5mm}
  S & = &
  \displaystyle\sum\limits_{n=0}^{+\infty} S^{(n)}, &

  S^{(n)} & = &
  \displaystyle\frac{i^{n}}{n!}\,
  \displaystyle\int
    d^{4}x_{1}\: d^{4}x_{2} \ldots d^{4}x_{n}\;
    T\, \Bigl\{ L_{I}(x_{1})\; L_{I}(x_{2}) \ldots L_{I}(x_{n}) \Bigr\},
\end{array}
\label{eq.4.4.1}
\end{equation}
where
$L_{I}$ is Lagrangian part of interaction in QED having form
[see Eq.~(3.2.6), Ref.~\cite{Ahiezer.1981}, p.~138,
Eq.~(3.1.12), Ref.~\cite{Ahiezer.1981}, p.~126]
\begin{equation}
\begin{array}{llll}
  L_{I} (x) = j_{\mu}(x)\, A_{\mu}(x), &
  j_{\mu} (x) = \displaystyle\frac{i\, e}{2}\, \bigl[ \bar{\psi} (x),\, \gamma_{\mu}\, \psi (x) \bigr].
\end{array}
\label{eq.4.4.2}
\end{equation}
Here,
$j_{\mu}$ is flux of fermion,
$A_{\mu}$ is electromagnetic field.
In the final form we obtain:
\begin{equation}
\begin{array}{llll}
  S^{(n)} & = &
  \displaystyle\frac{i^{n}}{n!}\,
  \displaystyle\int
    d^{4}x_{1}\: d^{4}x_{2} \ldots d^{4}x_{n}\;
    {\mathcal T}\, \Bigl\{ j_{\mu}(x_{1})\, A_{\mu}(x_{1})\; j_{\mu}(x_{2})\, A_{\mu}(x_{2}) \ldots j_{\mu}(x_{n})\, A_{\mu}(x_{n}) \Bigr\}.
\end{array}
\label{eq.4.4.3}
\end{equation}

The Lagrangian of interaction (\ref{eq.4.4.2}) can be rewritten via wave functions of fermion and electromagnetic field, using normal multiplication as
[see Ref.~\cite{Ahiezer.1981}, p.~142]
\begin{equation}
\begin{array}{llll}
  L_{I} (x) = j_{\mu}(x)\, A_{\mu}(x) =
  ie\, {\mathcal N}\, \bigl[ \bar{\psi}_{p'}(x)\, \hat{A}(x)\, \psi_{p} (x) \bigr], &
  \hat{A} (x) = \gamma_{\mu}\, A_{\mu} (x).
\end{array}
\label{eq.4.4.4}
\end{equation}
Then component of S-matrix of order $n$ is
\begin{equation}
\begin{array}{llll}
  S^{(n)} & = &
  \displaystyle\frac{(-1)^{n} e^{n}}{n!}\,
  \displaystyle\int
    d^{4}x_{1}\: d^{4}x_{2} \ldots d^{4}x_{n}\;
    {\mathcal T}\, \Bigl\{
      {\mathcal N}\, \bigl[ \bar{\psi}(x_{1})\, \hat{A}(x_{1})\, \psi (x_{1}) \bigr]\;
      {\mathcal N}\, \bigl[ \bar{\psi}(x_{2})\, \hat{A}(x_{2})\, \psi (x_{2}) \bigr] \ldots
      {\mathcal N}\, \bigl[ \bar{\psi}(x_{n})\, \hat{A}(x_{n})\, \psi (x_{n}) \bigr] \Bigr\}.
\end{array}
\label{eq.4.4.5}
\end{equation}
In particular, for S-matrix of the second order we obtain:
\begin{equation}
\begin{array}{llll}
\vspace{1.0mm}
  S^{(2)} & = &
  \displaystyle\frac{e^{2}}{2}\,
  \displaystyle\int
    d^{4}x_{1}\, d^{4}x_{2}\;
    {\mathcal T}\, \Bigl\{
      {\mathcal N}\, \bigl[ \bar{\psi}(x_{1})\, \hat{A}(x_{1})\, \psi (x_{1}) \bigr]\;
      {\mathcal N}\, \bigl[ \bar{\psi}(x_{2})\, \hat{A}(x_{2})\, \psi (x_{2}) \bigr] \Bigr\} = \\

  & = &
  \displaystyle\frac{- 1}{2}\,
  \displaystyle\int
    d^{4}x_{1}\, d^{4}x_{2}\;
    {\mathcal T}\, \Bigl\{ j_{\mu}(x_{1})\, A_{\mu}(x_{1})\; j_{\mu}(x_{2})\, A_{\mu}(x_{2}) \Bigr\}.
\end{array}
\label{eq.4.4.6}
\end{equation}
One can see that the first line in this equation coincides with Eq.~(\ref{eq.4.1.3}) completely.
The second line in this equation represents formula with fluxes for fermions.

For production of leptons pair we should take into account only connection between two wave functions of virtual photon
[as it is indicated in Eq.~(\ref{eq.4.2.1}) and should correspond to process (2) in Eqs.~(\ref{eq.3.1.1})].
Using Wick's theorem, mixed $T$-multiplication is transformed to $N$-multiplication with the chosen process:
\begin{equation}
\begin{array}{llll}
\vspace{1.0mm}
  S^{(2)} & = &
  \displaystyle\frac{e^{2}}{2}\,
  \displaystyle\int
    d^{4}x_{1}\, d^{4}x_{2}\;
   {\mathcal N}\, \Bigl\{
      \bar{\psi}_{\rm p}(x_{1})\, \hat{A}^{a}(x_{1})\, \psi_{\rm p} (x_{1})\;
      \bar{\psi}_{\rm e}(x_{2})\, \hat{A}^{a}(x_{2})\, \psi_{\rm e} (x_{2}) \Bigr\} = \\

  & = &
  \displaystyle\frac{- 1}{2}\,
  \displaystyle\int
    d^{4}x_{1}\, d^{4}x_{2}\;
    {\mathcal T}\, \Bigl\{ j_{\mu}(x_{1})\, A_{\mu}^{a}(x_{1})\; j_{\nu}(x_{2})\, A_{\nu}^{a}(x_{2}) \Bigr\}.
\end{array}
\label{eq.4.4.7}
\end{equation}
%-----------------------------------------------------------------------------------------------------------------------

%-----------------------------------------------------------------------------------------------------------------------
% \subsection{Transition from operators to matrix elements
% \label{sec.4.5}}

From formula of operator one can obtain expression for matrix element.
One can follow logic in Ref.~\cite{Ahiezer.1981} (see p.~223 in that book) where for proton it needs to change wave function for proton instead of its operator,
for photon it needs to change wave functions for photon instead the corresponding operators.
We obtain matrix element as
\begin{equation}
\begin{array}{lcl}
  \langle f\, |\, S^{(2)}\, |\, i \rangle & = &
  \displaystyle\frac{1}{2}\,
  \displaystyle\int
    (j_{\mu} (x_{1})\, A_{\mu}^{a} (x_{1}))_{\rm p} \cdot
    (j_{\nu} (x_{2})\, A_{\nu}^{a} (x_{2}))_{\rm e}\;
    d^{4}x_{1}\: d^{4}x_{2}.
\end{array}
\label{eq.4.5.1}
\end{equation}
%-----------------------------------------------------------------------------------------------------------------------

%-----------------------------------------------------------------------------------------------------------------------
\subsection{Connection with formalism for many-nucleon nuclear system in scattering
\label{sec.4.6}}

We will find way how to describe production of dileptons from system of nucleons in nuclear scattering.
Matrix element in form (\ref{eq.4.5.1}) can be rewritten as
%
% \begin{equation}
% \begin{array}{lcl}
%   \langle f\, |\, S^{(2)}\, |\, i \rangle & = &
%   \displaystyle\frac{1}{2}\,
%   \displaystyle\int
%     (j_{\mu} (x_{1})\, A_{\mu}^{a} (x_{1}))_{\rm p} \cdot
%     (j_{\nu} (x_{2})\, A_{\nu}^{a} (x_{2}))_{\rm e}\;
%     d^{4}x_{1}\: d^{4}x_{2}.
% \end{array}
% \label{eq.4.1.6.1}
% \end{equation}
%
% We calculate this expression:
%
\begin{equation}
\begin{array}{lcl}
  \langle f\, |\, S^{(2)}\, |\, i \rangle & = &
%   \displaystyle\frac{1}{2}\,
%   \displaystyle\int
%     (j_{\nu} (x_{1})\, A_{\nu}^{a} (x_{1}))_{\rm p} \cdot
%     (j_{\mu} (x_{2})\, A_{\mu}^{a} (x_{2}))_{\rm e}\;
%     d^{4}x_{1}\: d^{4}x_{2} = \\
%   & = &
  \displaystyle\frac{1}{2}\,
  \displaystyle\int
    (j_{\nu} (x_{1})\, A_{\nu}^{a} (x_{1}))_{\rm p}d^{4}\; x_{1} \cdot
  \displaystyle\int
    (j_{\mu} (x_{2})\, A_{\mu}^{a} (x_{2}))_{\rm e}\; d^{4}x_{2} =
    M_{\rm p}^{a} \cdot M_{\rm e}^{a},
\end{array}
\label{eq.4.6.1}
\end{equation}
where
\begin{equation}
\begin{array}{lll}
  M_{\rm p}^{a} =
  \displaystyle\frac{1}{\sqrt{2}}\,
  \displaystyle\int
    (j_{\nu} (x_{1})\, A_{\nu}^{a} (x_{1}))_{\rm p}\; d^{4} x_{1}, &
  M_{\rm e}^{a} =
  \displaystyle\frac{1}{\sqrt{2}}\,
  \displaystyle\int
    (j_{\mu} (x_{2})\, A_{\mu}^{a} (x_{2}))_{\rm e}\; d^{4} x_{2}.
\end{array}
\label{eq.4.6.2}
\end{equation}
Now we see that the first integral $M_{\rm p}^{a}$ represents matrix element of emission of photon from one nucleon.
%-----------------------------------------------------------------------------------------------------------------------

%-----------------------------------------------------------------------------------------------------------------------
% \subsubsection{Relativ. and non-relativ. forms of matrix element of emission of photon
% \label{sec.4.6.2}}

We summarize different formulations of matrix element of emission of photon
[see Eq.~(2.5.35) in p.~124, Eq.~(4.2.2) in p.~223 Ref.~\cite{Ahiezer.1981}].

\vspace{1.5mm}
1) Relativistic formulation for one nucleon
(for matrix element in form~(\ref{eq.4.6.1}), flux is simplied):
\begin{equation}
\begin{array}{llllllll}
\vspace{1.0mm}
  \begin{array}{llllllll}
  j_{\mu} (x) = ie\, (\bar{\psi}_{f}(x)\, \gamma_{\mu}\, \psi_{i} (x)), &
  \psi_{\alpha} (x) = \displaystyle\frac{1}{\sqrt{2\, \varepsilon}}\; u\, e^{ipx}, &
  A_{\mu}^{a} (x) = \displaystyle\frac{1}{\sqrt{2\, w_{\rm ph}}}\; e_{\mu}^{a}\, e^{ikx},
\end{array} \\

  M_{\rm p}^{a} =
  \displaystyle\frac{1}{\sqrt{2}}\,
  \displaystyle\int
    (j_{\mu} (x_{1})\, A_{\mu}^{a} (x_{1}))_{\rm n}\; d^{4} x_{1} =

  \displaystyle\frac{ie}{\sqrt{2}}\,
  \displaystyle\int
    \bigl[ \bar{\psi}_{f}(x_{1})\, \gamma_{\mu}\, \psi_{i} (x_{1}) \bigr]\, A_{\mu}^{a} (x_{1}))_{\rm n}\; d^{4} x_{1}.
\end{array}
\label{eq.4.6.2.1}
\end{equation}

2) Non-relativistic formulation for one nucleon (without magnetic moment of nucleon):
\begin{equation}
\begin{array}{llllllll}
\vspace{1.0mm}
  \begin{array}{llllllll}
  j_{\mu} (x) = e\, \displaystyle\frac{i\, \hbar}{2m_{\rm p}} \Bigl[ \psi_{f} (x) \nabla \psi_{i}^{*} (x) - c.c. \Bigr], &
  % \psi_{\alpha} (x) = \displaystyle\frac{1}{\sqrt{2\, \varepsilon}}\; u\, e^{ipx}, &
  A_{\mu}^{a} (x) = \displaystyle\frac{1}{\sqrt{2\, w_{\rm ph}}}\; e_{\mu}^{a}\, e^{ikx},
\end{array} \\

  M_{\rm p}^{a} =
  \displaystyle\frac{1}{\sqrt{2}}\,
  \displaystyle\int
    (j_{\mu} (x_{1})\, A_{\mu}^{a} (x_{1}))_{\rm n}\; d^{4} x_{1} =

  \displaystyle\frac{1}{\sqrt{2}}\,
  \displaystyle\frac{i\, e\, \hbar}{2m_{\rm p}}
  \displaystyle\int
    (\psi_{f} (x_{1}) \nabla \psi_{i}^{*} (x_{1})\, A_{\mu}^{a} (x_{1}) - c.c.)_{\rm n}\; d^{3} x_{1}.
\end{array}
\label{eq.4.6.2.2}
\end{equation}
% -----------------------------------------------------------------------------------------------------------------------

% -----------------------------------------------------------------------------------------------------------------------
3) Non-relativistic formulation for the proton-nucleus scattering where nucleus is composed on protons with number $Z$ and neutrons with number $N$ and magnetic moments of nucleons are included.
Calculation of such a matrix element is straightforward. Results and details of such calculations are given in App.~\ref{sec.app.2}.
% *******************************************************************************************************************

% *******************************************************************************************************************
\subsection{Connection of matrix element with one nucleon and matrix element with proton-nucleus system in scattering
\label{sec.4.8}}

To understand clearly solution of this problem with inclusion of many-nucleon nuclear system in scattering,
at first, let us study problem on the basis the coherent matrix elements.
After calculations [see Eqs.~(\ref{eq.app.2.1.2.10}), (\ref{eq.app.2.resultingformulas.1}), (\ref{eq.app.2.resultingformulas.3}),
App.~\ref{sec.app.2} for details, definitions, explanations],
we write down the final formulas of matrix element for many-nucleon nuclear system in scattering as
\begin{equation}
\begin{array}{lll}
  \langle \Psi_{f} |\, \hat{H}_{\gamma} |\, \Psi_{i} \rangle^{(a)} \;\; = \;\;
  \sqrt{\displaystyle\frac{2\pi\, c^{2}}{\hbar w_{\rm ph}}}\,
  M_{\rm full}^{(a)}, &
  M_{\rm full}^{(a)} = M_{p}^{(E), a},
\end{array}
\label{eq.4.8.1}
\end{equation}
\begin{equation}
\begin{array}{lll}
  M_{p}^{(E), a} \simeq
  M_{p}^{(E,\, {\rm mon},0), a} & = &
  i \hbar\, (2\pi)^{3} \displaystyle\frac{2\, \mu_{N}\,  m_{\rm p}}{\mu}\;
  Z_{\rm eff}^{\rm (mon,\, 0)}\;
  % \displaystyle\sum\limits_{\alpha=1,2}
    \vb{e}^{(a)} \cdot \vb{I}_{1}, \\
\end{array}
\label{eq.4.8.2}
\end{equation}
\begin{equation}
\begin{array}{lll}
  \vb{I}_{1} & = &
    \Bigl\langle\: \Phi_{{\rm p - nucl},\, f} (\vb{r})\; \Bigl|\, e^{-i\, \vb{k}_{\rm ph} \vb{r}}\; \vb{\displaystyle\frac{d}{dr}}
    \Bigr|\: \Phi_{{\rm p - nucl}, \, i} (\vb{r})\: \Bigr\rangle,
\end{array}
\label{eq.4.8.3}
\end{equation}
where
$\mu_{N} = e\hbar / (2m_{\rm p}c)$ is nuclear magneton,
$\mu = m_{\rm p} m_{A} / (m_{\rm p} + m_{A})$ is reduced mass of proton and nucleus,
$m_{\rm p}$ and $m_{A}$ are masses of proton and nucleus,
$Z_{\rm eff}^{\rm (mon,\, 0)}$ is effective electric charge of proton-nucleus system defined in Eq.~(\ref{eq.app.2.resultingformulas.5}),
$\vb{e}^{(\alpha)}$ is spacial parts of vectors of polarization of photon,
$\vb{k}_{\rm ph}$ is wave vector of photon,
$\vb{r}$ is vector of relative distance between the scattered proton and center-of-mass of nucleus-target,
$\Psi_{i}$ and $\Psi_{f}$ are the wave functions of the full nuclear system in states before emission of photon ($i$-state) and after such an emission ($f$-state)
(see Eqs.~(\ref{eq.app.2.1.2.3})--(\ref{eq.app.2.1.2.5}), details in App.~\ref{sec.app.2}),
$\Phi_{\rm p - nucl} (\vb{r})$ is the function describing relative motion between proton and nucleus in the scattering
(without description of internal relative motions of nucleons in nucleus).
Now, instead of the full matrix element of production of leptons pair from one nucleon as
\begin{equation}
\begin{array}{lcl}
  \langle f\, |\, S^{(2)}\, |\, i \rangle & = &

  \displaystyle\frac{1}{2}\,
  \displaystyle\int
    (j_{\nu} (x_{1})\, A_{\nu}^{a} (x_{1}))_{\rm p}d^{4}\; x_{1} \cdot
  \displaystyle\int
    (j_{\mu} (x_{2})\, A_{\mu}^{a} (x_{2}))_{\rm e}\; d^{4}x_{2} =
    M_{\rm p}^{a} \cdot M_{\rm e}^{a},
\end{array}
\label{eq.4.8.4}
\end{equation}
we find new full matrix element of production of leptons pair based on proton-nucleus scattering where nucleus is composed on
protons with number $Z$ and neutrons with number $N$:
\begin{equation}
\begin{array}{lcl}
  \langle f\, |\, \hat{O}\, |\, i \rangle & = &
  \displaystyle\frac{1}{2}\,
   \langle \Psi_{f} |\, \hat{H}_{\gamma} |\, \Psi_{i} \rangle^{a} \cdot
  \displaystyle\int
    (j_{\mu} (x_{2})\, A_{\mu}^{a} (x_{2}))_{\rm e}\; d^{4}x_{2} =
    \displaystyle\frac{1}{\sqrt{2}}\,
   \langle \Psi_{f} |\, \hat{H}_{\gamma} |\, \Psi_{i} \rangle^{a}
   \cdot M_{\rm e}^{a}.
\end{array}
\label{eq.4.8.5}
\end{equation}
%
% \textcolor[rgb]{1.00,0.00,0.00}{\textbf{%
Calculations of integral (\ref{eq.4.8.3}) are straightforward
[see Eqs.~(56), (59) in Ref.~\cite{Maydanyuk_Vasilevsky.2023.fold.arXiv}]:
% Appendix~\ref{sec.app.2}, TO CHECK!!!]. % [see Eqs.~(), Appendix~, p.~ ].
% }}
%
\begin{equation}
\begin{array}{llllll}
\vspace{0.5mm}
  \vb{I}_{1} =
  \vb{e}^{(2)} \cdot i\, \displaystyle\frac{\sqrt{3}}{6}\,
  \Bigl\{
    J_{1}(0,1,0) -
    \displaystyle\frac{47}{40} \sqrt{\displaystyle\frac{1}{2}} \cdot J_{1}(0,1,2)
  \Bigr\}, \\

  \vb{e}^{(1)} \cdot \vb{I}_{1} = 0, \\

  \vb{e}^{(2)} \cdot \vb{I}_{1} =
  \displaystyle\sum\limits_{\alpha=1,2} \vb{e}^{(\alpha)} \cdot \vb{I}_{1} =
  i\, \displaystyle\frac{\sqrt{3}}{6}\,
  \Bigl\{
    J_{1}(0,1,0) -
    \displaystyle\frac{47}{40} \sqrt{\displaystyle\frac{1}{2}} \cdot J_{1}(0,1,2)
  \Bigr\},
\end{array}
% \label{eq.app.integrals.3}
\label{eq.4.8.6}
% \label{eq.4.6.8}
\end{equation}
where
$J_{1}(0,1,0)$ and $J_{1}(0,1,2)$ are the radial integrals defined in Eqs.~(\ref{eq.app.2.multiple.3})
and is subject of numeric integration in this problem
(which gives properties of the proton-nucleus scattering in study of dilepton production).

% \textcolor[rgb]{1.00,0.00,0.00}{\textbf{%
Note that integral $\vb{I}_{1}$ is calculated as summation over vectors $\xibf_{- 1}$ and $\xibf_{+ 1}$ of circular polarization of the emitted photon
(see Eq.~(56), App.~C in Ref.~\cite{Maydanyuk_Vasilevsky.2023.fold.arXiv}).
Following to Ref.~\cite{Maydanyuk_Vasilevsky.2023.fold.arXiv} [see Eq.~(C1) in that paper],
this summation is expressed via only one vector $\vb{e^{(2)}}$ of linear polarization.
By such a reason, integral  $\vb{I}_{1}$ is parallel to vector $\vb{e^{(2)}}$ and we obtain formulas (\ref{eq.4.8.6}).
% }}
%
Using such solutions,
from Eqs.~(\ref{eq.4.8.1})--(\ref{eq.4.8.3}) we rewrite the matrix element as
($\hbar=1$, $c=1$)
\begin{equation}
\begin{array}{rllll}
  \langle \Psi_{f} |\, \hat{H}_{\gamma} |\, \Psi_{i} \rangle^{(a)} =
  \vb{e}^{(a)}\, \vb{e}^{(2)} \cdot f, &
  f =
  -\, \sqrt{\displaystyle\frac{2\pi}{3 \hbar w_{\rm ph}}}\,
  (2\pi)^{3} \displaystyle\frac{\mu_{N}\,  m_{\rm p}}{\mu}\:
  Z_{\rm eff}^{\rm (mon,\, 0)}\,
  \Bigl\{
    J_{1}(0,1,0) -
    \displaystyle\frac{47}{40} \sqrt{\displaystyle\frac{1}{2}} \cdot J_{1}(0,1,2)
  \Bigr\}.
\end{array}
\label{eq.4.8.8}
% \label{eq.resultingformulas.1}
% \label{eq.result.formulas.3}
\end{equation}
In case of just proton scattering in field of another proton (instead of nucleus) we have
% (TO CHECK FORMULA!!!)
% }}
%
\begin{equation}
\begin{array}{llllllll}
\vspace{0.5mm}
  M_{\rm p}^{a} =
%   \varepsilon_{\mu}^{a, *} \varepsilon_{\mu}^{(2)}\; f, &
  e_{\mu}^{a, *} e_{\mu}^{(2)}\; f, &

  f =
  -\; \displaystyle\frac{e\, \hbar\, \sqrt{3}}{24m_{\rm p}\, \sqrt{w_{\rm ph}}}\;
  \Bigl\{
    J_{1}(0,1,0) - \displaystyle\frac{47}{40} \sqrt{\displaystyle\frac{1}{2}} \cdot J_{1}(0,1,2)
  \Bigr\} .
\end{array}
\label{eq.4.8.9}
% \label{eq.4.6.10}
\end{equation}
%-----------------------------------------------------------------------------------------------------------------------

%-----------------------------------------------------------------------------------------------------------------------
\section{Case of proton-nucleon scattering
\label{sec.6-pp}}

\subsection{Separation of tensor associated with production of leptons pair in square of $S$-matrix element
\label{sec.6-pp.1}}

We write down square of the matrix element (\ref{eq.4.6.1}):
\begin{equation}
\begin{array}{lcl}
  \Bigl| \langle f\, |\, S^{(2)}\, |\, i \rangle \Bigr|^{2} & = &
  \Bigl| M_{\rm p}^{a} \cdot M_{\rm e}^{a} \Bigr|^{2} =
   M_{\rm p}^{a} \cdot M_{\rm e}^{a} \cdot M_{\rm p}^{b, *} \cdot M_{\rm e}^{b, *}.
\end{array}
\label{eq.6-pp.1.1}
\end{equation}
Let's calculate term associated for leptons.
%
% \begin{equation}
% \begin{array}{lcl}
%    M_{\rm lep}^{a} \cdot M_{\rm lep}^{b, *}.
% \end{array}
% \label{eq.6-pp.1.2}
% \end{equation}
%
We write down
\begin{equation}
\begin{array}{lllll}
\vspace{1.5mm}
   M_{\rm lep}^{a} \cdot M_{\rm lep}^{b, *} & = &

%   \displaystyle\frac{1}{\sqrt{2}}\,
%   \displaystyle\int
%     (j_{\mu} (x_{2})\, A_{\mu}^{a} (x_{2}))_{\rm e}\; d^{4} x_{2}\,
%   \Bigl\{
%     \displaystyle\frac{1}{\sqrt{2}}\,
%     \displaystyle\int
%       (j_{\nu} (x'_{2})\, A_{\nu}^{b} (x'_{2}))_{\rm e}\; d^{4} x'_{2}
%   \Bigr\}^{*}\; = \\

%  & = &
  \displaystyle\frac{1}{2}\,
  \displaystyle\int
    (j_{\mu} (x_{2})\, A_{\mu}^{a} (x_{2}))_{\rm e}\; d^{4} x_{2}\,
  \displaystyle\int
    (j_{\nu}^{*} (x'_{2})\, A_{\nu}^{b,\, *} (x'_{2}))_{\rm e}\; d^{4} x'_{2}.
\end{array}
\label{eq.6-pp.1.3}
\end{equation}
We use explicit formulas for wave functions for leptons
[see Eqs.~(\ref{eq.4.2.3}), we use $V=1$, $\varepsilon$ is energy of lepton]
\begin{equation}
\begin{array}{llllllll}
%   j_{\mu} (x) = ie\, N (\bar{\psi}(x)\, \gamma_{\mu}\, \psi (x)), &
  \psi_{k\, a} (x)         = \displaystyle\frac{1}{\sqrt{2\,V\, \varepsilon}}\; u_{a}\,(+k)\, e^{ikx}, &
  \bar{\psi}_{k\, a} (x) = \displaystyle\frac{1}{\sqrt{2\,V\, \varepsilon}}\; \bar{u}_{a}\,(-k)\, e^{-ikx}, &
%   \psi_{\alpha} (x) = \displaystyle\frac{1}{\sqrt{2\, \varepsilon}}\; u\, e^{ipx}, &
%   A_{\mu}^{a} (x) = \displaystyle\frac{1}{\sqrt{2\, w}}\; \varepsilon_{\mu}^{a}\, e^{ikx},
\end{array}
\label{eq.6-pp.1.4}
\end{equation}
and calculate corresponding flux as
\begin{equation}
\begin{array}{llllllll}
%  j_{\mu} (x) & = &
  j_{\mu} (x) & = & ie\, (\bar{\psi}_{k'}(x)\, \gamma_{\mu}\, \psi_{k} (x)) =
%   ie\,
%   \Bigl(
%     \displaystyle\frac{1}{\sqrt{2\, \varepsilon}}\; \bar{u}_{k'}\, e^{-ik'x}\,
%     \gamma_{\mu}\, \displaystyle\frac{1}{\sqrt{2\, \varepsilon}}\; u_{k}\, e^{ikx}
%  \Bigr) = % \\

  \displaystyle\frac{ie}{2\, \varepsilon}\,
  \Bigl( \bar{u}\, \gamma_{\mu}\, u\, e^{-i\, (k'-k) x} \Bigr).
\end{array}
\label{eq.6-pp.1.5}
\end{equation}
Then Eq.~(\ref{eq.6-pp.1.3}) is transformed as
\begin{equation}
\begin{array}{lcl}
% \vspace{1.5mm}
   M_{\rm lep}^{a} \cdot M_{\rm lep}^{b, *} & = &
%   \displaystyle\frac{1}{2}\,
%   \displaystyle\int
%     \displaystyle\frac{ie}{2\, \varepsilon}\,
%     \Bigl( \bar{u}_{k'}\, \gamma_{\mu}\, u_{k}\, \Bigr)\,e^{-i\,(k'-k) x_{2})}
%     A_{\mu}^{a} (x_{2}))_{\rm e}\; d^{4} x_{2}\,
%   \displaystyle\int
%     \Bigl( \displaystyle\frac{ie}{2\, \varepsilon} \Bigr)^{*}\,
%     \Bigl( \bar{u}_{k'}\, \gamma_{\nu}\, u_{k}\, e^{-i\, (k'-k) x'_{2})} \Bigr)^{*}\,
%     A_{\nu}^{b,\, *} (x'_{2}))_{\rm e}\; d^{4} x'_{2} = \\
%
% \vspace{1.5mm}
%   & = &
%   \displaystyle\frac{1}{2}\,
%   \Bigl| \displaystyle\frac{ie}{2\, \varepsilon} \Bigr|^{2}\,
%   \displaystyle\int d^{4} x_{2}
%   \displaystyle\int d^{4} x'_{2}
%     \Bigl( \bar{u}_{k'}\, \gamma_{\mu}\, u_{k}\, \Bigr)
%     \Bigl( \bar{u}_{k'}\, \gamma_{\nu}\, u_{k} \Bigr)^{*}\;
%     A_{\mu}^{a} (x_{2})_{\rm e}\;
%     A_{\nu}^{b,\, *} (x'_{2})_{\rm e}\,
%       e^{-i\, (k'-k) x_{2}} e^{+i\, (k'-k) x'_{2}} = \\

% \vspace{1.5mm}
%   & = &
  \displaystyle\frac{1}{2}\,
  \Bigl| \displaystyle\frac{ie}{2\, \varepsilon} \Bigr|^{2}\,
  \displaystyle\int d^{4} x_{2}
  \displaystyle\int d^{4} x'_{2}
    \biggl\{ \Bigl[ \bar{u}_{k'}\, \gamma_{\mu}\, u_{k}\, \Bigr]\, \Bigl[ \bar{u}_{k'}\, \gamma_{\nu}\, u_{k} \Bigr]^{*} \biggr\}\;
    A_{\mu}^{a} (x_{2})_{\rm e}\;
    A_{\nu}^{b,\, *} (x'_{2})_{\rm e}\,
    e^{-i\, (k'-k) x_{2}} e^{+i\, (k'-k) x'_{2}}.
\end{array}
\label{eq.6-pp.1.6}
\end{equation}
From (\ref{eq.3.2.5}) we have
%
% \textcolor[rgb]{1.00,0.00,0.00}{\textbf{%
\begin{equation}
\begin{array}{llllllll}
  L_{\rm p}^{\mu\nu}\, =
  \displaystyle\frac{1}{2}\,
  \displaystyle\sum\limits_{\rm spin\, states\, for\, electron}
    \Bigl[\bar{u}_{C} (k')\, \gamma^{\mu}\, u_{A}(k) \Bigr]\,
    \Bigl[\bar{u}_{C} (k')\, \gamma^{\nu}\, u_{A}(k) \Bigr]^{*}.
\end{array}
\label{eq.6-pp.1.7}
\end{equation}
% }}
%
Summation in this formula is performed over all states of spinors $u$ and $\bar{u}$
(see (6.19)--(6.20) in book~\cite{Halzen.book.1987}, p.~152--153).
Then Eq.~(\ref{eq.6-pp.1.6}) can be rewritten as
\begin{equation}
\begin{array}{lcl}
% \vspace{1.5mm}
   M_{\rm lep}^{a} \cdot M_{\rm lep}^{b, *} & = &
%   \displaystyle\frac{e^{2}}{4\, \varepsilon^{2}}\,
%   \displaystyle\int d^{4} x_{2}\,
%   \displaystyle\int d^{4} x'_{2}\,
%     L_{\mu\nu}^{\rm lep}\;
%     A_{\mu}^{a} (x_{2})_{\rm e}\;
%     A_{\nu}^{b,\, *} (x'_{2})_{\rm e}\,
%     e^{-i\,(k'-k) x_{2})} e^{+i\, (k'-k) x'_{2})}\; = \\

%   & = &
  \displaystyle\frac{e^{2}}{4\, \varepsilon^{2}}\,
    L_{\mu\nu}^{\rm lep}\;
  \displaystyle\int d^{4} x_{2}\,
  \displaystyle\int d^{4} x'_{2}\,
    A_{\mu}^{a} (x_{2})_{\rm e}\;
    A_{\nu}^{b,\, *} (x'_{2})_{\rm e}\,
    e^{-i\, (k'-k) x_{2}} e^{+i\, (k'-k) x'_{2}}
\end{array}
\label{eq.6-pp.1.8}
\end{equation}
and $L_{\mu\nu}^{\rm lep}$ is calculated in Eq.~(\ref{eq.3.2.9}).
We substitute explicit form of wave function of photon (\ref{eq.4.3.1})
% (we use $V=1$)
%
% \begin{equation}
% \begin{array}{llllllll}
%   A_{\mu}^{a} (x) = \displaystyle\frac{1}{\sqrt{2\, w}}\; \varepsilon_{\mu}^{a}\, e^{ik_{\rm ph}x}
% \end{array}
% \label{eq.6-pp.1.9}
% \end{equation}
%
and obtain
\begin{equation}
\begin{array}{lcl}
\vspace{1.5mm}
   M_{\rm lep}^{a} \cdot M_{\rm lep}^{b, *} & = &
%   \displaystyle\frac{e^{2}}{4\, \varepsilon^{2}}\,
%     L_{\mu\nu}^{\rm lep}\;
%   \displaystyle\int d^{4} x_{2}
%   \displaystyle\int d^{4} x'_{2}\:
%     \displaystyle\frac{1}{\sqrt{2\, w}}\; \varepsilon_{\mu}^{a}\, e^{ik_{\rm ph}x_{2}}\,
%     \displaystyle\frac{1}{\sqrt{2\, w}}\; \varepsilon_{\nu}^{b,\, *}\, e^{-ik_{\rm ph}x'_{2}}\;
%     e^{-i\, (k'-k) x_{2}} e^{+i\, (k'-k) x'_{2}} = \\

% \vspace{1.5mm}
%   & = &
  \displaystyle\frac{e^{2}}{4\, \varepsilon^{2}}\;
    L_{\mu\nu}^{\rm lep}\;
  \displaystyle\frac{1}{2\, w_{\rm ph}}\; e_{\mu}^{a}\, e_{\nu}^{b,\, *}\,
  \displaystyle\int d^{4} x_{2}
  \displaystyle\int d^{4} x'_{2}\:
    e^{-i\, (k' - k - k_{\rm ph}) x_{2}} e^{+i\, (k' - k - k_{\rm ph}) x'_{2}}.
\end{array}
\label{eq.6-pp.1.10}
\end{equation}
Using definition of delta function:
\begin{equation}
\begin{array}{lcl}
% \vspace{1.5mm}
  \delta (t) =
  \displaystyle\frac{1}{2\pi}
  \displaystyle\int_{-\infty}^{+\infty}
    e^{iw_{\rm ph}t}\; dt, &

  \delta^{4} (k'-k - k_{\rm ph}) =
  \displaystyle\frac{1}{(2\pi)^{4}}
  \displaystyle\int_{-\infty}^{+\infty}
    e^{i(k' - k - k_{\rm ph})x}\; d^{4}x,
\end{array}
\label{eq.6-pp.1.11}
\end{equation}
we obtain
\begin{equation}
\begin{array}{lcl}
\vspace{1.5mm}
   M_{\rm lep}^{a} \cdot M_{\rm lep}^{b, *} & = &
%   \displaystyle\frac{e^{2}}{4\, \varepsilon^{2}}\;
%     L_{\mu\nu}^{\rm lep}\;
%   \displaystyle\frac{1}{2\, w}\; \varepsilon_{\mu}^{a}\, \varepsilon_{\nu}^{b,\, *}\;
%   (2\pi)^{8}\, \Bigl[ \delta^{4} (k' - k - k_{\rm ph}) \Bigr]^{2}\; = \\

% \vspace{1.5mm}
%   & = &
  \displaystyle\frac{e^{2}}{8\, \varepsilon^{2}\, w_{\rm ph}}\:
  (2\pi)^{8}\;
  L_{\mu\nu}^{\rm lep}\;
  e_{\mu}^{a}\, e_{\nu}^{b,\, *}\;
  \Bigl[ \delta^{4} (k' - k - k_{\rm ph}) \Bigr]^{2},
\end{array}
\label{eq.6-pp.1.12}
\end{equation}
where leptonic tensor is given in Eq.~(\ref{eq.3.2.9})
\begin{equation}
\begin{array}{llllllll}
% \vspace{1.5mm}
  L^{\rm lep}_{\mu\nu}\, & = &
  2\, \Bigl[ k'_{\mu}\, k_{\nu} + k'_{\nu}\, k_{\mu} - \bigl(k' \cdot k - m_{\rm lep}^{2} \bigr)\, g_{\mu\nu} \Bigr]
\end{array}
\label{eq.6-pp.1.13}
\end{equation}
and $m_{\rm lep}$ is mass of lepton.
%
% \textcolor[rgb]{0.00,0.00,1.00}{\textbf{%
Note that $\delta$-function is appeared in matrix elements (\ref{eq.6-pp.1.12}) in momentum representation after integration
[for example, see Ref.~\cite{Ahiezer.1981}, (4.2.3), p.~223].
This is general property of calculation of such matrix elements.
Ways how to work with such $\delta$-functions are described, for example, in Ref.~\cite{Ahiezer.1981} [see p.~225 in that book].
% }}
%-----------------------------------------------------------------------------------------------------------------------

%-----------------------------------------------------------------------------------------------------------------------
% \subsection{Calculation of square of $S$-matrix elements
% \label{sec.6-pp.2}}

Let us consider the solutions for the matrix multiplication of leptonic matrix elements in Eq.~(\ref{eq.6-pp.1.12}) and
nucleon matric element in Eq.~(\ref{eq.4.8.9}).
We substitute these solutions to the squared of $S$-matrix element in Eq.~(\ref{eq.6-pp.1.1}) and find
\begin{equation}
\begin{array}{rllll}
\vspace{1.5mm}
  \Bigl| \langle f\, |\, S^{(2)}\, |\, i \rangle \Bigr|^{2} & = &

  |f|^{2}\,
  \displaystyle\frac{e^{2}\, (2\pi)^{8}}{8\, \varepsilon^{2}\, w_{\rm ph}}\:
  L_{\mu\nu}^{\rm lep}\;
  e_{\mu}^{(2)}\, e_{\nu}^{(2),\, *}\;
  \Bigl[ \delta^{4} (k' - k - k_{\rm ph}) \Bigr]^{2}, \\

  f & = &
  -\; \displaystyle\frac{e\, \hbar\, \sqrt{3}}{24\, m_{\rm p}\, \sqrt{w_{\rm ph}}}\;
  \Bigl\{
    J_{1}(0,1,0) -
    \displaystyle\frac{47}{40} \sqrt{\displaystyle\frac{1}{2}} \cdot J_{1}(0,1,2)
  \Bigr\}.
\end{array}
\label{eq.6-pp.2.5}
\end{equation}
%-----------------------------------------------------------------------------------------------------------------------

%-----------------------------------------------------------------------------------------------------------------------
% % \newpage
\subsection{Calculation of leptonic tensor with vectors of polarizations
\label{sec.6-pp.3}}

We will find multiplication of leptonic tensor in form~(\ref{eq.6-pp.1.13}) on vectors of polarization
\begin{equation}
\begin{array}{llllllll}
% \vspace{1.5mm}
  L^{\rm lep}_{\mu\nu}\,
  e_{\mu}^{(2)}\,
  e_{\nu}^{(2),\, *}
%   & = &
%   2\, \Bigl[
%     (k'_{\mu}\, \varepsilon_{\mu}^{(2)}\,)\, ( k_{\nu}\, \varepsilon_{\nu}^{(2),\, *}) +
%     (k'_{\nu}\, \varepsilon_{\nu}^{(2),\, *})\, (k_{\mu}\, \varepsilon_{\mu}^{(2)}) -
%     \bigl(k' \cdot k - m_{\rm lep}^{2} \bigr) \Bigr]\; = \\

% \vspace{1.5mm}
%   & = &
%   2\, \Bigl[
%     (k'_{\mu}\, \varepsilon_{\mu}^{(2)}\,)\, ( k_{\nu}\, \varepsilon_{\nu}^{(2)}) +
%     (k'_{\nu}\, \varepsilon_{\nu}^{(2)})\, (k_{\mu}\, \varepsilon_{\mu}^{(2)}) -
%     \bigl(k' \cdot k - m_{\rm lep}^{2} \bigr) \Bigr]\; = \\

  & = &
  2\, \Bigl[
    2\, (k'_{\mu}\, e_{\mu}^{(2)}\,)\, ( k_{\nu}\, e_{\nu}^{(2)}) -
    \bigl(k' \cdot k - m_{\rm lep}^{2} \bigr) \Bigr],
\end{array}
\label{eq.6-pp.3.5}
\end{equation}
where we take into account properties of vectors of polarizations as
\begin{equation}
\begin{array}{llllllll}
% \vspace{1.5mm}
  e_{\mu}^{(2)}\, e_{\nu}^{(2),\, *}\; = \delta_{\mu\nu}, &
  \bigl| e_{\mu}^{(2)} \bigr|^{2} =
  e_{\mu}^{(2)} e_{\mu}^{(2),\, *} = 1, &
  e_{\mu}^{(2)}\, = e_{\mu}^{(2),\, *}.
\end{array}
\label{eq.6-pp.3.3}
\end{equation}
We calculate multiplication (we omit bottom indexes $f, i$ for leptons)
\begin{equation}
\begin{array}{llllllll}
% \vspace{1.5mm}
%   k\, e^{(2)} =
  k_{\mu}\, \vb{e}_{\mu}^{(2)} =
%   k_{0}\, \vb{e}_{0}^{(2)} - \vb{k}\, \vb{e}^{(2)} =
%   E_{e}\, e_{0}^{(2)} - \vb{k}_{\rm e}\, \vb{e}^{(2)} =
  \sqrt{m_{e}^{2} + k_{\rm e}^{2}}\, e_{0}^{(2)} - \vb{k}_{\rm e}\, \vb{e}^{(2)}.
\end{array}
\label{eq.6-pp.3.6}
\end{equation}
We take into account values of vectors of polarization as
% $e_{0}^{(i)} = 0$, $e_{0}^{(0)} = 1$,
% $e^{0}_{\nu} = \delta_{0 \nu}$,
% $i = 1,2,3$, $\nu = 0,1,2,3$
[see Ref.~\cite{Bogoliubov.1980}, p.~34]:
\begin{equation}
\begin{array}{llllllll}
% \vspace{1.5mm}
  e_{0}^{(i)} = 0, & e_{0}^{(0)} = 1, &
  e^{0}_{\nu} = \delta_{0 \nu}, &
  \vb{e}^{(3)} = \displaystyle\frac{\vb{k_{ph}}}{|\vb{k}_{\rm ph}|}, &
  [ \vb{k_{ph}}^{(i)} \times \vb{k_{ph}}^{(j)}] = \vb{k_{ph}}^{(k)}, &
  i, j, k = 1,2,3, & \nu = 0,1,2,3,
\end{array}
\label{eq.6-pp.3.7}
\end{equation}
and obtain
%
% \textcolor[rgb]{1.00,0.00,0.00}{\textbf{%
\begin{equation}
\begin{array}{llllllll}
% \vspace{1.5mm}
  k_{e}\, e^{(2)} =
%   \sqrt{m_{e}^{2} + k_{\rm e}^{2}}\, e_{0}^{(2)} - \vb{k}_{\rm e}\, \vb{e}^{(2)}
  k_{\rm e,0}^{2}\, e_{0}^{(2)} - \vb{k}_{\rm e}\, \vb{e}^{(2)} =
  - \vb{k}_{\rm e}\, \vb{e}^{(2)}.
\end{array}
\label{eq.6-pp.3.8}
\end{equation}
% }}
%-----------------------------------------------------------------------------------------------------------------------

%-----------------------------------------------------------------------------------------------------------------------
Let us assume that three vectors $\vb{k}_{i}$, $\vb{e}^{(2)}$ and $\vb{k}_{\rm ph}$ are in the same space plane.
We obtain property:
\begin{equation}
\begin{array}{llllllll}
% \vspace{1.5mm}
  \vb{k}_{i}\, \vb{e}^{(2)} =
  - \vb{k}_{f}\, \vb{e}^{(2)}.
\end{array}
\label{eq.6-pp.3.9}
\end{equation}
Then
\begin{equation}
\begin{array}{llllllll}
\vspace{1.5mm}
  \bigl(k_{f, \mu}\, e_{\mu}^{(2)} \bigr)\,
  \bigl(k_{i, \nu}\, e_{\nu}^{(2)} \bigr) =

%   \bigl(k_{f}\, e_{0}^{(2)} - \vb{k}_{f}\, \vb{e}^{(2)} \bigr)\,
%   \bigl(k_{i}\, e_{0}^{(2)} - \vb{k}_{i}\, \vb{e}^{(2)} \bigr) =

%   \bigl(k_{i}\, e_{0}^{(2)} + \vb{k}_{i}\, \vb{e}^{(2)} \bigr)\,
%   \bigl(k_{i}\, e_{0}^{(2)} - \vb{k}_{i}\, \vb{e}^{(2)} \bigr)\; = \\

%   =
%   \bigl(k_{i}\, e_{0}^{(2)}\bigr)^{2} - \bigl(\vb{k}_{i}\, \vb{e}^{(2)} \bigr)^{2} =
  - \bigl(\vb{k}_{i}\, \vb{e}^{(2)} \bigr)^{2}.
\end{array}
\label{eq.6-pp.3.10}
\end{equation}
Let's $\theta$ is angle between direction of virtual photon emission (defined by $\vb{k}_{\rm ph}/|\vb{k}_{\rm ph}|$) and
direction of lepton emission (with index $i$; defined by $\vb{k}_{i} / |\vb{k}_{i}|$).
We find
\begin{equation}
\begin{array}{llllllll}
\vspace{1.5mm}
  \theta =
  {\rm angle}\, (\vb{k}_{i},\, \vb{k}_{\rm ph}) =
  \displaystyle\frac{\pi}{2} - {\rm angle}\, (\vb{k}_{i},\, \vb{e}^{(2)}), &

% \vspace{1.5mm}
  \cos{(\vb{k}_{i},\, \vb{e}^{(2)})} =
%   \cos{\Bigl( \displaystyle\frac{\pi}{2} - {\rm angle}\, (\vb{k}_{i},\, \vb{k}_{\rm ph}) \Bigr)} =
%   \cos{\Bigl( \displaystyle\frac{\pi}{2} - \theta \Bigr)} =
  \sin{\theta}
\end{array}
\label{eq.6-pp.3.11}
\end{equation}
and
\begin{equation}
\begin{array}{llllllll}
\vspace{1.5mm}
  \vb{k}_{i}\, \vb{e}^{(2)} =
  |\vb{k}_{i}| \cdot |\vb{e}^{(2)}| \cdot \cos( \vb{k}_{i},\, \vb{e}^{(2)}) =
  k_{i} \cdot \sin{\theta}, &
  |\vb{e}^{(2)}| = 1, &

  \bigl( \vb{k}_{i}\, \vb{e}^{(2)} \bigr)^{2} =
  k_{i}^{2} \cdot \sin^{2} \theta,
\end{array}
\label{eq.6-pp.3.12}
\end{equation}
%
% and
%
\begin{equation}
\begin{array}{llllllll}
% \vspace{1.5mm}
  \bigl(k_{f, \mu}\, e_{\mu}^{(2)} \bigr)\, \bigl(k_{i, \nu}\, e_{\nu}^{(2)} \bigr) =
  - \bigl(\vb{k}_{i}\, \vb{e}^{(2)} \bigr)^{2} =
  - k_{i}^{2} \cdot \sin^{2} \theta.
\end{array}
\label{eq.6-pp.3.13}
\end{equation}
Then leptonic tensor (\ref{eq.6-pp.3.5}) obtains form
\begin{equation}
\begin{array}{llllllll}
% \vspace{1.5mm}
  L^{\rm lep}_{\mu\nu}\,
  \varepsilon_{\mu}^{(2)}\,
  \varepsilon_{\nu}^{(2),\, *}
  & = &
  2\, \Bigl[
    - 2\, k_{i}^{2}\, \sin^{2} \theta - \bigl(k' \cdot k - m_{\rm lep}^{2} \bigr) \Bigr].
\end{array}
\label{eq.6-pp.3.14}
\end{equation}
We calculate
\begin{equation}
\begin{array}{llllllll}
% \vspace{1.5mm}
  k_{f} \cdot k_{i} =
%   k_{f, \mu}\, k_{i, \mu} =
%   k_{f, 0}\, k_{i, 0} - \vb{k}_{f}\, \vb{k}_{i} =
%   E_{e, f}\, E_{e, i} - \vb{k}_{f}\, \vb{k}_{i} =
  E_{e, i}^{2} - \vb{k}_{f}\, \vb{k}_{i}, &
% \end{array}
% \label{eq.6-pp.3.15}
% \end{equation}
%
% \begin{equation}
% \begin{array}{llllllll}
  \Bigl( E_{{\rm e},\, i} = E_{{\rm e},\, f} \equiv E_{e} \Bigr) \to

%   \Bigl(
%     m_{e}^{2} + (\vb{k}_{{\rm e},\, i})^{2} =
%     E_{{\rm e},\, i}^{2} =
%     E_{{\rm e},\, f}^{2} =
%     m_{e}^{2} + (\vb{k}_{{\rm e},\, f})^{2}
%   \Bigr) \to

  \Bigl( |\vb{k}_{{\rm e},\, i}| = |\vb{k}_{{\rm e},\, f}| \Bigr),
\end{array}
\label{eq.6-pp.3.16}
\end{equation}
Also we have
\begin{equation}
\begin{array}{llllllll}
\vspace{1.5mm}
  \vb{k}_{i}\, \vb{k}_{f} =
  |\vb{k}_{i}| \cdot |\vb{k}_{f}| \cdot \cos( \vb{k}_{i},\, \vb{k}_{f}) =
  k_{i}^{2} \cdot \cos{\theta_{2}}.
%  \bigl( \vb{k}_{i}\, \vb{e}^{(2)} \bigr)^{2} =
%   k_{i}^{2} \cdot \sin^{2} \theta,
\end{array}
\label{eq.6-pp.3.17}
\end{equation}
%-----------------------------------------------------------------------------------------------------------------------

%-----------------------------------------------------------------------------------------------------------------------
% \vspace{3.0mm}
% \textcolor[rgb]{0.00,0.00,1.00}{\textbf{\underline{Approximation of zero mass of electron.}}}

% \vspace{2.0mm}
%v In approximation of small mass of electron is
%
% \begin{equation}
% \begin{array}{llllllll}
% \vspace{1.5mm}
%   m_{\rm e} = 0.5~MeV.
% \end{array}
% \label{eq.6-pp.3.18}
% \end{equation}
%
% So,
One can neglect
% \textcolor[rgb]{1.00,0.00,0.00}{\textbf{the}}
the mass of
% \textcolor[rgb]{1.00,0.00,0.00}{\textbf{the}}
the electron
in calculations and we obtain
\begin{equation}
\begin{array}{llllllll}
%v\vspace{1.5mm}
  E_{e,\, i}^{2} = k_{i}^{2}, &

% \vspace{1.5mm}
  \vb{k}_{i}\, \vb{k}_{f} =
  E_{i}^{2}\, \cos{2\theta}, &

  k_{f} \cdot k_{i} =
%  k_{f, \mu}\, k_{i, \mu} =
%   k_{f, 0}\, k_{i, 0} - \vb{k}_{f}\, \vb{k}_{i} =
%   E_{e, f}\, E_{e, i} - \vb{k}_{f}\, \vb{k}_{i} =
  E_{e, i}^{2} - \vb{k}_{f}\, \vb{k}_{i} =
%   E_{e, i}^{2} - E_{i}^{2}\, \cos{\theta_{2}} =
  E_{e, i}^{2}\, ( 1 - \cos{2\theta}).
\end{array}
\label{eq.6-pp.3.19}
\end{equation}
From Eq.~(\ref{eq.6-pp.3.13}) the leptonic tensor is
\begin{equation}
\begin{array}{llllllll}
% \vspace{1.5mm}
  L^{\rm lep}_{\mu\nu}\,
  e_{\mu}^{(2)}\,
  e_{\nu}^{(2),\, *}
  & = &
%   2\, \Bigl[ - 2\, E_{e}^{2}\, \sin^{2} \theta - E_{e}^{2}\, ( 1 - \cos{\theta_{2}}) \Bigr] =
  -\, 2\, E_{e}^{2}\, \Bigl[ 2\, \sin^{2} \theta + (1 - \cos{2\theta}) \Bigr].
\end{array}
\label{eq.6-pp.3.20}
\end{equation}
Now we rewrite matrix element (\ref{eq.6-pp.2.5}) as
\begin{equation}
\begin{array}{rllll}
% \vspace{1.5mm}
  \Bigl| \langle f\, |\, S^{(2)}\, |\, i \rangle \Bigr|^{2} & = &
  -\, |f|^{2}\,
  \displaystyle\frac{e^{2}\, (2\pi)^{8}}{4\, \varepsilon^{2}\, w_{\rm ph}} \cdot
  E_{e}^{2}\, \Bigl[ 2\, \sin^{2} \theta + (1 - \cos{2\,\theta}) \Bigr] \cdot
  \Bigl[ \delta^{4} (k' - k - k_{\rm ph}) \Bigr]^{2}.
\end{array}
\label{eq.6-pp.3.25}
\end{equation}
%-----------------------------------------------------------------------------------------------------------------------
%
%-----------------------------------------------------------------------------------------------------------------------
% \subsection{Removing of $\delta$-functions
% \label{sec.6-pp.4}}
%
In formula (\ref{eq.6-pp.3.25}), we have 8 $\delta$-functions, that is not convenient for calculations.
As a next step, we have to remove these $\delta$-functions from the found matrix element.
% This technique is described in App.~\ref{sec.app.6-pp.delta} (see p.~\pageref{sec.app.6-pp.delta}).
Such calculations are straightforward but standard.
Finally, we obtain:
% [see Eq.~(\ref{eq.app.6-pp.delta.2.4}), p.~\pageref{eq.app.6-pp.delta.2.4}]:
%
% \begin{equation}
% \begin{array}{rllll}
% \vspace{1.5mm}
%   \Bigl| \langle f\, |\, S^{(2)}\, |\, i \rangle \Bigr|^{2} \cdot
%   \displaystyle\frac{d^{3}p_{e, f1}}{(2\pi)^{3}}\,
%   \displaystyle\frac{d^{3}p_{e, f2}}{(2\pi)^{3}} & \to &
%
%   -\, |f|^{2}\,
%   \displaystyle\frac{e^{2}\, (2\pi)^{6}}{4\, w} \cdot
%   \Bigl[ 2\, \sin^{2} \theta + (1 - \cos{2\,\theta}) \Bigr] \cdot
% %   \Bigl[ \delta^{4} (k' - k - k_{\rm ph}) \Bigr]^{2}, \\
%   d^{3}p_{e,f} \cdot
%   \delta \Bigl( E_{\rm ph} - 2\,E_{e,f} \Bigr).
% \end{array}
% \label{eq.6-pp.4.1}
% % \label{eq.6-pp.delta.2.4}
% \end{equation}
%
% After removing last $\delta$-function of energies, we obtain
% [see Eq.~(\ref{eq.app.6-pp.delta.5.9}), p.~\pageref{eq.app.6-pp.delta.5.9}]
%
\begin{equation}
\begin{array}{rllll}
% \vspace{1.5mm}
  \Bigl| \langle f\, |\, S^{(2)}\, |\, i \rangle \Bigr|^{2} \cdot
  \displaystyle\frac{d^{3}k_{e, f1}}{(2\pi)^{3}}\,
  \displaystyle\frac{d^{3}k_{e, f2}}{(2\pi)^{3}} & \to &

%   -\, |f|^{2}\,
%   \displaystyle\frac{e^{2}\, (2\pi)^{6}}{4\, w} \cdot
%   \Bigl[ 2\, \sin^{2} \theta + (1 - \cos{2\,\theta}) \Bigr] \cdot
%   d^{3}p_{e,f} \cdot
%   \delta \Bigl( E_{\rm ph} - 2\,E_{e,f} \Bigr) \to \\

%   & = \to &
  -\, |f|^{2}\,
  \displaystyle\frac{e^{2}\, (2\pi)^{6}}{4}\;
  |\vb{k}_{{\rm e}, f}|\, \Bigl[ 2\, \sin^{2} \theta + (1 - \cos{2\,\theta}) \Bigr]\;
  \displaystyle\frac{1 + \cos{2\, \theta}}{1 - \cos{2\, \theta}}\; do_{e,2}.
\end{array}
\label{eq.6-pp.4.2}
% \label{eq.6-pp.delta.4.20}
% \label{eq.6-pp.delta.3.12}
\end{equation}
%-----------------------------------------------------------------------------------------------------------------------

%-----------------------------------------------------------------------------------------------------------------------
\section{Case of proton-nucleus scattering
\label{sec.6-nucleus}}

We will calculate matrix element for production of lepton pair for the proton-nucleus scattering.
Square of the matrix element for the proton-nucleon scattering is defined in Eq.~(\ref{eq.6-pp.1.1}).
For proton-nucleus scattering we define square of the full matrix element as
\begin{equation}
\begin{array}{lcl}
  \Bigl| \langle f\, |\, \hat{O}\, |\, i \rangle \Bigr|^{2} & = &
  \Bigl| \langle \Psi_{f} |\, \hat{H}_{\gamma} |\, \Psi_{i} \rangle^{(a)} \cdot M_{\rm e}^{a} \Bigr|^{2} =
   \langle \Psi_{f} |\, \hat{H}_{\gamma} |\, \Psi_{i} \rangle^{(a)}  \cdot M_{\rm e}^{a} \cdot
   \langle \Psi_{f} |\, \hat{H}_{\gamma} |\, \Psi_{i} \rangle^{(b), *} \cdot M_{\rm e}^{b, *}.
\end{array}
\label{eq.6-nucleus.1.1}
\end{equation}
We will derive tensor associated with production of lepton pair.
Such calculations are similar to formalism in Sec.~\ref{sec.6-pp.1}
[see Eqs.(\ref{eq.6-pp.1.1})--(\ref{eq.6-pp.1.13})], and we obtain
\begin{equation}
\begin{array}{cllll}
\vspace{1.5mm}
   M_{\rm lep}^{a} \cdot M_{\rm lep}^{b, *} & = &
  \displaystyle\frac{e^{2}}{8\, \varepsilon^{2}\, w_{\rm ph}}\:
  (2\pi)^{8}\;
  L_{\mu\nu}^{\rm lep}\;
  e_{\mu}^{a}\, e_{\nu}^{b,\, *}\;
  \Bigl[ \delta^{4} (k' - k - k_{\rm ph}) \Bigr]^{2}, \\

  L^{\rm lep}_{\mu\nu}\, & = &
  2\, \Bigl[ k'_{\mu}\, k_{\nu} + k'_{\nu}\, k_{\mu} - \bigl(k' \cdot k - m_{\rm lep}^{2} \bigr)\, g_{\mu\nu} \Bigr].
\end{array}
\label{eq.6-nucleus.1.2}
% \label{eq.6-pp.1.12}
\end{equation}
%-----------------------------------------------------------------------------------------------------------------------
%
%-----------------------------------------------------------------------------------------------------------------------
We will find square of $O$-matrix element for production of lepton pair.
This is matrix multiplication of leptonic matrix elements in Eq.~(\ref{eq.6-nucleus.1.2}) and
nuclear matric element in Eq.~(\ref{eq.4.8.8}).
Calculations are similar to formalism in Sec.~\ref{sec.6-pp.1}
[see Eqs.~(\ref{eq.6-pp.1.1})--(\ref{eq.6-pp.2.5})],
just now for the function $f$ we should use solution~(\ref{eq.4.8.8}) for the proton-nucleus scattering:
\begin{equation}
\begin{array}{rllll}
\vspace{1.5mm}
  \Bigl| \langle f\, |\, \hat{O}\, |\, i \rangle \Bigr|^{2} & = &

  |f|^{2}\,
  \displaystyle\frac{e^{2}\, (2\pi)^{8}}{8\, \varepsilon^{2}\, w_{\rm ph}}\:
  L_{\mu\nu}^{\rm lep}\;
  e_{\mu}^{(2)}\, e_{\nu}^{(2),\, *}\;
  \Bigl[ \delta^{4} (k' - k - k_{\rm ph}) \Bigr]^{2}.
\end{array}
% \label{eq.6-pp.2.5}
\label{eq.6-nucleus.1.3}
\end{equation}
%-----------------------------------------------------------------------------------------------------------------------

%-----------------------------------------------------------------------------------------------------------------------
In this formula we should calculate multiplication of leptonic tensor on vectors of polarizations
$L^{\rm lep}_{\mu\nu}\, e_{\mu}^{(2)}\, e_{\nu}^{(2),\, *}$.
Such calculations are done in Sec.~\ref{sec.6-pp.3}
% [see Eqs.~(\ref{eq.6-pp.3.1})--(\ref{eq.6-pp.3.24})]
where solution is Eq.~(\ref{eq.6-pp.3.20}).
%
% \begin{equation}
% \begin{array}{llllllll}
%   L^{\rm lep}_{\mu\nu}\,
%   \varepsilon_{\mu}^{(2)}\,
%   \varepsilon_{\nu}^{(2),\, *}
%   & = &
% % 2\, \Bigl[ 2\, E_{i}^{2}\, \cos^{2} \theta - E_{i}^{2}\, \sin{\theta_{2}} \Bigr] =
%   -\, 2\, E_{e}^{2}\, \Bigl[ 2\, \cos^{2} \theta + (1 - \cos{2\,\theta}) \Bigr].
% \end{array}
% \label{eq.6-nucleus.1.4}
% % \label{eq.6-pp.3.24}
% \end{equation}
%
Substituting such a solution to Eq.~(\ref{eq.6-nucleus.1.3}), we find % final formula as
%-----------------------------------------------------------------------------------------------------------------------
%
%-----------------------------------------------------------------------------------------------------------------------
% \vspace{3.0mm}
% \textcolor[rgb]{0.00,0.00,1.00}{\textbf{\underline{Final formulas:}}}
%
\begin{equation}
\begin{array}{rllll}
\vspace{1.5mm}
  \Bigl| \langle f\, |\, \hat{O}\, |\, i \rangle \Bigr|^{2} & = &

  -\, |f|^{2}\,
  \displaystyle\frac{e^{2}\, (2\pi)^{8}}{4\, \varepsilon^{2}\, w_{\rm ph}} \cdot
  E_{e}^{2}\, \Bigl[ 2\, \sin^{2} \theta + (1 - \cos{2\,\theta}) \Bigr] \cdot
  \Bigl[ \delta^{4} (k' - k - k_{\rm ph}) \Bigr]^{2}, \\

  f & = &
  -\, \sqrt{\displaystyle\frac{2\pi}{3 \hbar w_{\rm ph}}}\,
  (2\pi)^{3} \displaystyle\frac{\mu_{N}\,  m_{\rm p}}{\mu}\:
  Z_{\rm eff}^{\rm (mon,\, 0)}\,
  \Bigl\{
    J_{1}(0,1,0) -
    \displaystyle\frac{47}{40} \sqrt{\displaystyle\frac{1}{2}} \cdot J_{1}(0,1,2)
  \Bigr\}.
\end{array}
\label{eq.6-nucleus.1.5}
% \label{eq.6-pp.3.26}
\end{equation}
Here, $\theta$ is angle between direction of virtual photon emission (defined by $\vb{k}_{\rm ph}/|\vb{k}_{\rm ph}|$) and
direction of one lepton emission (with index $i$; defined by $\vb{k}_{i} / |\vb{k}_{i}|$).

% \subsection{Reduction of $\delta$-functions in formula for square of matrix element
% \label{sec.6-nucleus.2}}

% For computer calculations we reduce $\delta$-functions in square of the matrix element.
In Eq.~(\ref{eq.6-nucleus.1.5}) we have 8 $\delta$-functions, which sould be reduced for computer calculation.
At first, we reduce the first 6 $\delta$-functions on momenta and one $\delta$-function on energy.
% following to the logic in Sec.~\ref{sec.app.6-pp.delta.2}.
For that we will use formula of reduction of $\delta$-functions as
%
% Далее следует устранить $\delta$-функцию в формуле (\ref{eq.6-nucleus.1.5}) ``вероятности'' рождения лептонов в изучаемом процессе.
% В начале мы устраним первые 6 $\delta$-функций по импульсам и 1 по энергии (следуя логике в Разд.~\ref{sec.6-pp.delta.2}).
% Для этого воспользуемся найденной выше формулой перехода (\ref{eq.6-pp.delta.2.3}) [см. стр.~\pageref{eq.6-pp.delta.2.3}]:
%
\begin{equation}
\begin{array}{llllll}
  L_{\mu\nu}^{\rm lep}\;
  e_{\mu}^{a}\, e_{\nu}^{b,\, *}\;
  \Bigl[ \delta^{4} (k' - k - k_{\rm ph}) \Bigr]^{2} \cdot
  \displaystyle\frac{d^{3}k_{e, f1}}{(2\pi)^{3}}\,
  \displaystyle\frac{d^{3}k_{e, f2}}{(2\pi)^{3}} & \to &

  L_{\mu\nu}^{\rm lep}\;
  e_{\mu}^{a}\, e_{\nu}^{b,\, *}\;
  \displaystyle\frac{d^{3}k_{f}}{(2\pi)^{2}} \cdot
  \delta \Bigl( E_{\rm ph} - 2\,E_{e,f} \Bigr).
\end{array}
\label{eq.6-nucleus.2.1}
% \label{eq.6-pp.2.2}
\end{equation}
Calculations are similar to case of proton-proton scattering.
%  and we obtain
% [see calculations in App.~\ref{sec.app.6-nucleus.delta.1}, p.~\pageref{eq.app.6-nucleus.delta.1.3}, Eq.~(\ref{eq.app.6-nucleus.delta.1.3})]
%-----------------------------------------------------------------------------------------------------------------------
%
%-----------------------------------------------------------------------------------------------------------------------
After reducing last $\delta$-function on energy we obtain
% the following formula
% [see App.~\ref{sec.app.6-nucleus.delta.1}, (\ref{eq.app.6-nucleus.delta.1.4})].
%
% Чтобы исключить последнюю $\delta$-функцию по энергии, можно воспользоваться формализмом в Разд.~\ref{sec.app.6-pp.delta.5}
% [см. стр.~\pageref{sec.app.6-pp.delta.5}--\pageref{eq.app.6-pp.delta.5.15}].
% Он нисколько не меняется для протон-ядерного рассечяния (кроме изменения функции $f$).
% В результат, мы получаем следующую формулу [см. (\ref{eq.app.6-pp.delta.5.15}), стр.~\pageref{eq.app.6-pp.delta.5.15}].
%
\begin{equation}
\begin{array}{rllll}
  \Bigl| \langle f\, |\, S^{(2)}\, |\, i \rangle \Bigr|^{2} \cdot
  \displaystyle\frac{d^{3}k_{e, f1}}{(2\pi)^{3}}\,
  \displaystyle\frac{d^{3}k_{e, f2}}{(2\pi)^{3}} & \to &
  |f|^{2}\,
  \displaystyle\frac{e^{2}\, (2\pi)^{6}}{4}\;
  |\vb{k}_{{\rm e}, f}|\, \Bigl[ 2\, \sin^{2} \theta + (1 - \cos{2\,\theta}) \Bigr]\;
  \displaystyle\frac{1 + \cos{2\, \theta}}{1 - \cos{2\, \theta}}\; do_{e,2}.
\end{array}
\label{eq.6-nucleus.2.3}
% \label{eq.app.6-nucleus.delta.1.4}
\end{equation}
% *******************************************************************************************************************

% *******************************************************************************************************************
\section{Definition of cross-section of production of leptons pair
\label{sec.model.bremprobability}}

Cross-sections of the production of leptons pair can be defined using logic of determination of emission of bremsstrahlung photons in nucleon-nucleus and nucleus-nucleus scattering
(see Refs.~\cite{Maydanyuk.2023.PRC.delta,Maydanyuk_Zhang_Zou.2016.PRC,Maydanyuk.2012.PRC,Maydanyuk_Zhang.2015.PRC}, reference therein).
We define cross-section of production of leptons pair on the basis of the full matrix element $p_{fi}$ % (\ref{eq.2.5.6.2})
in frameworks of formalism in Ref.~\cite{Maydanyuk_Zhang_Zou.2016.PRC}
(see Eq.~(22) in that paper, reference therein).
% and it is not repeated in this paper.
Finally, the cross-sections of production of dileptons is obtained
%
% \footnote{We obtain the formula (\ref{eq.2.6.1}) in dependence on mass of proton $m_{\rm p}$ while in Ref.~\cite{Maydanyuk.2012.PRC} we had the bremsstrahlung probability (49) in dependence on the reduced mass $\mu$.
% Such a difference is explained by that in the current paper we develop formalism on the basis of the emission operator of the many-nucleon system (\ref{eq.2.2.3})
% while in Ref.~\cite{Maydanyuk.2012.PRC} we started the formalism on the basis of the operator of emission (4) of the proton-nucleus system defined via the reduced mass of proton and nucleus.}
%
on the basis of matrix element of nuclear part (\ref{eq.app.2.1.2.10})--(\ref{eq.app.2.1.2.15}) as
[see Eqs.~(\ref{eq.app.2.bremprobability.1}), App.~\ref{sec.app.2} for details]
%
% \begin{equation}
% \begin{array}{lll}
%   \langle \Psi_{f} |\, \hat{H}_{\gamma} |\, \Psi_{i} \rangle \;\; = \;\;
%   \sqrt{\displaystyle\frac{2\pi\, c^{2}}{\hbar w_{\rm ph}}}\,
%   M_{\rm full}, &
%   % \Bigl\{ M_{P} + M_{p}^{(E)} + M_{p}^{(M)} + M_{k} + M_{\Delta E} + M_{\Delta M} \Bigr\}, &
%   M_{\rm full} = M_{P} + M_{p}^{(E)} + M_{p}^{(M)} + M_{k} + M_{\Delta E} + M_{\Delta M},
% \end{array}
% \label{eq.model.bremprobability.2}
% \end{equation}
%
% expressed via matrix element $M_{\rm full}$.
%
\begin{equation}
\begin{array}{llll}
  \displaystyle\frac{d \sigma}{dw_{\rm ph}} =
    \displaystyle\frac{1}{2\pi\,c^{5}}\: \displaystyle\frac{w_{\rm ph}\,E_{i}}{k_{i}}\:
    \bigl| M_{\rm full} \bigr|^{2}.
%   \displaystyle\frac{d^{2} \sigma}{dw_{\rm ph}\, d \cos \theta} =
%     \displaystyle\frac{e^{2}}{2\pi\,c^{5}}\: \displaystyle\frac{w_{\rm ph}\,E_{i}}{m_{\rm p}^{2}\,k_{i}}\:
%     \bigl\{ p_{fi} \displaystyle\frac{d\,p_{fi}^{*}}{d\, \cos \theta}  + c.\,c. \bigr\}, &
%   M_{\rm full} = - \displaystyle\frac{e}{m_{\rm p}}\, p_{fi},
\end{array}
\label{eq.model.bremprobability.3}
\end{equation}
%
%
% where $p_{fi}$ is proportional to the electrical component $p_{\rm el}$ in Eqs.~(10) in \cite{Maydanyuk.2012.PRC}
% [with the additional factor of $2\, e^{-\, (a^{2} k_{x}^{2} + b^{2} k_{y}^{2} + c^{2} k_{z}^{2})\,/4}$ and the included effective charge $\tilde{Z}_{\rm eff}^{\rm (dip)}$] and
% $d\,p_{fi} (\theta_{f})\, / d\,\cos{\theta_{f}}$ is defined by the same way as $d\,p\, (k_{i}, k_{f}, \theta_{f})\, / d\,\cos{\theta_{f}}$ in Ref.~\cite{Maydanyuk.2012.PRC}.
%
% The different contributions of the emitted photons to the full bremsstrahlung spectrum are calculated.
% For estimation of the interesting contribution the corresponding matrix element of emission are used.
%
In this paper the matrix elements are calculated on the basis of wave functions with quantum numbers $l_{i}=0$, $l_{f}=1$ and $l_{\rm ph}=1$
[here, $l_{i}$ and $l_{f}$ are orbital quantum numbers of wave function $\Phi_{\rm p - nucl} (\vb{r})$ % defined in Eq.~(\ref{eq.18.1.2})
for states before emission of photon and after this emission,
$l_{\rm ph}$ is orbital quantum number of photon in the multipole approach].
% defined in Eq.~(\ref{eq.app.integrals.3})]

\vspace{2.5mm}
We will formulate the following rule for such a transformation of formula for cross section.

\vspace{1.5mm}
\noindent
% \textcolor[rgb]{0.00,0.00,1.00}{\textbf{%
\emph{Number of all final states for process of emission of bremsstrahlung photon
should be changed into number of all final states for process of production of lepton pairs.}
% }}

\vspace{1.5mm}
In particular, for current problem we have:
\begin{itemize}
\item
For the previous formalism with emission of bremsstrahlung photons,
the final states is state of the emitted bremsstrahlung real photon,

\item
For the new formalism with production of lepton pair,
the final states are states of the produced leptons. % (in final states).
\end{itemize}
So, we find the following change:
\begin{equation}
\begin{array}{llll}
  \displaystyle\frac{\vb{d^{3}k}_{\rm ph, f}}{(2\pi)^{3}\, (2\, E_{\rm ph, f})} \to
  \prod\limits_{f}
  \displaystyle\frac{\vb{d^{3}k}_{e,f}}{(2\pi)^{3}\, (2\, E_{e,f})}.
\end{array}
\label{eq.model.bremprobability.4}
\end{equation}
Applying such a rule, we will rewrite Eq.~(\ref{eq.model.bremprobability.3}) and obtain
%
% \begin{equation}
% \begin{array}{lllllll}
%   \Bigl(
%     d^{2} \sigma^{\rm (br)} =
%     \displaystyle\frac{1}{2\pi\,c^{5}}\, \displaystyle\frac{w_{\rm ph}\,E_{i}}{k_{i}}\:
%       \bigl| M_{\rm full} \bigr|^{2}\: dw_{\rm ph}\, do_{\rm ph} =
%
%     \displaystyle\frac{1}{2\pi\,c^{5}}\, \displaystyle\frac{E_{i}}{w_{\rm ph}\, k_{i}}\:
%     \bigl| M_{\rm full} \bigr|^{2}\: w_{\rm ph}^{2} dw_{\rm ph}\, do_{\rm ph} =
%
%     \displaystyle\frac{(2\pi)^{2}}{c^{5}}\, \displaystyle\frac{2E_{i}}{k_{i}}\,
%     \bigl| M_{\rm full} \bigr|^{2}\:
%     \displaystyle\frac{\vb{d^{3}k}_{\rm ph}}{(2\pi)^{3}\, (2E_{\rm ph})}
%   \Bigr).
% \end{array}
% \label{eq.model.bremprobability.5}
% \end{equation}
%
% After applying rule ~(\ref{eq.model.bremprobability.4}), we obtain:
%
\begin{equation}
\begin{array}{lllllll}
  \Bigl(
    d^{2} \sigma^{\rm (br)} =
    \displaystyle\frac{(2\pi)^{2}}{c^{5}}\, \displaystyle\frac{2E_{i}}{k_{i}}\,
    \bigl| M_{\rm full} \bigr|^{2}\:
    \displaystyle\frac{\vb{d^{3}k}_{\rm ph}}{(2\pi)^{3}\, (2E_{\rm ph})}
  \Bigr) &
    \to &

  \Bigl(
    d^{2} \sigma^{\rm (lep)} =
    \displaystyle\frac{(2\pi)^{2}}{c^{5}}\, \displaystyle\frac{2E_{i}}{k_{i}}\,
    \bigl| M_{\rm full}^{\rm (lep)} \bigr|^{2}\:
    \displaystyle\frac{\vb{d^{3}k}_{e, f1}}{(2\pi)^{3}\, (2 E_{\rm e, f1})}\,
    \displaystyle\frac{\vb{d^{3}k}_{e, f2}}{(2\pi)^{3}\, (2 E_{\rm e, f2})}
  \Bigr).
\end{array}
\label{eq.model.bremprobability.6}
\end{equation}
Taking into account square of the coherent matrix element after reduction of all $\delta$-functions in form (\ref{eq.6-nucleus.2.3}),
%
% \begin{equation}
% \begin{array}{rllll}
%   \Bigl| \langle f\, |\, S^{(2)}\, |\, i \rangle \Bigr|^{2} \cdot
%   \displaystyle\frac{d^{3}p_{e, f1}}{(2\pi)^{3}}\,
%   \displaystyle\frac{d^{3}p_{e, f2}}{(2\pi)^{3}} & = \to &
%   |f|^{2}\,
%   \displaystyle\frac{e^{2}\, (2\pi)^{6}}{4}\;
%   |\vb{p}_{{\rm e}, f}|\, \Bigl[ 2\, \sin^{2} \theta + (1 - \cos{2\,\theta}) \Bigr]\;
%   \displaystyle\frac{1 + \cos{2\, \theta}}{1 - \cos{2\, \theta}}\; do_{e,2},
% \end{array}
% \label{eq.model.bremprobability.7}
% \end{equation}
%
we write down the final solution as
\begin{equation}
\begin{array}{rllll}
% \vspace{0.5mm}
  d^{2} \sigma^{\rm (lep)} =
  \displaystyle\frac{(2\pi)^{8} e^{2}}{8\, c^{5}} \cdot

  |f|^{2}\,
  \displaystyle\frac{E_{i}}{k_{i}\, E_{\rm e, f1}^{2}}\;
  |\vb{k}_{{\rm e}, f}|\, \Bigl[ 2\, \sin^{2} \theta + (1 - \cos{2\,\theta}) \Bigr]\;
  \displaystyle\frac{1 + \cos{2\, \theta}}{1 - \cos{2\, \theta}}\; do_{e,2},
\end{array}
\label{eq.resultingformula.4}
% \label{eq.model.bremprobability.9}
% \label{eq.6-nucleus.2.2}
\end{equation}
%-----------------------------------------------------------------------------------------------------------------------

%-----------------------------------------------------------------------------------------------------------------------
%
\begin{equation}
\begin{array}{rllll}
\vspace{1.5mm}
  f & = &
  -\, \sqrt{\displaystyle\frac{2\pi}{3 \hbar w_{\rm ph}}}\,
  (2\pi)^{3} \displaystyle\frac{\mu_{N}\,  m_{\rm p}}{\mu}\:
  Z_{\rm eff}^{\rm (mon,\, 0)}\,
  \Bigl\{
    J_{1}(0,1,0) -
    \displaystyle\frac{47}{40} \sqrt{\displaystyle\frac{1}{2}} \cdot J_{1}(0,1,2)
  \Bigr\}, \\
% \vspace{1.5mm}
  f^{vir} & = &
  -\, \sqrt{\displaystyle\frac{2\pi}{3 \hbar w_{\rm ph}}}\,
  (2\pi)^{3} \displaystyle\frac{\mu_{N}\,  m_{\rm p}}{\mu}\:
  Z_{\rm eff}^{\rm (mon,\, 0)}\,
  \Bigl\{
    J_{1}^{vir}(0,1,0; 0) -
    \displaystyle\frac{47}{40} \sqrt{\displaystyle\frac{1}{2}} \cdot J_{1}^{vir}(0,1,2;0)
  \Bigr\}.
\end{array}
\label{eq.resultingformula.5}
% \label{eq.6-nucleus.1.5}
% \label{eq.app.6-pp.3.26}
\end{equation}
Here,
$E_{i}$ is energy of relative motion between proton in beam and nucleus as target in the center-of-mass frame of these objects
(before emission of virtual photon),
$E_{{\rm e},\, f1}$ is energy of one produced lepton,
$\mu = m_{A}\, m_{\rm p}/(m_{A} + m_{\rm p})$ is reduced mass,
$\theta$ is angle between direction of virtual photon emission (defined by $\vb{k}_{\rm ph}/|\vb{k}_{\rm ph}|$) and
direction of one lepton emission (with index $i$; defined by $\vb{k}_{e, i} / |\vb{k}_{e, i}|$).
% Instead of this angle one can use another angle $\theta_{2} = 2\, \theta$
% which is the angle between directions of emission of two leptons.
%-----------------------------------------------------------------------------------------------------------------------

%-----------------------------------------------------------------------------------------------------------------------
Radial integrals are
[see Eqs.~(\ref{eq.virtual.2.10})]
\begin{equation}
\begin{array}{llllll}
  J_{1}^{real} (l_{i},l_{f},n) & = &
    \displaystyle\int\limits^{+\infty}_{0} \displaystyle\frac{dR_{i}(r, l_{i})}{dr}\: R^{*}_{f}(l_{f},r)\, j_{n}(k_{\rm ph}r)\; r^{2} dr, \\
  \tilde{J}^{real}\,(c, l_{i},l_{f},n) & = &
    \displaystyle\int\limits^{+\infty}_{0} R_{i}(l_{i}, r)\, R^{*}_{f}(l_{f},r)\, j_{n}(c\, k_{\rm ph}r)\; r^{2} dr,
%  \breve{J}\,(c_{A}, l_{i}, l_{f},n) & = & \displaystyle\int\limits^{+\infty}_{0} R_{i}(r)\, R^{*}_{l,f}(r)\, V(\mathbf{r})\, j_{n}(c_{A}\,kr)\; r^{2} dr.
\end{array}
\label{eq.resultingformula.6}
% \label{eq.multimple.4}
% \label{eq.resultingformulas.7}
\end{equation}
\begin{equation}
\begin{array}{llllll}
  J_{1}^{vir}(l_{i},l_{f},n,\; l^{\parallel}) & = &
  \displaystyle\int\limits^{+\infty}_{0}
    \displaystyle\frac{dR_{i}(r, l_{i})}{dr}\:
    R^{*}_{f}(l_{f},r)\,
    j_{n}(k_{\rm ph}^{\bot}r)\;
    j_{l^{\parallel}}(k^{\parallel} r)\; r^{2} dr, \\

  \tilde{J}^{vir}\,(c, l_{i},l_{f},n;\; l^{\parallel}) & = &
  \displaystyle\int\limits^{+\infty}_{0} R_{i}(l_{i}, r)\, R^{*}_{f}(l_{f},r)\,
  j_{n}(c\, k_{\rm ph}^{\bot}r)\;
  j_{l^{\parallel}}(k^{\parallel} r)\; r^{2} dr.

%  \breve{J}\,(c_{A}, l_{i}, l_{f},n) & = & \displaystyle\int\limits^{+\infty}_{0} R_{i}(r)\, R^{*}_{l,f}(r)\, V(\mathbf{r})\, j_{n}(c_{A}\,kr)\; r^{2} dr.
\end{array}
\label{eq.resultingformula.7}
% \label{eq.virtual.2.10}
% \label{eq.multimple.4}
% \label{eq.resultingformulas.7}
\end{equation}
Here,
$k_{\rm ph}^{\bot} = |\vb{k_{\rm ph}^{\bot}}|$ and
$k_{\rm ph}^{\parallel} = |\vb{k_{\rm ph}^{\parallel}}|$ are
perpendicular and longitudinal parts of wave vector $\vb{k_{\rm ph}}$ of photon
(we have $\vb{k_{\rm ph}^{\parallel}} \parallel \vb{e^{(3)}}$,
for real photon $\vb{k_{\rm ph}^{\parallel}} = 0$),
integrals in Eqs.~(\ref{eq.resultingformula.6}) are without inclusion of the longitudinal part of virtual photon,
integrals in Eqs.~(\ref{eq.resultingformula.7}) are with inclusion of the longitudinal part of virtual photon
(see App.~\ref{sec.virtual}),
$R_{i,f}$ is radial part of wave function $\Phi_{\rm p - nucl} (\vb{r})$ in $i$-state or $f$-state,
$j_{n}(k_{\rm ph}r)$ is spherical Bessel function of order $n$.
The effective electric charge is
% [see Eqs.~(\ref{eq.app.2.13.2.1}) in p.~\pageref{eq.app.2.13.2.1}, (\ref{eq.app.2.13.5.1}) in p.~\pageref{eq.app.2.13.5.1}]
%
\begin{equation}
\begin{array}{llll}
  Z_{\rm eff}^{\rm (mon)} (\vb{k}_{\rm ph}) = \displaystyle\frac{m_{\rm p}\, Z_{A}(\vb{k}_{\rm ph}) - m_{A}\, z_{\rm p}}{m_{A} + m_{\rm p}}.

%   \vb{M}_{\rm eff}^{\rm (mon)} (\vb{k}_{\rm ph}, \vb{r}) =
%   \displaystyle\frac{\mu}{m_{\rm p}}\, \Bigl[ \vb{F}_{A,\, {\rm mag}} (\vb{k}_{\rm ph}) - \vb{F}_{B,\, {\rm mag}} (\vb{k}_{\rm ph}) \Bigr],
% \label{eq.app.2.13.5.1}
\end{array}
\label{eq.resultingformula.8}
% \label{eq.resultingformulas.5}
% \label{eq.app.2.13.2.1}
\end{equation}
% *******************************************************************************************************************

% *******************************************************************************************************************
% \subsection{Cross-section of production of leptons pair in dependence on their invariant mass
% \label{sec.model.crosssection.2}}

One can redefine the cross section of production of pair of leptons, where dependence on energy of these leptons or their invariant mass is included:
% Such formalism is given in App.~\ref{sec.app.model.crosssection.2} and we obtain
% [see Eqs.~(\ref{eq.app.model.crosssection.2.9})--(\ref{eq.app.model.crosssection.2.11}) and details]:
%
\begin{equation}
\begin{array}{lllll}
\vspace{0.5mm}
  d^{2} \sigma^{\rm (lep)} =
  \displaystyle\frac{(2\pi)^{8} e^{2}}{8\, c^{5}} \cdot

  |f\, (M, \chi)|^{2}\,
  \displaystyle\frac{E_{i}}{k_{i}\, E_{\rm e, f1}^{2}}\;
  |\vb{k}_{{\rm e}, f}|\, \Bigl[ 2\, \sin^{2} \theta + (1 - \cos{2\,\theta}) \Bigr]\;
  \displaystyle\frac{1 + \cos{2\, \theta}}{1 - \cos{2\, \theta}}\; do_{e,2}, \\

%   |\vb{p}_{\rm e, 1}| =
%   \displaystyle\frac{M}{\sqrt{2\, (1 + \cos{2\, \theta})}}\,
%   \displaystyle\frac{1}{\sqrt{(1 + \chi^{2})}}, \quad

  |\vb{k}_{\rm e, 1}| = \displaystyle\frac{M}{2}, \quad

  E_{\rm e}^{2} = |\vb{k}_{{\rm e}, f}|^{2} + m_{\rm e}^{2}, \quad
  \chi = \displaystyle\frac{k_{\rm ph}^{\parallel}}{k_{\rm ph}^{\bot}}.
\end{array}
\label{eq.model.crosssection.2.9}
\end{equation}
%
% We find that parameter of virtuality $\chi$ is useful in analysis of properties of virtual photon and their influence on cross section of dilepton production.
%
It turns out that parameter $\chi$ is useful in analysis of properties of virtual photon and their influence on calculations of the cross section of dilepton production.
So, we call such a parameter $\chi$ as \emph{parameter of virtuality of photon}.
% *******************************************************************************************************************

% *******************************************************************************************************************
\section{Role of incoherent terms in production of leptons pair
\label{sec.model.incoh}}

In this Section we will construct formalist with inclusion of incoherent terms to the model above.
% For that, we rewrite the obtained formulas for ...
We obtained the cross section of dilepton production based on coherent term
in Eqs.~(\ref{eq.model.crosssection.2.9})
% [see Eqs.~(\ref{eq.app.model.crosssection.2.9}), (\ref{eq.app.model.crosssection.2.11})]
%
% \begin{equation}
% \begin{array}{lllll}
% \vspace{0.5mm}
%   d^{2} \sigma^{\rm (lep)} =
%   \displaystyle\frac{(2\pi)^{8} e^{2}}{8\, c^{5}} \cdot
%
%   |f\, (Q^{2}, \chi)|^{2}\,
%   \displaystyle\frac{E_{i}}{k_{i}\, E_{\rm e, f1}^{2}}\;
%   |\vb{p}_{{\rm e}, f}|\, \Bigl[ 2\, \sin^{2} \theta + (1 - \cos{2\,\theta}) \Bigr]\;
%   \displaystyle\frac{1 + \cos{2\, \theta}}{1 - \cos{2\, \theta}}\; do_{e,2}, \\
%
%   |\vb{p}_{\rm e, 1}| = \displaystyle\frac{M}{2}, \quad
%
% %   |\vb{p}_{\rm e, 1}| =
% %   \displaystyle\frac{M}{\sqrt{2\, (1 + \cos{2\, \theta})}}\,
% %   \displaystyle\frac{1}{\sqrt{(1 + \chi^{2})}}, \quad
%
%   E_{\rm e}^{2} = |\vb{p}_{{\rm e}, f}|^{2} + m_{\rm e}^{2}, \quad
%   \chi = \displaystyle\frac{k_{\rm ph}^{\parallel}}{k_{\rm ph}^{\bot}},
% \end{array}
% \label{eq.model.incoh.1.1}
% \end{equation}
%
and
% we define dilepton invariant mass $M$ as
% [see Eqs.~(\ref{eq.app.model.crosssection.2.10})]
%
\begin{equation}
\begin{array}{rllll}
  |\vb{k}_{\rm ph}|^{2} =
  (k_{\rm ph}^{\bot})^{2} + (k_{\rm ph}^{\parallel})^{2} =
  (k_{\rm ph}^{\bot})^{2} (1 + \chi^{2}).
\end{array}
\label{eq.model.incoh.1.2}
% \label{eq.app.model.crosssection.2.10}
\end{equation}
%-----------------------------------------------------------------------------------------------------------------------
%
%-----------------------------------------------------------------------------------------------------------------------
In such formulas numerical calculation is in function $f$ defined in Eq.~(\ref{eq.resultingformula.5}).
%
% \begin{equation}
% \begin{array}{rllll}
%   f & = &
%   -\, \sqrt{\displaystyle\frac{2\pi}{3 \hbar w_{\rm ph}}}\,
%   (2\pi)^{3} \displaystyle\frac{\mu_{N}\,  m_{\rm p}}{\mu}\:
%   Z_{\rm eff}^{\rm (mon,\, 0)}\,
%   \Bigl\{
%     J_{1}(0,1,0) -
%     \displaystyle\frac{47}{40} \sqrt{\displaystyle\frac{1}{2}} \cdot J_{1}(0,1,2)
%   \Bigr\}.
% \end{array}
% \label{eq.model.incoh.1.3}
% \end{equation}
%-----------------------------------------------------------------------------------------------------------------------
%
%-----------------------------------------------------------------------------------------------------------------------
The full matrix element of nuclear part
with included incoherent terms is found in Eqs.~(\ref{eq.app.2.multiple.1}) as
\begin{equation}
\begin{array}{lll}
\vspace{1.0mm}
  M_{p}^{(E,\, {\rm mon},\, 0)} =
  -\, \hbar\, (2\pi)^{3}\,
  \displaystyle\frac{\mu_{N}}{\sqrt{3}}\,
  \displaystyle\frac{m_{\rm p}\, Z_{\rm eff}^{\rm (mon,\, 0)}}{\mu}\,
  \Bigl(
    J_{1}(0,1,0) -
    \displaystyle\frac{47}{40} \sqrt{\displaystyle\frac{1}{2}}\, J_{1}(0,1,2)
  \Bigr), \\

\vspace{1.0mm}
  M_{p}^{(E,\, {\rm mon},\, 0)} + M_{p}^{(M,\, {\rm mon},\, 0)} =
%   M_{p}^{(E,\, {\rm mon},\, 0)}\,
%   \Bigl( 1 + i\: \displaystyle\frac{\alpha_{M}} {2\, m_{\rm p}\, Z_{\rm eff}^{\rm (mon,\, 0)}} \Bigr) =
%
  M_{p}^{(E,\, {\rm mon},\, 0)}\,
  \Bigl( 1 + i\: \displaystyle\frac{\mu^{2}} {2\, m_{\rm p}^{2}\, Z_{\rm eff}^{\rm (mon,\, 0)}}\; \alpha \Bigr), \\

\vspace{1.0mm}
  M_{\Delta M} =
  \hbar\, (2\pi)^{3}\,
  \displaystyle\frac{\sqrt{3}}{2}\,
  \mu_{N}\, k_{\rm ph}\,
  f_{A} \cdot Z_{\rm A} (\vb{k}_{\rm ph}) \cdot \tilde{J}\, (- c_{\rm p}, 0,1,1), \\

  M_{\Delta M} + M_{k} =
  - \hbar\, (2\pi)^{3}\,
  \displaystyle\frac{\sqrt{3}}{2}\,
  \mu_{N}\, k_{\rm ph}\,
  \Bigl\{
    \displaystyle\frac{A+1}{2A}\, \bar{\mu}_{\rm pn}^{(A)} \cdot Z_{\rm A} (\vb{k}_{\rm ph})\, \tilde{J}\, (- c_{\rm p}, 0,1,1) +
    \mu_{\rm p}\, z_{\rm p}\, \tilde{J}\, (c_{A}, 0,1,1)
  \Bigr\}.
\end{array}
\label{eq.model.incoh.1.4}
% \label{eq.app.2.multiple.1}
\end{equation}
where parameter $\alpha$ is derived in Eq.~(\ref{eq.app.2.resultingformulas.6}).
%-----------------------------------------------------------------------------------------------------------------------
%
%-----------------------------------------------------------------------------------------------------------------------
From Eqs.~(\ref{eq.model.incoh.1.4}) we find:
\[
\begin{array}{llllll}
  M_{p}^{(E,\, {\rm mon},\, 0)} & = &
  \hbar\, \sqrt{\displaystyle\frac{\hbar w_{\rm ph}}{2\pi}}
  \cdot\Bigl\{
  -\, \sqrt{\displaystyle\frac{2\pi}{3 \hbar w_{\rm ph}}}\,
  (2\pi)^{3} \displaystyle\frac{\mu_{N}\,  m_{\rm p}}{\mu}\:
  Z_{\rm eff}^{\rm (mon,\, 0)}\,
  \Bigl[
    J_{1}(0,1,0) -
    \displaystyle\frac{47}{40} \sqrt{\displaystyle\frac{1}{2}} \cdot J_{1}(0,1,2)
  \Bigr]
  \Bigr\}\; = % \\

%   & = &
  \hbar\, \sqrt{\displaystyle\frac{\hbar w_{\rm ph}}{2\pi}}
  \cdot
  f (Q, \chi)
\end{array}
\]
or
\begin{equation}
\begin{array}{llllll}
  M_{p}^{(E,\, {\rm mon},\, 0)} =
  c_{1}(w_{\rm ph}) \cdot f (Q, \chi), \quad \quad
  c_{1} (w_{\rm ph}) = \hbar\, \sqrt{\displaystyle\frac{\hbar w_{\rm ph}}{2\pi}}.
\end{array}
\label{eq.model.incoh.1.6}
\end{equation}
In order to perform transition to formalism with included incoherent terms, we use the following change
($M_{\Delta E}=0$):
\begin{equation}
\begin{array}{llllll}
  M_{p}^{(E,\, {\rm mon},\, 0)} \to
  M_{p}^{\rm (full)} =
  M_{p}^{(E,\, {\rm mon},\, 0)} +
  M_{p}^{(M,\, {\rm mon},\, 0)} +
  M_{\Delta M} + M_{k}
\end{array}
\label{eq.model.incoh.1.7}
\end{equation}
or
\begin{equation}
\begin{array}{llllll}
  f (Q, \chi) =
  \displaystyle\frac{1}{c_{1}} \cdot M_{p}^{(E,\, {\rm mon},\, 0)} \to
  \displaystyle\frac{1}{c_{1}} \cdot M_{p}^{\rm full}.
\end{array}
\label{eq.model.incoh.1.8}
\end{equation}

%-----------------------------------------------------------------------------------------------------------------------

%-----------------------------------------------------------------------------------------------------------------------

% *******************************************************************************************************************
% \end{document}

% *******************************************************************************************************************
% \newpage
\section{Analisys
\label{sec.analysis}}

\subsection{The coherent spectra of production of dilepton pair in the scattering $p + \isotope[9]{Be}$ and experimental data
\label{sec.analysis.1}}

We start calculations
% \textcolor[rgb]{1.00,0.00,0.00}{\textbf{%
of nuclei and energies where experimental data exist.
Such
% \textcolor[rgb]{1.00,0.00,0.00}{\textbf{%
experimental data
% }}
(for proton-nucleus scattering) were obtained by DLS Collaboration for \isotope[]{Be} at proton beam energies of
1.0~GeV, 2.1~GeV and 4~GeV~\cite{Naudet.1989.PRL} and
by HADES Collaboration for \isotope[]{Nb} at proton beam energies of 3.5~GeV~\cite{Agakishiev.HADEScollab.2012.plb,Weber.HADEScollab.2011.JPConfSer}.
Those data were analyzed in details in different approaches
(for example, see Refs.~\cite{Wolf.1993.PPNP,Wolf.1990.NPA}).

At first, we begin calculations for \isotope[9]{Be}.
We calculate wave function of relative motion between proton and center-of-mass of nucleus numerically
concerning to the proton-nucleus potential in form of $V (r) = v_{c}(r) + v_{N}(r) + v_{\rm so}(r) + v_{l} (r)$,
where $v_{c}(r)$, $v_{N}(r)$, $v_{\rm so}(r)$, and $v_{l} (r)$ are Coulomb, nuclear, spin-orbital, and centrifugal components, respectively.
Parameters of the potential are defined in Eqs.~(46)--(47) in Ref.~\cite{Maydanyuk_Zhang.2015.PRC}
%
% \textcolor[rgb]{1.00,0.00,0.00}{\textbf{%
(see also App.~\ref{sec.app.4}).
% }}
At first, we will analyze coherent processes
% \textcolor[rgb]{1.00,0.00,0.00}{\textbf{%
for production of dileptons.
We calculate the coherent
% \textcolor[rgb]{1.00,0.00,0.00}{\textbf{%
cross section using
% }}
Eq.~(\ref{eq.model.crosssection.2.9}).
% where we include matrix elements of coherent emission $M_{p}^{(E,\, {\rm dip})}$, $M_{p}^{(M,\, {\rm dip})}$ in Eqs.~(\ref{eq.resultingformulas.1}),
% and matrix elements of incoherent emission $ M_{\Delta M}$, $M_{k}$ in Eqs.~(\ref{eq.resultingformulas.2}).
% The boundary conditions and normalization are used in form of~(B.1)--(B.9) in  \cite{Maydanyuk.2011.JPG}.
% \footnote{One proton from beam can transfer energy to one nucleon of nucleus for transition $NN \to \Delta N$ with formation of $\Delta$-resonance in nucleus.
% Other protons of beam with nucleons of nucleus-target and $\Delta$-resonance can emit bremsstrahlung photons.
%
% \textcolor[rgb]{1.00,0.00,0.00}{\textbf{%
Through considering after in the product channels like Dalitz-decay, and decays for vector mesons,
new normalization would be necessary.
% }}

\begin{figure}[htbp]
\centerline{\includegraphics[width=90mm]{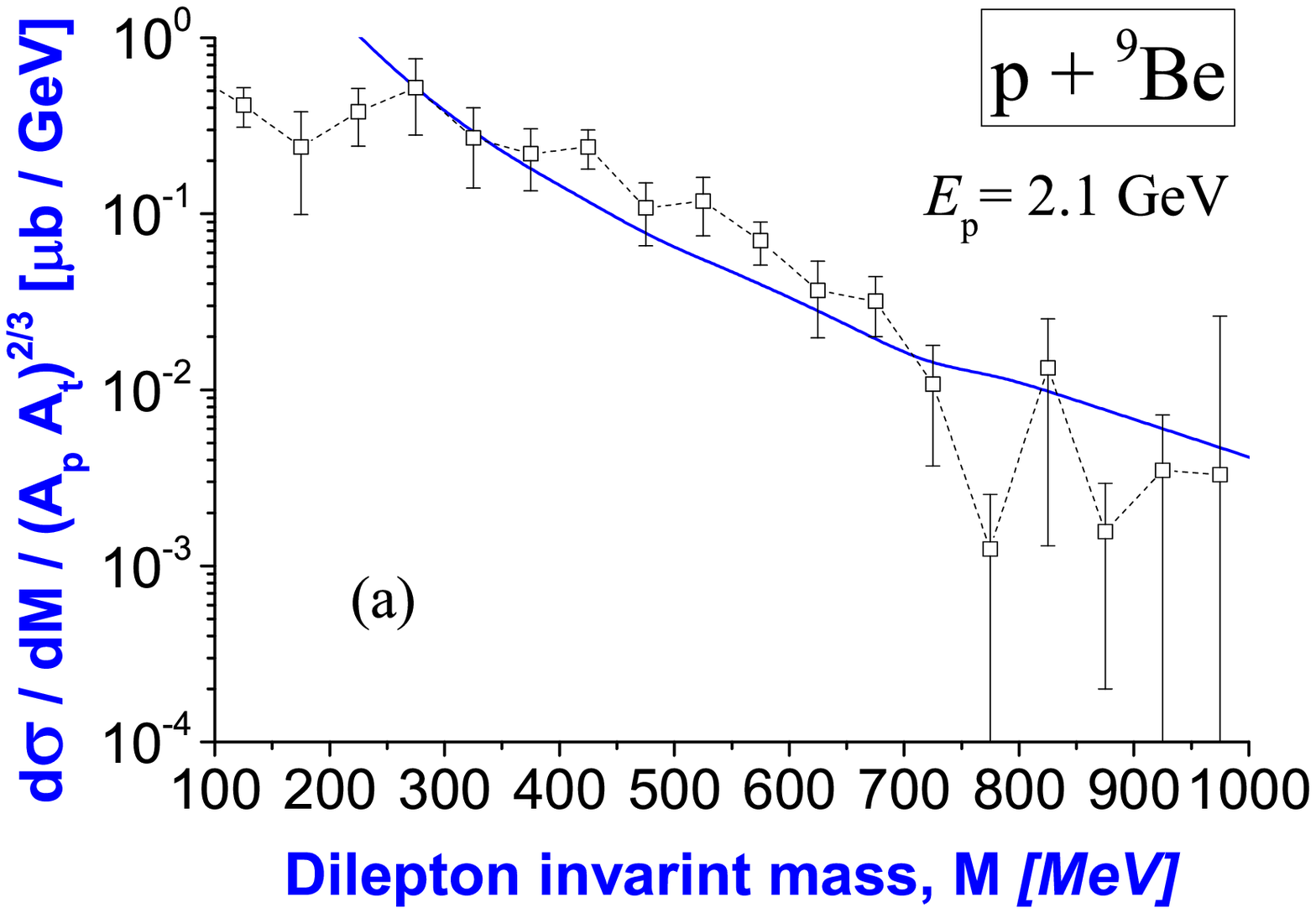}
\hspace{-1mm}\includegraphics[width=90mm]{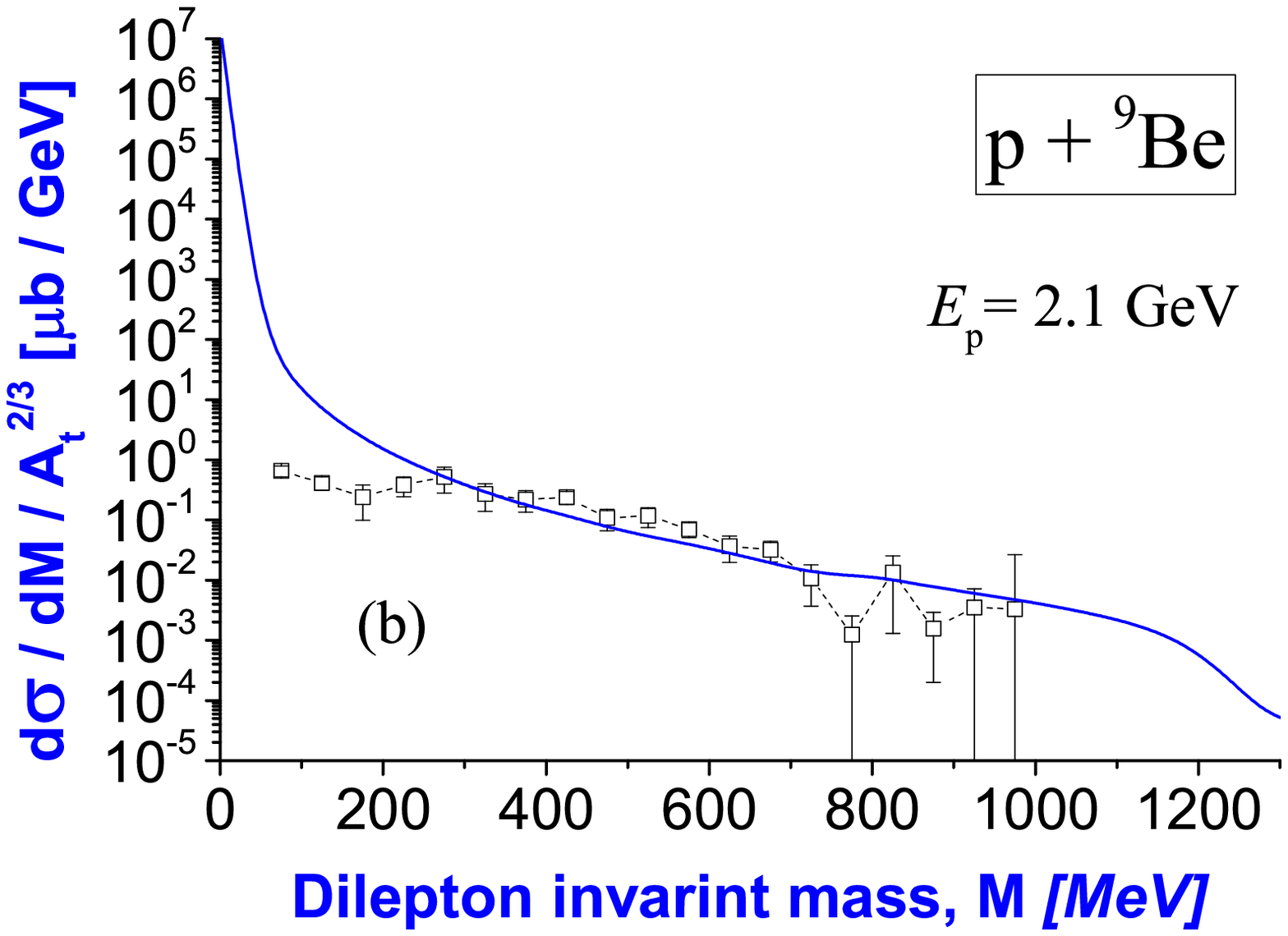}}
\vspace{-4mm}
\caption{\small (Color online)
The calculated coherent cross section of production of leptons pair % (with coherent and incoherent terms)
in the scattering of protons off the \isotope[9]{Be} nuclei at energy of proton beam of $E_{\rm p}=2.1$~GeV
in comparison with experimental data~\cite{Naudet.1989.PRL}
[% matrix elements are defined in Eqs.~(\ref{eq.resultingformulas.1})--(\ref{eq.resultingformulas.2}),
cross section is defined in Eq.~(\ref{eq.model.crosssection.2.9}),
function $f^{}$ is defined in Eqs.~(\ref{eq.resultingformula.5}),
radial integrals are defined in Eqs.~(\ref{eq.resultingformula.6}),
% $Z_{A} (k_{\rm ph})\simeq Z_{A}$ is electric charge of nucleus,
we normalize
% \textcolor[rgb]{1.00,0.00,0.00}{\textbf{%
the
calculated spectrum on one point of experimental data;
time of computer calculations is 26--30 min for 40 points of each calculated spectrum].
Here,
experimental data given by open rectangles (Naudet 1989)
are extracted from Ref.~\cite{Naudet.1989.PRL},
blue solid line is the calculated coherent cross section.
% defined by $M_{p}^{(E,\, {\rm dip})}$.
% red solid line is full spectrum with coherent and incoherent contributions defined by $M_{p}^{(E,\, {\rm dip})}$, $M_{p}^{(M,\, {\rm dip})}$, $M_{\Delta M}$ and $M_{k}$.
%
Panel (a):
Picture of comparison between the calculated spectra and experimental data.
% The calculated bremsstrahlung spectra (with coherent and incoherent terms)
% in the scattering of protons off the \isotope[197]{Au} nuclei at energy of proton beam of $E_{\rm p}=190$~MeV
% in comparison with experimental data~\cite{Goethem.2002.PRL}
% [matrix elements are defined in Eqs.~(\ref{eq.resultingformulas.1})--(\ref{eq.resultingformulas.2}),
% $Z_{A} (k_{\rm ph})\simeq Z_{A}$ is electric charge of nucleus].
% blue dashed line is coherent contribution defined by $M_{p}^{(E,\, {\rm dip})}$,
% red solid line is full spectrum with coherent and incoherent contributions defined by $M_{p}^{(E,\, {\rm dip})}$, $M_{p}^{(M,\, {\rm dip})}$, $M_{\Delta M}$ and $M_{k}$.
%
% \vspace{1.5mm}
% \newline
Panel (b):
The same calculated spectra shown in larger region of mass
% \textcolor[rgb]{1.00,0.00,0.00}{\textbf{%
(for more complete picture).
% New calculated spectrum of full bremsstrahlung for \isotope[197]{Au} at energy of proton beam of $E_{\rm p}=800$~MeV
%
% Ration between incoherent and coherent contributions in dependence on energy of emitted bremsstrahlung photon
% [we define ratio as $\varepsilon = \sigma_{\rm incoh}/\sigma_{\rm coh}$, in calculations we use factor of incoherence of $f_{\rm incoh}=1.0$]
% related to full spectrum shown by dashed brown line in figure (a)].
% One can see that
% (a) incoherent emission is essentially more intensive than the coherent emission,
% (b) role of incoherent processes is increased at increasing of energy of photon,
% (c) better agreement with experimental data at $f_{\rm incoh}=0.001$ (than at $f_{\rm incoh}=1$) indicates on presence of some unknown effect,
% which highly suppresses the incoherent processes.
%
% factor $f_{\rm incoh}$ which suppresses the intensity of incoherent processes.
% One can see clear changes of the full spectrum in dependence on factor $f_{\rm incoh}$,
% red dash-dotted solid line (at $f=0.001$) corresponds to the best agreement with experimental data.
% This result confirms the important (and not small) role of incoherent emission in bremsstrahlung.
% (b) Contributions of the bremsstrahlung emission given by term $p_{\rm q,1}$ with different $Q$ in comparison with electric and magnetic emissions.
% Можно видеть, что магнитное излучение вносит вклад около 28 процентов в диапазоне энергий 50--300~кэВ.
\label{fig.2}}
\end{figure}

% Results of previous study of bremsstrahlung emission
% \cite{Maydanyuk_Zhang.2015.PRC,Maydanyuk.2012.PRC,Maydanyuk_Zhang_Zou.2016.PRC,Liu_Maydanyuk_Zhang_Liu.2019.PRC.hypernuclei}
% \cite{Maydanyuk_Zhang.2015.PRC,Maydanyuk.2012.PRC,Maydanyuk_Zhang_Zou.2016.PRC,Liu_Maydanyuk_Zhang_Liu.2019.PRC.hypernuclei,Maydanyuk_Zhang_Zou.2019.PRC.microscopy}
% show that incoherent emission is essentially larger than coherent one.
Results of such calculations for \isotope[9]{Be} in comparison with experimental data are presented in Fig.~\ref{fig.2}.
%
% \cite{Maydanyuk_Zhang.2015.PRC,Maydanyuk.2012.PRC,Maydanyuk_Zhang_Zou.2016.PRC,Liu_Maydanyuk_Zhang_Liu.2019.PRC.hypernuclei}
% \cite{Maydanyuk_Zhang.2015.PRC,Maydanyuk.2012.PRC,Maydanyuk_Zhang_Zou.2016.PRC,Liu_Maydanyuk_Zhang_Liu.2019.PRC.hypernuclei,Maydanyuk_Zhang_Zou.2019.PRC.microscopy}
% show that incoherent emission is essentially larger than coherent one.
% By such a reason, we start calculations for \isotope[197]{Au}, where experimental data exist.
% Results of such calculations with inclusion of coherent and incoherent contributions in comparison with experimental data are presented in Fig.~\ref{fig.1}~(a).
%
The calculated spectra are normalized on one point of these experimental data.
% These calculations are shown in Fig.~1~(a).
% All other calculated spectra for different nuclei and energies of proton beam use the same coefficient of normalization (this gives possibility to predict new spectra).
This shows that our model provides the spectrum in good agreement with experimental data (with exception of few points).

% These calculations are performed for test of our model, before its next use for analysis in the paper.
% Note that the first point in experimental data is not explained in satisfactory way by model with included incoherent bremsstrahlung contribution.
% Without the incoherent contribution, the calculated renormalized spectrum is in worse agreement with experimental data.
% Presence of such a point could indicate on existence of some unknown processes forming more intensive coherent bremsstrahlung emission at low energies of photons.
% That problem can motivate on next investigations and developments of the model or more precise measurements of emission of photons at low energies in possible future experiments.
% But, in current research we omit analysis of this problem.
% *******************************************************************************************************************

% *******************************************************************************************************************
In the next Fig.~\ref{fig.3} we show
% \textcolor[rgb]{1.00,0.00,0.00}{\textbf{%
the results of calculation of the coherent cross sections of dilepton production for different nuclei-targets
in region from low up to large masses at the same energy of proton beam of 2.1~GeV.
%
% \cite{Maydanyuk_Zhang.2015.PRC,Maydanyuk.2012.PRC,Maydanyuk_Zhang_Zou.2016.PRC,Liu_Maydanyuk_Zhang_Liu.2019.PRC.hypernuclei}
% \cite{Maydanyuk_Zhang.2015.PRC,Maydanyuk.2012.PRC,Maydanyuk_Zhang_Zou.2016.PRC,Liu_Maydanyuk_Zhang_Liu.2019.PRC.hypernuclei,Maydanyuk_Zhang_Zou.2019.PRC.microscopy}
% show that incoherent emission is essentially larger than coherent one.
% Results of such calculations with inclusion of coherent and incoherent contributions in comparison with experimental data are presented in Fig.~\ref{fig.1}~(a).
%
\begin{figure}[htbp]
\centerline{\includegraphics[width=90mm]{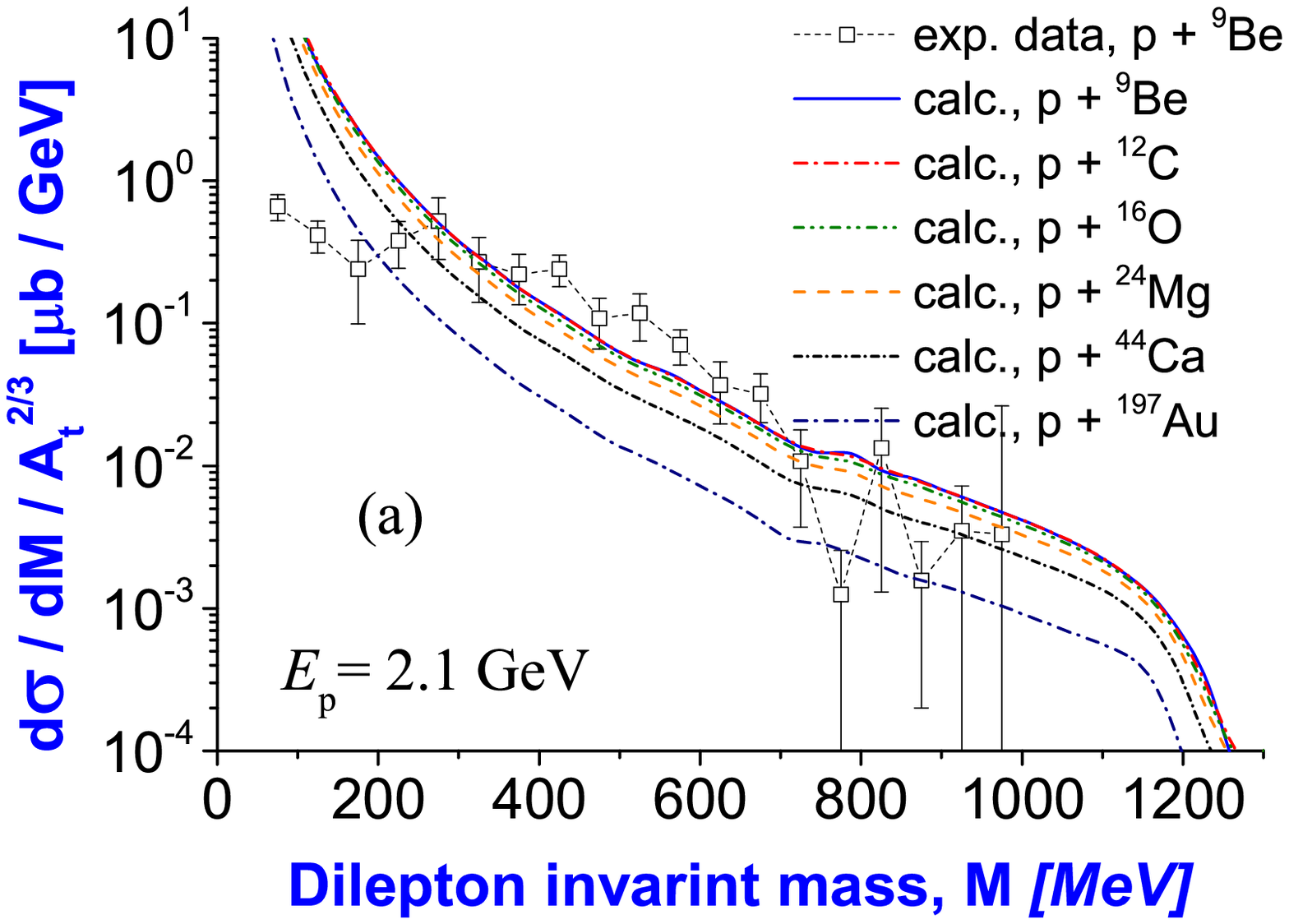}
\hspace{-1mm}\includegraphics[width=90mm]{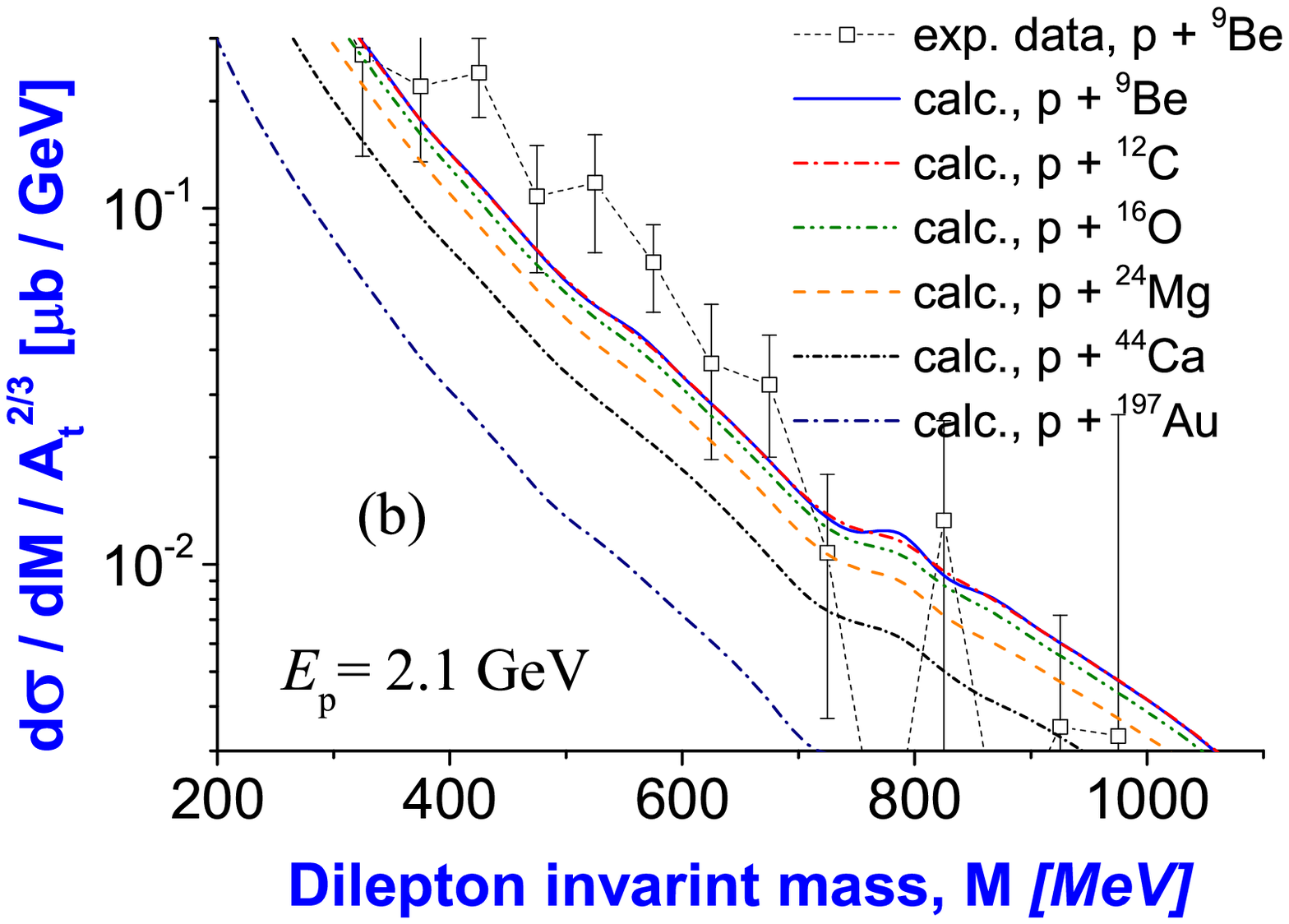}}
\vspace{-4mm}
\caption{\small (Color online)
The calculated coherent cross section of production of leptons pair % (with coherent and incoherent terms)
in the scattering of protons off
the \isotope[9]{Be}, \isotope[12]{C}, \isotope[16]{O}, \isotope[24]{Mg}, \isotope[44]{Ca}, \isotope[197]{Au} nuclei at energy of proton beam of $E_{\rm p}=2.1$~GeV
in comparison with experimental data~\cite{Naudet.1989.PRL} for \isotope[9]{Be}
[% matrix elements are defined in Eqs.~(\ref{eq.resultingformulas.1})--(\ref{eq.resultingformulas.2}),
cross section is defined in Eq.~(\ref{eq.model.crosssection.2.9}),
function $f^{}$ is defined in Eqs.~(\ref{eq.resultingformula.5}),
radial integrals are defined in Eqs.~(\ref{eq.resultingformula.6}),
% $Z_{A} (k_{\rm ph})\simeq Z_{A}$ is electric charge of nucleus,
all calculated spectra have the same normalization factor found in previous Fig.~\ref{fig.2}].
Here,
experimental data given by open rectangles (Naudet 1989)
are extracted from Ref.~\cite{Naudet.1989.PRL},
blue solid line is the calculated coherent cross section.
\label{fig.3}}
\end{figure}
From this figure one can see that cross section of dilepton production is smaller for nuclei-target with larger mass.
I.e. this is a new effect of suppressing of production of dileptons in the proton-nucleus scattering (by nuclear matter of nucleus-target).
There is a general tendency of monotonous decreasing of production of dileptons with increasing of mass of the nucleus-target.
Very close similarities in shapes of each calculated spectrum are observed inside the full studied region of masses
that confirms stability of
% \textcolor[rgb]{1.00,0.00,0.00}{\textbf{%
the calculations.
This
% \textcolor[rgb]{1.00,0.00,0.00}{\textbf{%
is a good test of the developed model and computer algorithms in numerical calculations.
% *******************************************************************************************************************

% *******************************************************************************************************************
Next question can be in how much the spectra are different for different isotopes of nuclei-target with the same chosen charge number.
Which isotope for the given nucleus with fixed charge number is better for more easy future experimental measurements?
% Calculations to clarify this question are shown in Fig.~\ref{fig.4}.
However, our detailed calculations and analysis have shown that the spectra are not
% \textcolor[rgb]{1.00,0.00,0.00}{\textbf{%
very sensitive
% \textcolor[rgb]{1.00,0.00,0.00}{\textbf{%
on the
choice of different isotope with the same charge number,
both for light nuclei and heavy nuclei.
This phenomenon is observed for the proton nucleus scattering,
but we estimate that for nucleus-nucleus scattering (heavy-ion collisions) such dependencies should be much larger.
This is clearly explained by dependencies of effective charge $Z_{\rm eff}^{\rm (mon)} (\vb{k}_{\rm ph})$ on different masses and charges of two participating nuclei
[see Eq.~(\ref{eq.resultingformula.8}), p.~\pageref{eq.resultingformula.8}] in reaction in formula for matrix elements.
But, we transfer this analysis for the nucleus-nucleus scattering for future investigations.
% *******************************************************************************************************************

\begin{figure}[htbp]
\centerline{\includegraphics[width=90mm]{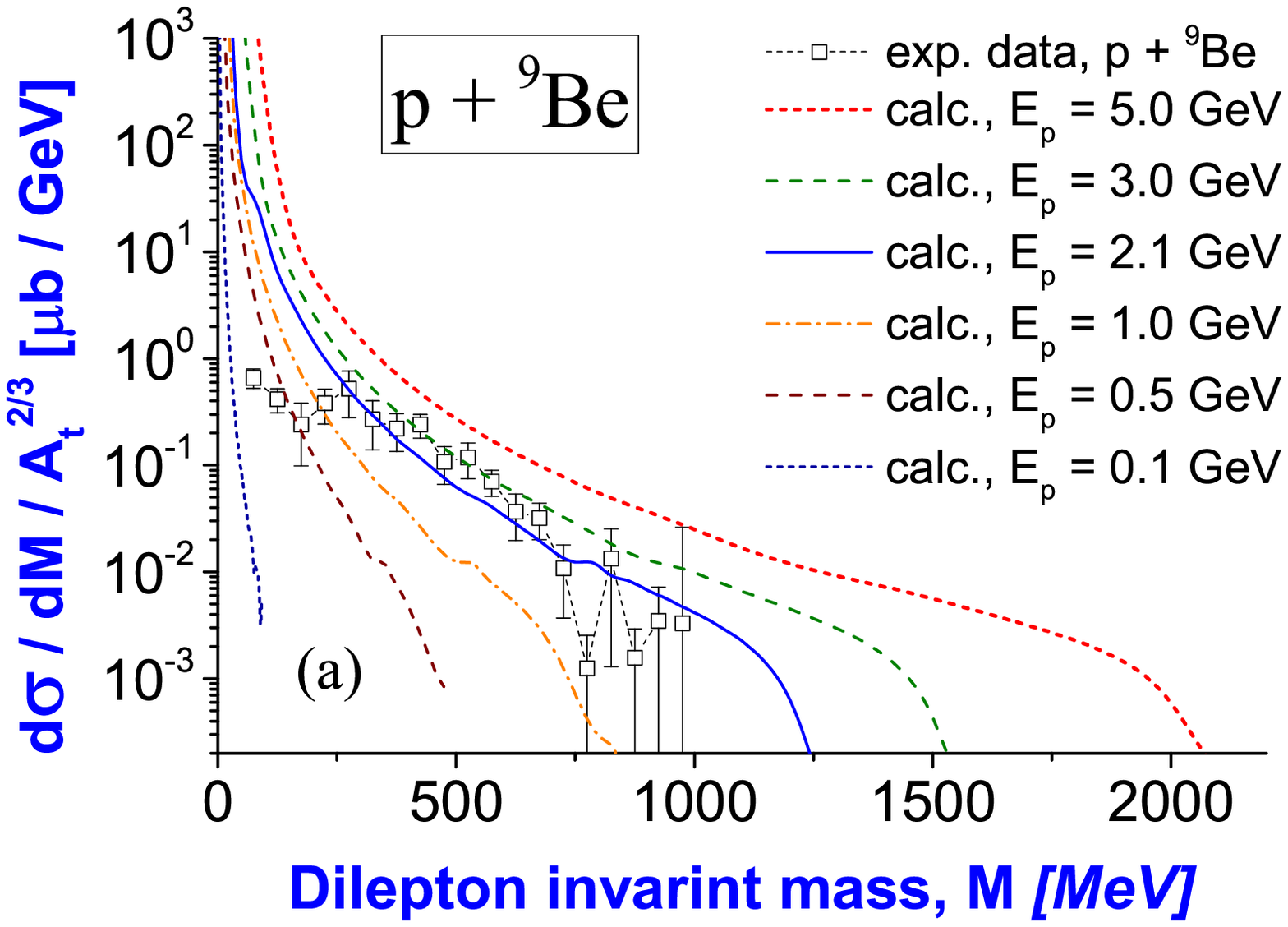}}
\vspace{-3mm}
\caption{\small (Color online)
The calculated coherent cross sections of production of leptons pair % (with coherent and incoherent terms)
in the scattering of protons off the \isotope[9]{Be} nuclei
% \textcolor[rgb]{1.00,0.00,0.00}{\textbf{%
at different proton beam energies, $E_{\rm p}$
% }}
% in comparison with experimental data~\cite{Naudet.1989.PRL} for \isotope[9]{Be}
[% matrix elements are defined in Eqs.~(\ref{eq.resultingformulas.1})--(\ref{eq.resultingformulas.2}),
cross section is defined in Eq.~(\ref{eq.model.crosssection.2.9}),
function $f^{}$ is defined in Eqs.~(\ref{eq.resultingformula.5}),
radial integrals are defined in Eqs.~(\ref{eq.resultingformula.6}),
% % $Z_{A} (k_{\rm ph})\simeq Z_{A}$ is electric charge of nucleus,
all calculated spectra have the same normalization factor found in Fig.~\ref{fig.2}].
\label{fig.4}}
\end{figure}

% *******************************************************************************************************************
Now we
% \textcolor[rgb]{1.00,0.00,0.00}{\textbf{%
show the
spectra the model gives at different energies of proton beam for the same fixed nucleus-target.
In Fig.~\ref{fig.4} we show
% \textcolor[rgb]{1.00,0.00,0.00}{\textbf{%
the results of such calculations for \isotope[9]{Be}.
%
% \textcolor[rgb]{1.00,0.00,0.00}{\textbf{%
From these figures one can see that
at larger proton beam energies, $E_{\rm p}$, the production of lepton pairs is more intensive.
% }}
One can see that each spectrum at increasing of invariant mass $M$ tends to some fixed maximum of this mass, that is explained by kinematic limit
from relation between energies of relative motion of proton-nucleus scattering before emission of virtual photon, after such an emission and
energy (invariant mass) of this emitted photon.
% *******************************************************************************************************************

% *******************************************************************************************************************
\subsection{Role of incoherent processes in production of pairs of leptons
\label{sec.analysis.incoh}}

In this Section we will
% \textcolor[rgb]{1.00,0.00,0.00}{\textbf{%
study the question how important is the incoherent processes
% }}
in problem of production of pairs of leptons in proton-nucleus scattering?
A clear answer can be obtained from estimation of the coherent and incoherent contributions to the full cross section of dilepton production.

\begin{figure}[htbp]
\centerline{\includegraphics[width=90mm]{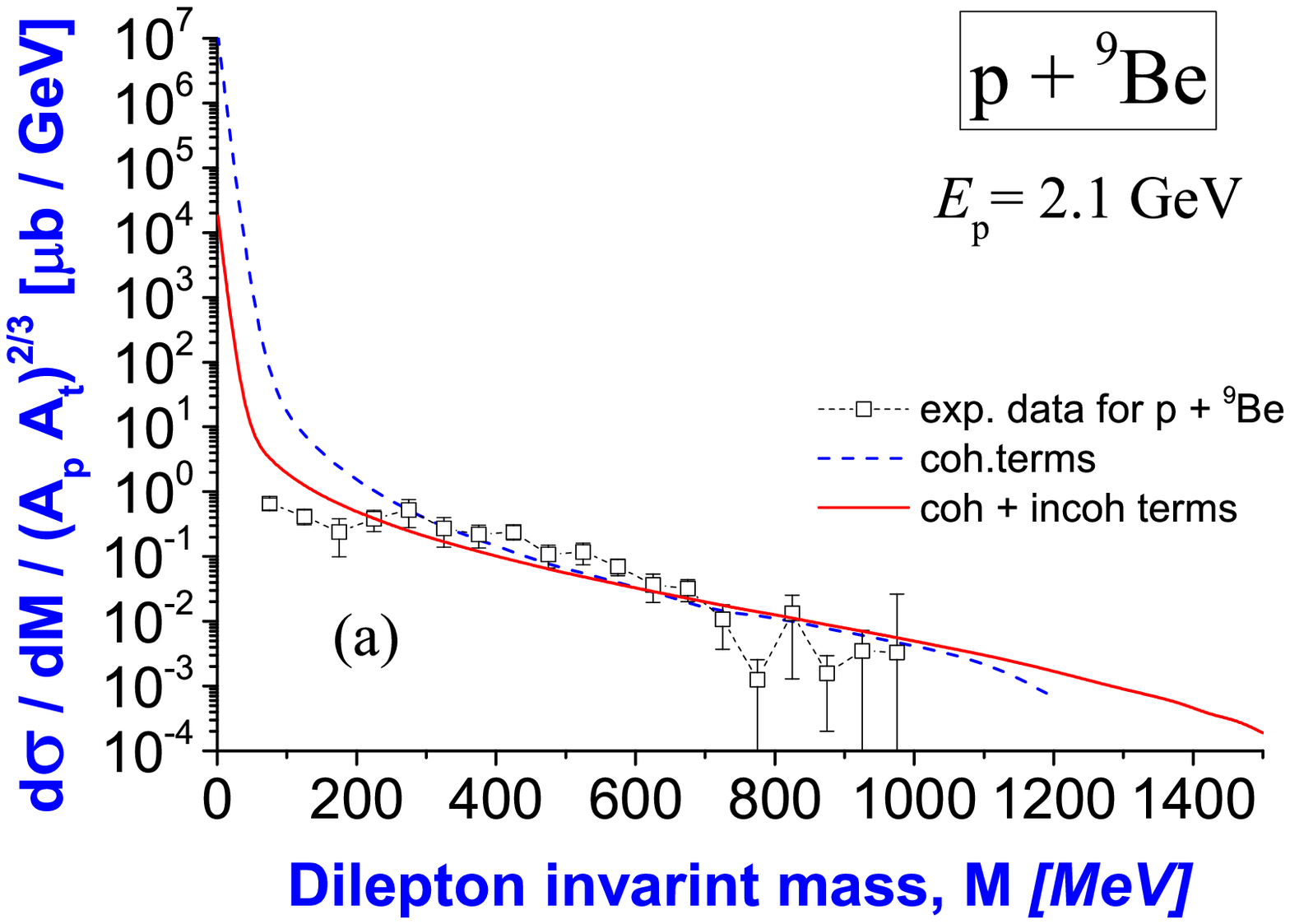}
\hspace{-1mm}\includegraphics[width=90mm]{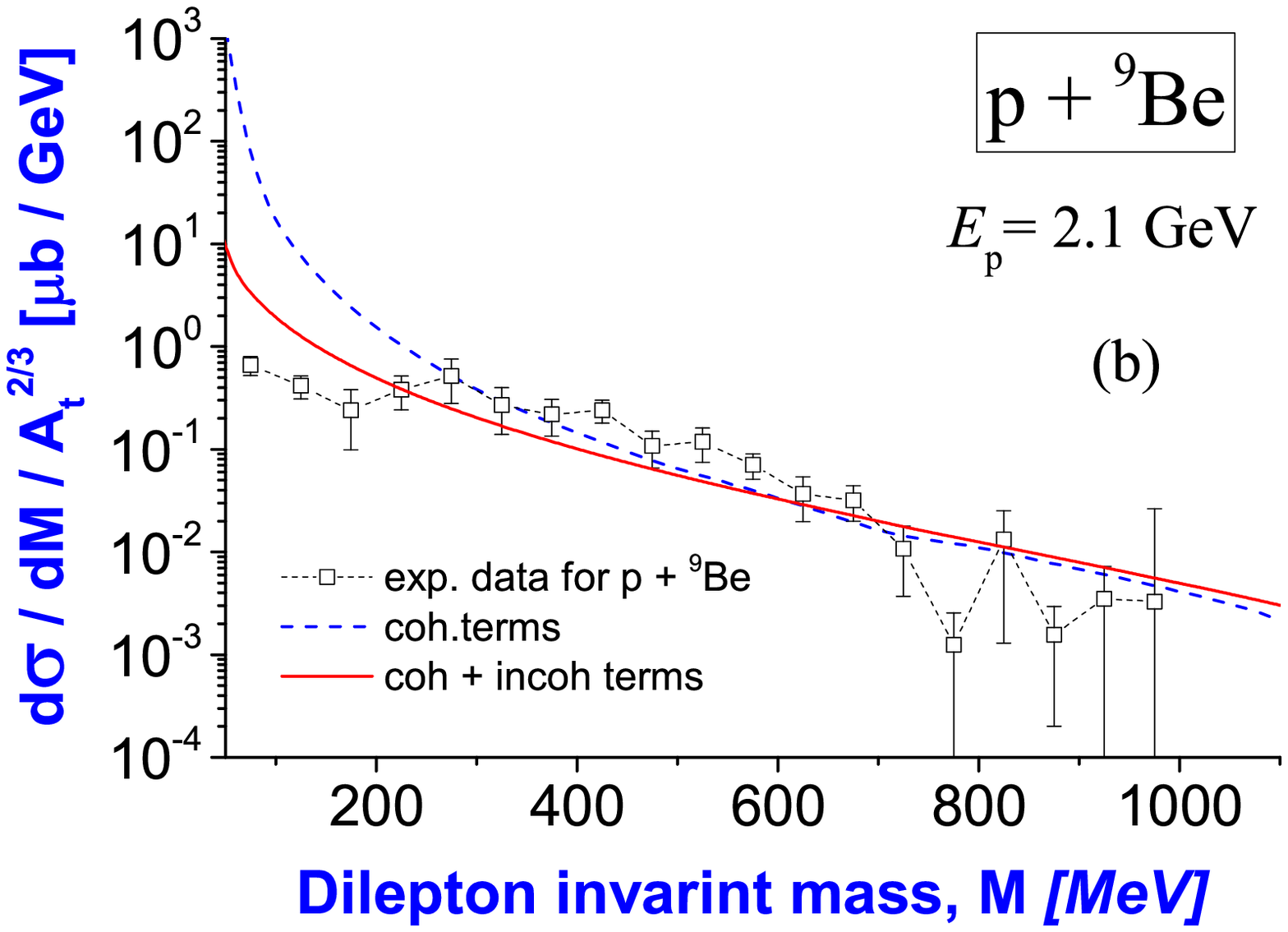}}
\vspace{-3mm}
\caption{\small (Color online)
The calculated full cross sections of production of leptons pair (with coherent and incoherent terms)
in the scattering of protons off the \isotope[9]{Be} nuclei at energy of proton beam of $E_{\rm p}=2.1$~GeV
in comparison with experimental data~\cite{Naudet.1989.PRL} % for \isotope[9]{Be}
[cross section is defined in Eq.~(\ref{eq.model.crosssection.2.9}),
function $f^{}$ is defined in Eqs.~(\ref{eq.model.incoh.1.8}),
radial integrals are defined in Eqs.~(\ref{eq.resultingformula.6}),
all calculated spectra have own normalization factors on experimental data].
Here,
experimental data given by open rectangles (Naudet 1989) are extracted from Ref.~\cite{Naudet.1989.PRL},
blue dashed line is the calculated cross section based on the coherent matrix elements
$M_{p}^{(E,\, {\rm mon},\, 0)}$ and $M_{p}^{(M,\, {\rm mon},\, 0)}$ only,
red solid line is the full spectrum with coherent and incoherent contributions
on the basis of matrix elements $M_{p}^{(E,\, {\rm mon},\, 0)}$, $M_{p}^{(M,\, {\rm mon},\, 0)}$, $M_{\Delta M}$ and $M_{k}$.
% Panel (a):
% \vspace{1.5mm}
% \newline
% Panel (b):
% Ration between incoherent and coherent contributions in dependence on energy of emitted bremsstrahlung photon
% [we define ratio as $\varepsilon = \sigma_{\rm incoh}/\sigma_{\rm coh}$, in calculations we use factor of incoherence of $f_{\rm incoh}=1.0$]
% related to full spectrum shown by dashed brown line in figure (a)].
% One can see that
% (a) incoherent emission is essentially more intensive than the coherent emission,
% (b) role of incoherent processes is increased at increasing of energy of photon,
% (c) better agreement with experimental data at $f_{\rm incoh}=0.001$ (than at $f_{\rm incoh}=1$) indicates on presence of some unknown effect,
% which highly suppresses the incoherent processes.
% factor $f_{\rm incoh}$ which suppresses the intensity of incoherent processes.
% One can see clear changes of the full spectrum in dependence on factor $f_{\rm incoh}$,
% red dash-dotted solid line (at $f=0.001$) corresponds to the best agreement with experimental data.
% This result confirms the important (and not small) role of incoherent emission in bremsstrahlung.
% (b) Contributions of the bremsstrahlung emission given by term $p_{\rm q,1}$ with different $Q$ in comparison with electric and magnetic emissions.
% Можно видеть, что магнитное излучение вносит вклад около 28 процентов в диапазоне энергий 50--300~кэВ.
\label{fig.5}}
\end{figure}

For analysis we use the scattering of $p + \isotope[9]{Be}$ at $E_{\rm p}=2.1$~GeV
% \textcolor[rgb]{1.00,0.00,0.00}{\textbf{%
analyzed previously.
% }}
Results of such calculation of the spectrum based on the coherent contribution in comparison with
the full spectrum (including coherent and incoherent terms) and experimental data
are presented in Fig.~\ref{fig.5}.
From these figures one can see that inclusion of incoherent processes to the model and calculations improves agreement with experimental data a little.
In particular, it allows to achieve better description of experimental data at low energies smaller
% \textcolor[rgb]{1.00,0.00,0.00}{\textbf{%
than 250~MeV.
% }}
% Our predictions in regions of low energies $M <250$~MeV and high energies $M > 1000$~MeV can be recommended for tests in possible experimental study in future.

\begin{figure}[htbp]
\centerline{\includegraphics[width=90mm]{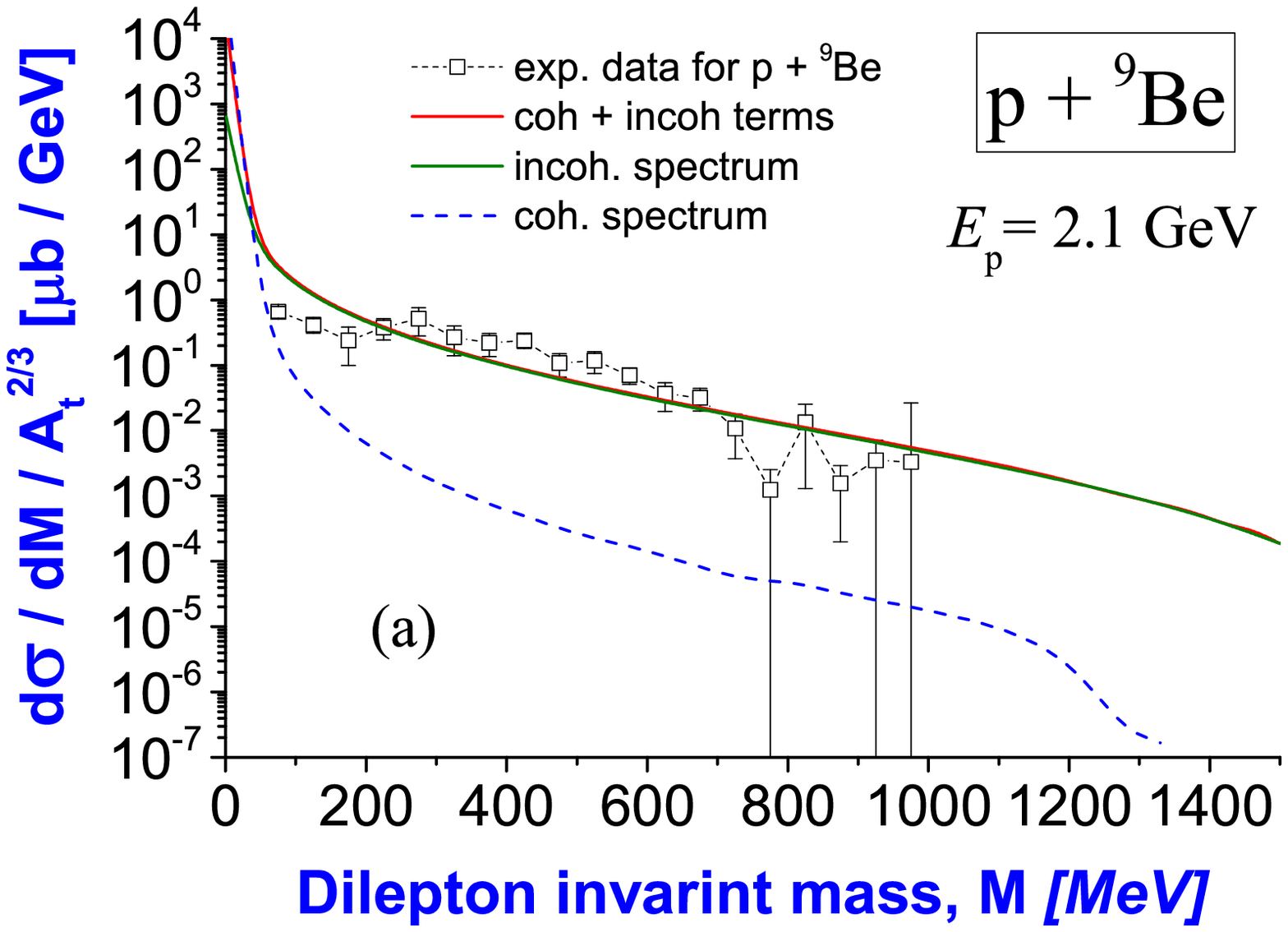}
\hspace{-1mm}\includegraphics[width=90mm]{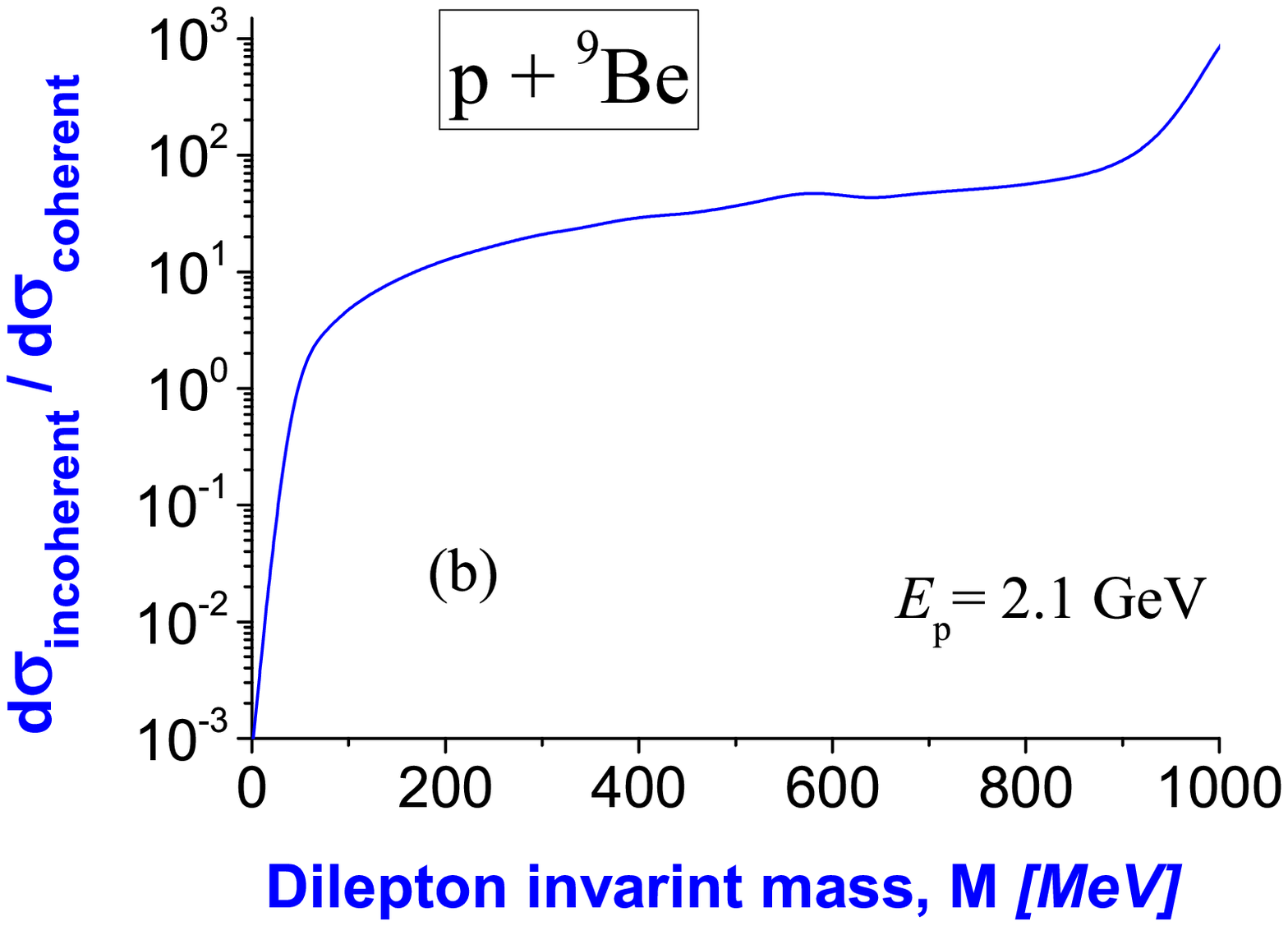}}
\vspace{-3mm}
\caption{\small (Color online)
The calculated incoherent and coherent contributions to the full cross sections of production of leptons pairs
in the scattering of protons off the \isotope[9]{Be} nuclei at energy of proton beam of $E_{\rm p}=2.1$~GeV
in comparison with experimental data~\cite{Naudet.1989.PRL} (a) and
% \textcolor[rgb]{1.00,0.00,0.00}{\textbf{%
the ratio between the incoherent and coherent contributions for this reaction (b)
[cross section is defined in Eq.~(\ref{eq.model.crosssection.2.9}),
function $f^{}$ is defined in Eqs.~(\ref{eq.model.incoh.1.8}),
radial integrals are defined in Eqs.~(\ref{eq.resultingformula.6}),
all calculated spectra have own normalization factors on experimental data].
Here,
experimental data given by open rectangles (Naudet 1989) are extracted from Ref.~\cite{Naudet.1989.PRL},
blue dashed line is the calculated cross section based on the coherent matrix elements
$M_{p}^{(E,\, {\rm mon},\, 0)}$ and $M_{p}^{(M,\, {\rm mon},\, 0)}$ only,
green solid line is the calculated cross section based on the incoherent matrix elements
$M_{\Delta M}$ and $M_{k}$,
red solid line is the full spectrum with coherent and incoherent contributions
on the basis of matrix elements $M_{p}^{(E,\, {\rm mon},\, 0)}$, $M_{p}^{(M,\, {\rm mon},\, 0)}$, $M_{\Delta M}$ and $M_{k}$.
\label{fig.6}}
\end{figure}

Result above can be explained if the incoherent contribution is larger than the coherent one. If this is so, then the incoherent terms are important to be included to the model and calculations.
In Fig.~\ref{fig.6} the different contributions in comparison with the full spectrum (a) and ratio between the incoherent contribution and coherent one (b) are presented.
From figure~(a) one can see that incoherent contribution is essentially more important than the coherent one.
By other words, the incoherent contribution has a leading role in production of pairs of leptons for the studied reaction.
From figure~(b) one can see that
\emph{the ratio between incoherent and coherent contributions is about 10--100 inside main region of invariant mass values}.

% \emph{Also we observe a new phenomenon of suppression of production of lepton pairs at low energy region due to incoherent processes.}
% This is explained by that at very low energies coherent contribution is dominant in comparison with incoherent one.
% This is confirmed by calculations in Fig.~\ref{fig.6}.

\begin{figure}[htbp]
\centerline{\includegraphics[width=90mm]{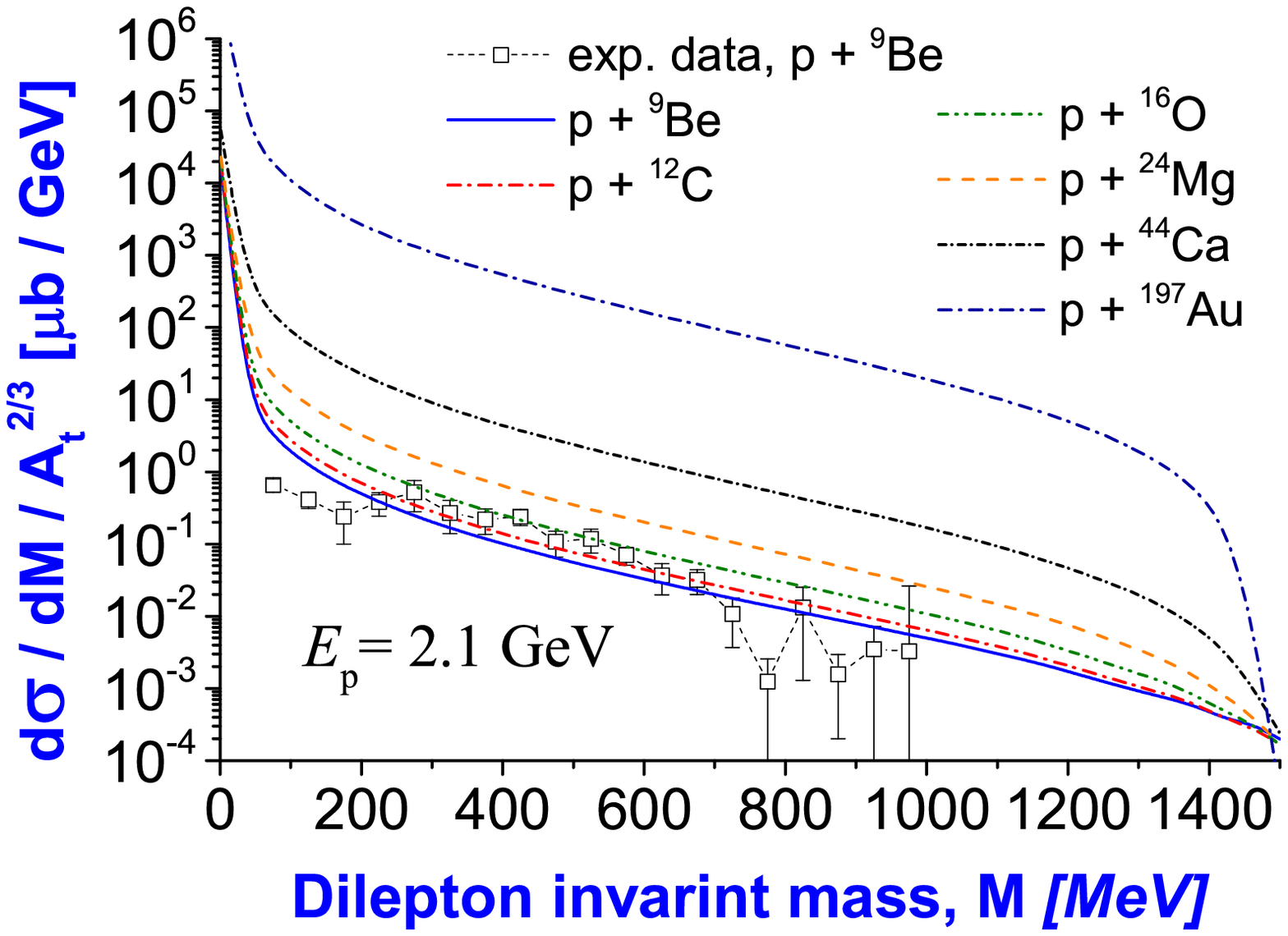}}
\vspace{-3mm}
\caption{\small (Color online)
The calculated full cross sections of production of leptons pair
in the scattering of protons off the \isotope[9]{Be} nuclei
at energy of proton beam of $E_{\rm p}=2.1$~GeV
in comparison with experimental data for \isotope[9]{Be}~\cite{Naudet.1989.PRL}
[time of computer calculations is 26--40 min for 40 points of each calculated spectrum].
% (a) and ratio between the incoherent and coherent contributions for this reaction (b)
% [cross section is defined in Eq.~(\ref{eq.model.crosssection.2.9}),
% function $f^{}$ is defined in Eqs.~(\ref{eq.model.incoh.1.8}),
% radial integrals are defined in Eqs.~(\ref{eq.resultingformula.6}),
% all calculated spectra have own normalization factors on experimental data].
% Here,
% experimental data given by open rectangles (Naudet 1989) are extracted from Ref.~\cite{Naudet.1989.PRL},
% blue dashed line is the calculated cross section based on the coherent matrix elements
% $M_{p}^{(E,\, {\rm mon},\, 0)}$ and $M_{p}^{(M,\, {\rm mon},\, 0)}$ only,
% green solid line is the calculated cross section based on the incoherent matrix elements
% $M_{\Delta M}$ and $M_{k}$,
% red solid line is the full spectrum with coherent and incoherent contributions
% on the basis of matrix elements $M_{p}^{(E,\, {\rm mon},\, 0)}$, $M_{p}^{(M,\, {\rm mon},\, 0)}$, $M_{\Delta M}$ and $M_{k}$.
\label{fig.6add}}
\end{figure}

In Fig.~~\ref{fig.6add} we add results of calculations of full spectra at $E_{\rm p}=2.1$~GeV.
From these calculations one can see that full cross section of production of dileptons is larger for heavier nucleus-target
(in contrast to calculations of the coherent contributions).
This is explained by that role of incoherent contribution is increased for larger mass of nucleus-target.

% *******************************************************************************************************************

% *******************************************************************************************************************
\subsection{Role of longitudinal part of virtual photons in production of pairs of leptons
\label{sec.analysis.virtual}}

Let us analyze influence of
% \textcolor[rgb]{1.00,0.00,0.00}{\textbf{%
the longitudinal part of virtual photon on the obtained result of production of dilepton pairs.
% \textcolor[rgb]{1.00,0.00,0.00}{\textbf{%
For simplicity, let's reduce ourselves by the coherent processes only.
% We will be interesting in question if is it important to study incoherent processes in problem of production of pairs of leptons in the proton-nucleus scattering?
We write down the final formulas for calculation of cross section of dilepton production based on coherent term
with taking into account
% \textcolor[rgb]{1.00,0.00,0.00}{\textbf{%
the longitudinal part of virtual photon~\cite{Zetenyi_Wolf.2012.PRC}:
%
% [see Eqs.~(\ref{eq.app.model.crosssection.2.9}), (\ref{eq.app.model.crosssection.2.11}), p.~\pageref{eq.app.model.crosssection.2.9};
% Eq.~(\ref{eq.resultingformula.5}), p.~\pageref{eq.resultingformula.5}]
%
\begin{equation}
\begin{array}{lllll}
\vspace{0.5mm}
  d^{2} \sigma^{\rm (lep)} =
  \displaystyle\frac{(2\pi)^{8} e^{2}}{8\, c^{5}} \cdot

  |f\, (M, \chi)|^{2}\,
  \displaystyle\frac{E_{i}}{k_{i}\, E_{\rm e, f1}^{2}}\;
  |\vb{k}_{{\rm e}, f}|\, \Bigl[ 2\, \sin^{2} \theta + (1 - \cos{2\,\theta}) \Bigr]\;
  \displaystyle\frac{1 + \cos{2\, \theta}}{1 - \cos{2\, \theta}}\; do_{e,2}, \\

\vspace{0.5mm}
  f\, (M, \chi) =
  -\, \sqrt{\displaystyle\frac{2\pi}{3 \hbar w_{\rm ph}}}\,
  (2\pi)^{3} \displaystyle\frac{\mu_{N}\,  m_{\rm p}}{\mu}\:
  Z_{\rm eff}^{\rm (mon,\, 0)}\,
  \Bigl\{
    J_{1}^{vir}(0,1,0, 0) -
    \displaystyle\frac{47}{40} \sqrt{\displaystyle\frac{1}{2}} \cdot J_{1}^{vir}(0,1,2, 0)
  \Bigr\}, \\

% \vspace{0.5mm}
  |\vb{k}_{\rm e, 1}| = \displaystyle\frac{M}{2}, \quad

%   |\vb{p}_{\rm e, 1}| =
%   \displaystyle\frac{M}{\sqrt{2\, (1 + \cos{2\, \theta})}}\,
%   \displaystyle\frac{1}{\sqrt{(1 + \chi^{2})}}, \quad

  E_{\rm e}^{2} = |\vb{k}_{{\rm e}, f}|^{2} + m_{\rm e}^{2}, \quad
  \chi = \displaystyle\frac{k_{\rm ph}^{\parallel}}{k_{\rm ph}^{\bot}},
\end{array}
\label{eq.model.virtual.1.1}
% \label{eq.app.model.crosssection.2.9}
\end{equation}
and dilepton invariant mass $M$ is defined as
[% see Eqs.~(\ref{eq.app.model.crosssection.2.10}), p.~\pageref{eq.app.model.crosssection.2.10},
$E_{\rm ph} = w_{\rm ph}$]
\begin{equation}
\begin{array}{rllll}
  M_{\rm lep}^{2} \simeq
  w_{\rm ph}^{2}, &
%  k_{\rm ph}^{2} =
  (k_{\rm ph}^{\bot})^{2} + (k_{\rm ph}^{\parallel})^{2} =
  (k_{\rm ph}^{\bot})^{2} (1 + \chi^{2}), &

  k_{\rm ph}^{\bot} = \displaystyle\frac{|\vb{k_{\rm ph}}| }{\sqrt{1 + \chi^{2}}}, &

  k_{\rm ph}^{\parallel} = \chi \cdot k_{\rm ph}^{\bot}.
\end{array}
\label{eq.model.virtual.1.2}
% \label{eq.app.model.crosssection.2.10}
\end{equation}
%-----------------------------------------------------------------------------------------------------------------------
%
%-----------------------------------------------------------------------------------------------------------------------
Radial integrals for the coherent terms are
[see Eqs.~(\ref{eq.resultingformula.6})]
\begin{equation}
\begin{array}{llllll}
  J_{1}^{real} (l_{i},l_{f},n) & = &
    \displaystyle\int\limits^{+\infty}_{0} \displaystyle\frac{dR_{i}(r, l_{i})}{dr}\: R^{*}_{f}(l_{f},r)\, j_{n}(k_{\rm ph}r)\; r^{2} dr, \\

  J_{1}^{vir}(l_{i},l_{f},n,\; l^{\parallel}) & = &
  \displaystyle\int\limits^{+\infty}_{0}
    \displaystyle\frac{dR_{i}(r, l_{i})}{dr}\:
    R^{*}_{f}(l_{f},r)\,
    j_{n}(k_{\rm ph}^{\bot} r)\;
    j_{l^{\parallel}}(k^{\parallel} r)\; r^{2} dr.
\end{array}
\label{eq.model.virtual.1.3}
% \label{eq.resultingformula.6}
\end{equation}
Here,
the first integrals (\ref{eq.resultingformula.6}) are without inclusion of the longitudinal part of virtual photon,
the second integrals (\ref{eq.resultingformula.7}) are with inclusion of the longitudinal part of virtual photon;
$R_{i,f}$ is radial part of wave function $\Phi_{\rm p - nucl} (\vb{r})$ in $i$-state or $f$-state,
$j_{n}(k_{\rm ph}r)$ is spherical Bessel function of order $n$.

\begin{figure}[htbp]
\centerline{\includegraphics[width=90mm]{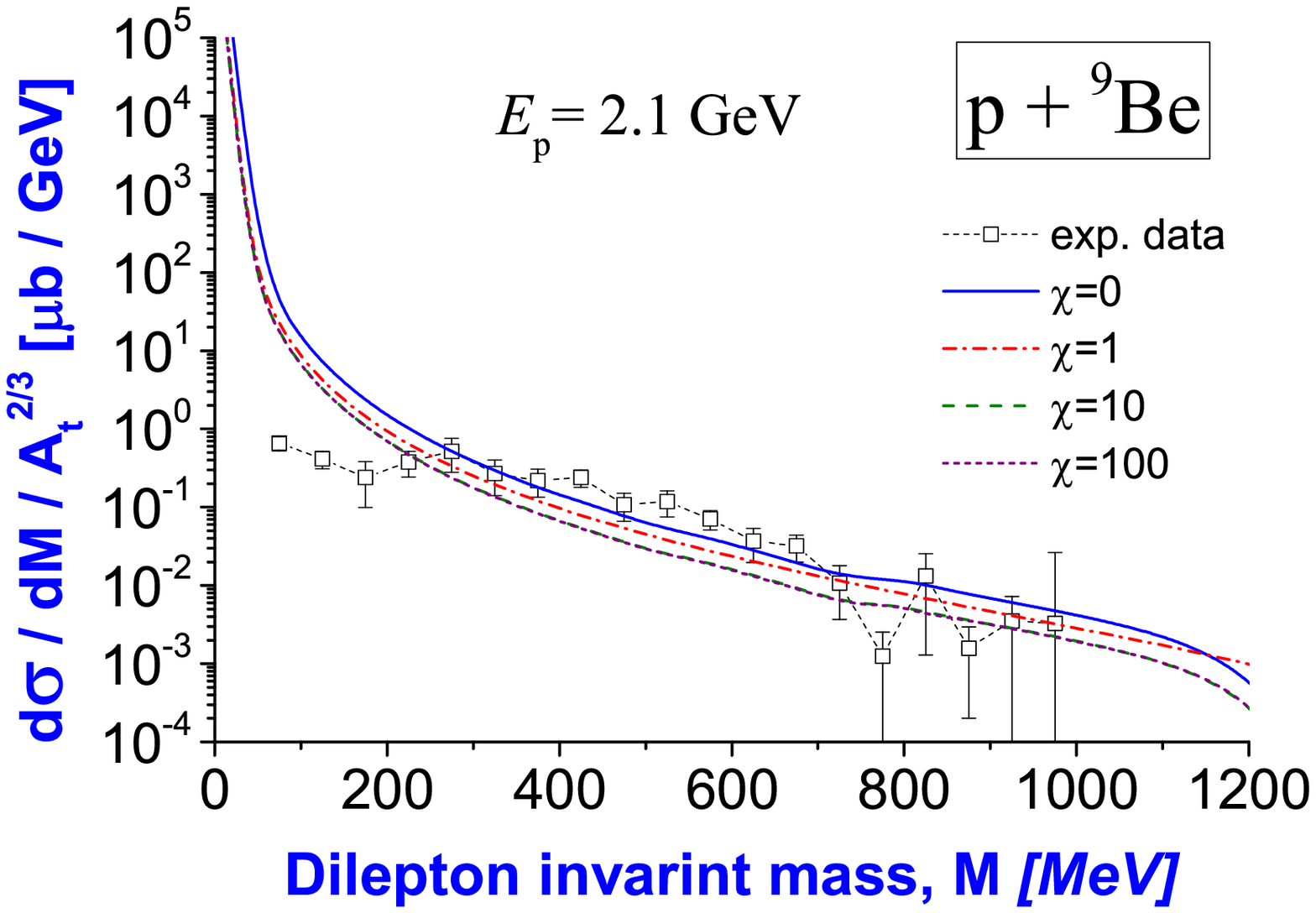}}
\vspace{-3mm}
\caption{\small (Color online)
The calculated cross sections of production of leptons pair (with coherent terms)
in the scattering of protons off the \isotope[9]{Be} nuclei at energy of proton beam of $E_{\rm p}=2.1$~GeV
for different virtualities $\chi$ of photon
in comparison with experimental data~\cite{Naudet.1989.PRL}
[cross section and matrix elements are defined in Eq.~(\ref{eq.model.virtual.1.1})--(\ref{eq.model.virtual.1.3}),
all calculated spectra have the same normalization factors on experimental data found from results in Fig.~\ref{fig.2}].
Here,
experimental data given by open rectangles (Naudet 1989) are extracted from Ref.~\cite{Naudet.1989.PRL},
blue solid line is the calculated cross section at $\chi=0$,
red dash-dotted line is the calculated cross section at $\chi=1.0$,
green dashed line is the calculated cross section at $\chi=10$,
purple short-dashed line is the calculated cross section at $\chi=100$.
% One can see that
% (a) incoherent emission is essentially more intensive than the coherent emission,
% (b) role of incoherent processes is increased at increasing of energy of photon,
% (c) better agreement with experimental data at $f_{\rm incoh}=0.001$ (than at $f_{\rm incoh}=1$) indicates on presence of some unknown effect,
% factor $f_{\rm incoh}$ which suppresses the intensity of incoherent processes.
% This result confirms the important (and not small) role of incoherent emission in bremsstrahlung.
\label{fig.7}}
\end{figure}

For analysis we use the scattering of $p + \isotope[9]{Be}$ at $E_{\rm p}=2.1$~GeV as above.
Results of such calculation of the spectra in comparison with experimental data
are presented in Fig.~\ref{fig.7}.
From these figures one can see that spectra at different $\chi$ have similar shapes of monotonous type.
Increasing of the longitudinal part of virtual photon suppresses the cross section of dilepton production.
However, difference between the spectra is small.
So, in some approximation, similar result can be obtained with minimal inclusion of longitudinal part of the virtual photon
(this makes formalism and calculations simpler, but shape of the resulted cross section is not deformed much,
and it will be closer to experimental data after renormalization).

Note that one can apply method in Ref.~\cite{Maydanyuk_Zhang_Zou.2019.PRC.microscopy} in order to find the most probable value of virtuality $\chi$ which allows to describe experimental data~\cite{Naudet.1989.PRL} on the basis of the developed model with the highest agreement (precision).
But, we omit such an analysis in this paper, postponing it to the future investigations.

% From these figures one can see that inclusion of incoherent processes to the model and calculations improves agreement with experimental data a little.
% In particular, it allows to achieve better description of experimental data at low energies smaller 250~MeV.
% Our predictions in regions of low energies $M <250$~MeV and high energies $M > 1000$~MeV can be recommended for tests in possible experimental study in future.
% Result above can be explained if the incoherent contribution is larger than the coherent one.
% If this is so, then the incoherent terms are important to be included to the model and calculations.
% In Fig.~\ref{fig.7} the different contributions in comparison with the full spectrum (a) and ratio between the incoherent contribution and coherent one (b) are presented.
% From figure~(a) one can see that %
% incoherent contribution is essentially more important than the coherent one.
% \emph{ratio between incoherent and coherent contributions is about 10--100 inside main region of invariant mass values}.
% \emph{Also we observe a new phenomenon of suppression of production of lepton pairs at low energy region due to incoherent processes.}
% This is explained by that at very low energies coherent contribution is dominant in comparison with incoherent one.
% This is confirmed by calculations in Fig.~\ref{fig.6}.
% *******************************************************************************************************************

% *******************************************************************************************************************
\subsection{Production of dileptons in the scattering $p + \isotope[93]{Nb}$
% and experimental data
\label{sec.analysis.expNb}}

\begin{figure}[htbp]
\centerline{\includegraphics[width=90mm]{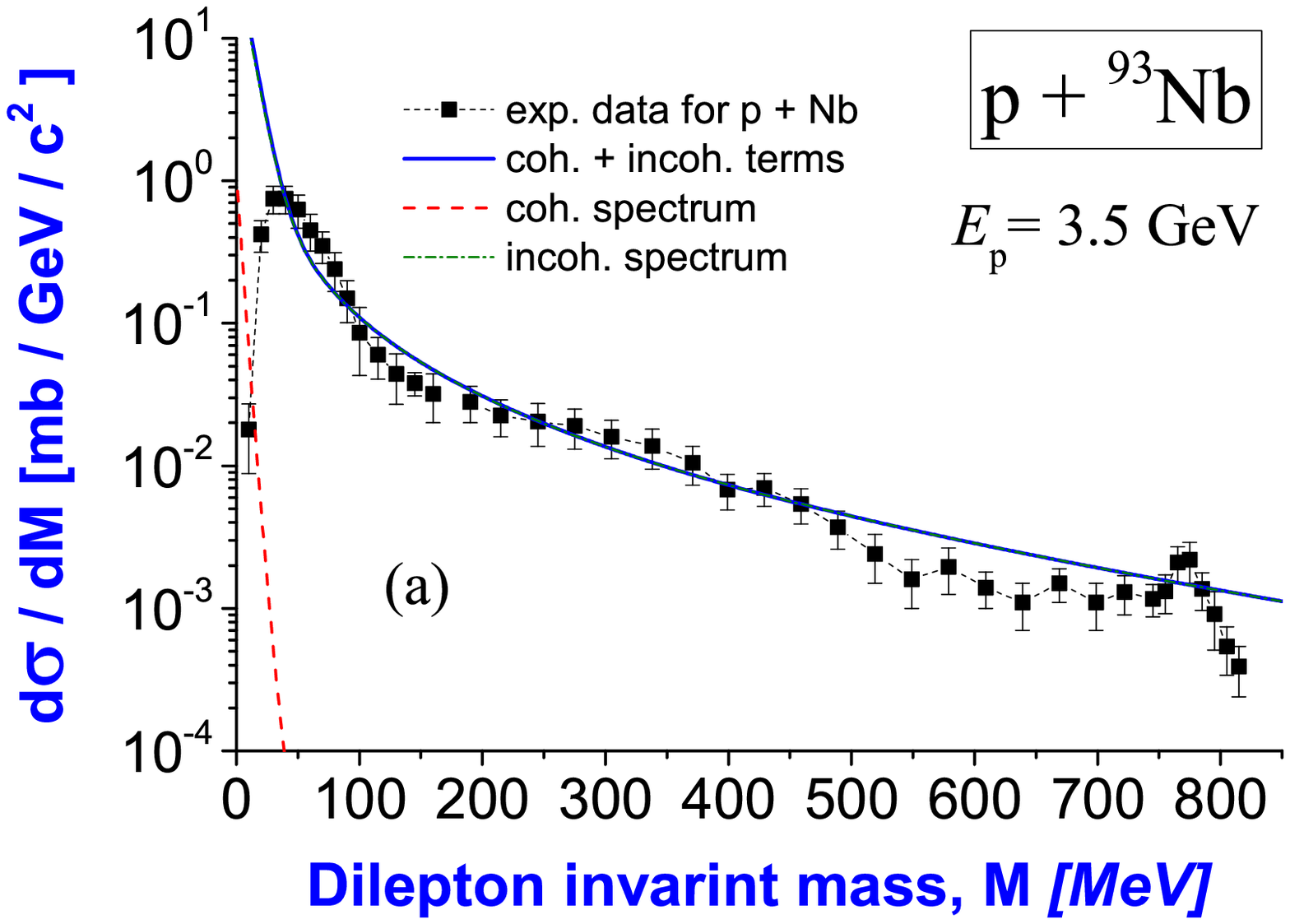}
\hspace{-1mm}\includegraphics[width=90mm]{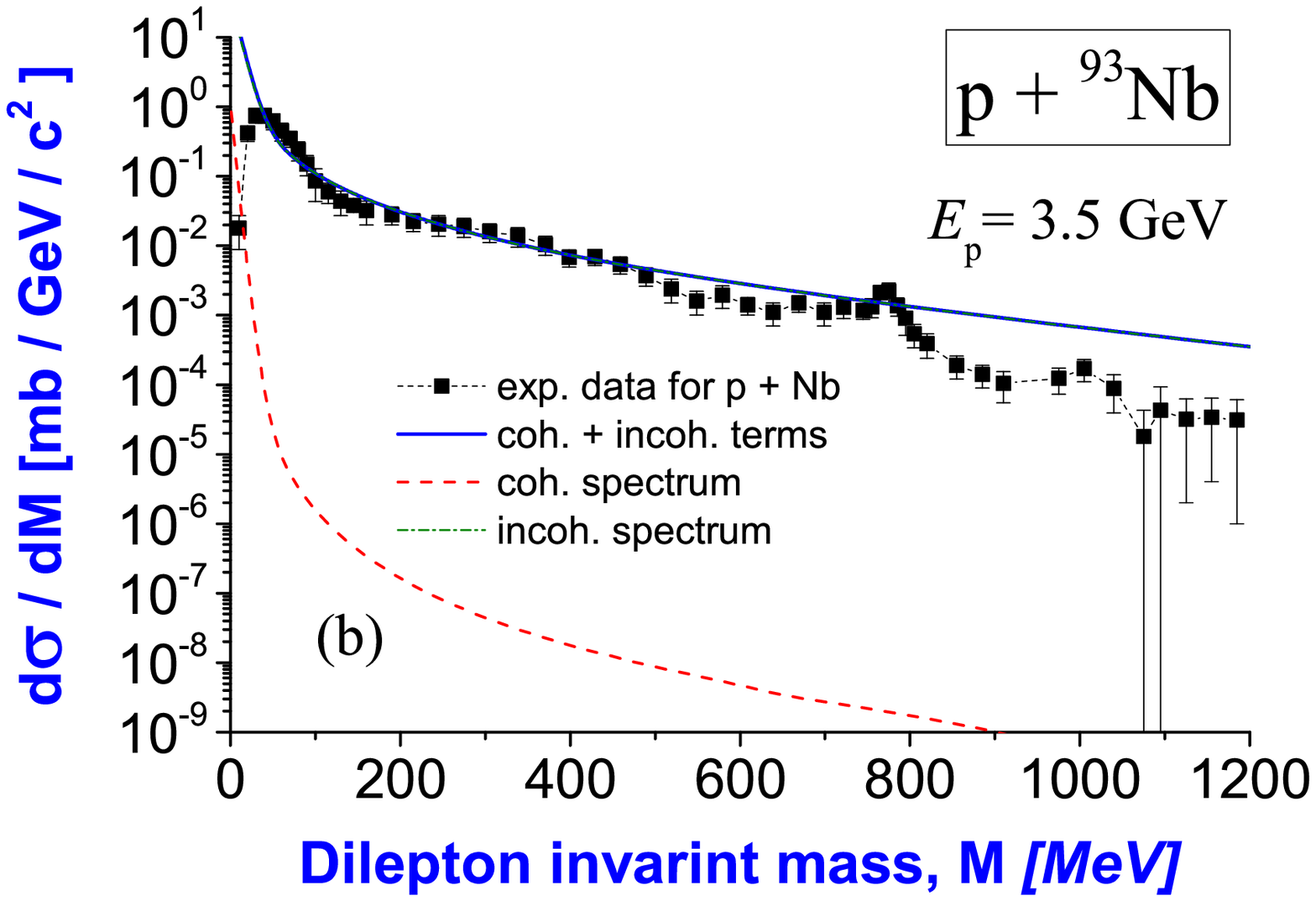}}
\vspace{-4mm}
\caption{\small (Color online)
The calculated cross section of production of leptons pair % (with coherent and incoherent terms)
in the scattering of protons off the \isotope[93]{Nb} nuclei at energy of proton beam of $E_{\rm p}=3.5$~GeV
in comparison with experimental data obtained by HADES collaboration~\cite{Agakishiev.HADEScollab.2012.plb}
[% matrix elements are defined in Eqs.~(\ref{eq.resultingformulas.1})--(\ref{eq.resultingformulas.2}),
% cross section is defined in Eq.~(\ref{eq.app.model.crosssection.2.9}),
% function $f^{}$ is defined in Eqs.~(\ref{eq.resultingformula.5}),
% radial integrals are defined in Eqs.~(\ref{eq.resultingformula.6}),
% $Z_{A} (k_{\rm ph})\simeq Z_{A}$ is electric charge of nucleus,
we normalize the full calculated spectrum on one point of experimental data,
obtain factor of normalization,
then renormalize the coherent and incoherent contributions on this one factor
% time of computer calculations is 26 min for 40 points of each calculated spectrum
].
Panel (a):
The coherent, incoherent contributions and full spectrum in comparison with experimental data.
Here,
experimental data given by full rectangles are extracted from Ref.~\cite{Agakishiev.HADEScollab.2012.plb},
blue solid line is the calculated full cross section (which includes all coherent and incoherent terms),
red dashed line is coherent contribution defined by $M_{p}^{(E,\, {\rm dip})}$, $M_{p}^{(M,\, {\rm dip})}$,
green dash-dotted line is incoherent contribution defined by $M_{\Delta M}$ and $M_{k}$
(it is almost coincides with blue solid line of the full spectrum).
%
% The calculated bremsstrahlung spectra (with coherent and incoherent terms)
% in the scattering of protons off the \isotope[197]{Au} nuclei at energy of proton beam of $E_{\rm p}=190$~MeV
% in comparison with experimental data~\cite{Goethem.2002.PRL}
% [matrix elements are defined in Eqs.~(\ref{eq.resultingformulas.1})--(\ref{eq.resultingformulas.2}),
% $Z_{A} (k_{\rm ph})\simeq Z_{A}$ is electric charge of nucleus].
% blue dashed line is coherent contribution defined by $M_{p}^{(E,\, {\rm dip})}$,
% red solid line is full spectrum with coherent and incoherent contributions defined by $M_{p}^{(E,\, {\rm dip})}$, $M_{p}^{(M,\, {\rm dip})}$, $M_{\Delta M}$ and $M_{k}$.
%
% \vspace{1.5mm}
% \newline
Panel (b):
The same calculated spectra shown in larger region of mass.
% (for more complete picture, and for possible further experimental tests).
% New calculated spectrum of full bremsstrahlung for \isotope[197]{Au} at energy of proton beam of $E_{\rm p}=800$~MeV
%
% Ration between incoherent and coherent contributions in dependence on energy of emitted bremsstrahlung photon
% [we define ratio as $\varepsilon = \sigma_{\rm incoh}/\sigma_{\rm coh}$, in calculations we use factor of incoherence of $f_{\rm incoh}=1.0$]
% related to full spectrum shown by dashed brown line in figure (a)].
% One can see that
% (a) incoherent emission is essentially more intensive than the coherent emission,
% (b) role of incoherent processes is increased at increasing of energy of photon,
% (c) better agreement with experimental data at $f_{\rm incoh}=0.001$ (than at $f_{\rm incoh}=1$) indicates on presence of some unknown effect,
% which highly suppresses the incoherent processes.
% factor $f_{\rm incoh}$ which suppresses the intensity of incoherent processes.
% One can see clear changes of the full spectrum in dependence on factor $f_{\rm incoh}$,
% red dash-dotted solid line (at $f=0.001$) corresponds to the best agreement with experimental data.
% This result confirms the important (and not small) role of incoherent emission in bremsstrahlung.
\label{fig.8}}
\end{figure}

In this Section we will analyze experimental data of the dilepton productions
obtained for $p + \isotope[]{Nb}$ at $E_{\rm p}=3.5$~GeV
by HADES collaboration~\cite{Agakishiev.HADEScollab.2012.plb,Weber.HADEScollab.2011.JPConfSer}.
Unified procedure for determination of the proton-nucleus potential is described in Sec.~\ref{sec.analysis.1}.
Wave functions of relative motion between proton and center-of-mass of the \isotope[93]{Nb} nucleus are calculated numerically concerning to this potential above.
%
% We calculate wave function of relative motion between proton and center-of-mass of nucleus numerically
% concerning to the proton-nucleus potential in form of $V (r) = v_{c}(r) + v_{N}(r) + v_{\rm so}(r) + v_{l} (r)$,
% where $v_{c}(r)$, $v_{N}(r)$, $v_{\rm so}(r)$, and $v_{l} (r)$ are Coulomb, nuclear, spin-orbital, and centrifugal components, respectively.
% Parameters of the potential are defined in Eqs.~(46)--(47) in Ref.~\cite{Maydanyuk_Zhang.2015.PRC}.
% We calculate the cross section by Eq.~(\ref{eq.app.model.crosssection.2.9}).
% where we include matrix elements of coherent emission $M_{p}^{(E,\, {\rm dip})}$, $M_{p}^{(M,\, {\rm dip})}$ in Eqs.~(\ref{eq.resultingformulas.1}),
% and matrix elements of incoherent emission $ M_{\Delta M}$, $M_{k}$ in Eqs.~(\ref{eq.resultingformulas.2}).
% The boundary conditions and normalization are used in form of~(B.1)--(B.9) in  \cite{Maydanyuk.2011.JPG}.
% \footnote{One proton from beam can transfer energy to one nucleon of nucleus for transition $NN \to \Delta N$ with formation of $\Delta$-resonance in nucleus.
% Other protons of beam with nucleons of nucleus-target and $\Delta$-resonance can emit bremsstrahlung photons.
%
% show that incoherent emission is essentially larger than coherent one.
Results of calculations of cross sections of dilepton production for \isotope[93]{Nb} by our model in comparison with experimental data above
are presented in Fig.~\ref{fig.8}.
% \cite{Maydanyuk_Zhang.2015.PRC,Maydanyuk.2012.PRC,Maydanyuk_Zhang_Zou.2016.PRC,Liu_Maydanyuk_Zhang_Liu.2019.PRC.hypernuclei}
% \cite{Maydanyuk_Zhang.2015.PRC,Maydanyuk.2012.PRC,Maydanyuk_Zhang_Zou.2016.PRC,Liu_Maydanyuk_Zhang_Liu.2019.PRC.hypernuclei,Maydanyuk_Zhang_Zou.2019.PRC.microscopy}
% show that incoherent emission is essentially larger than coherent one.
% By such a reason, we start calculations for \isotope[197]{Au}, where experimental data exist.
% Results of such calculations with inclusion of coherent and incoherent contributions in comparison with experimental data are presented in Fig.~\ref{fig.1}~(a).
%
%
% The calculated spectra are normalized on one point of these experimental data.
% These calculations are shown in Fig.~1~(a).
% All other calculated spectra for different nuclei and energies of proton beam use the same coefficient of normalization (this gives possibility to predict new spectra).

This result shows that our model provides the tendencies of the spectrum in good agreement with experimental data (with exception of few points) up to 750~MeV.
Such a result also shows that role of incoherent processes is really large, while the coherent processes are very small.
The second conclusion is in that the incoherent processes are highly increased
% \textcolor[rgb]{1.00,0.00,0.00}{\textbf{%
with increasing atomic mass of the nucleus-target.
Existence of peaks in experimental data can be explained by other processes, which are omitted in this paper.
These processes are also important in production of dileptons and can be further studied
% \textcolor[rgb]{1.00,0.00,0.00}{\textbf{%
in the future.
% }}

\begin{figure}[htbp]
\centerline{\includegraphics[width=90mm]{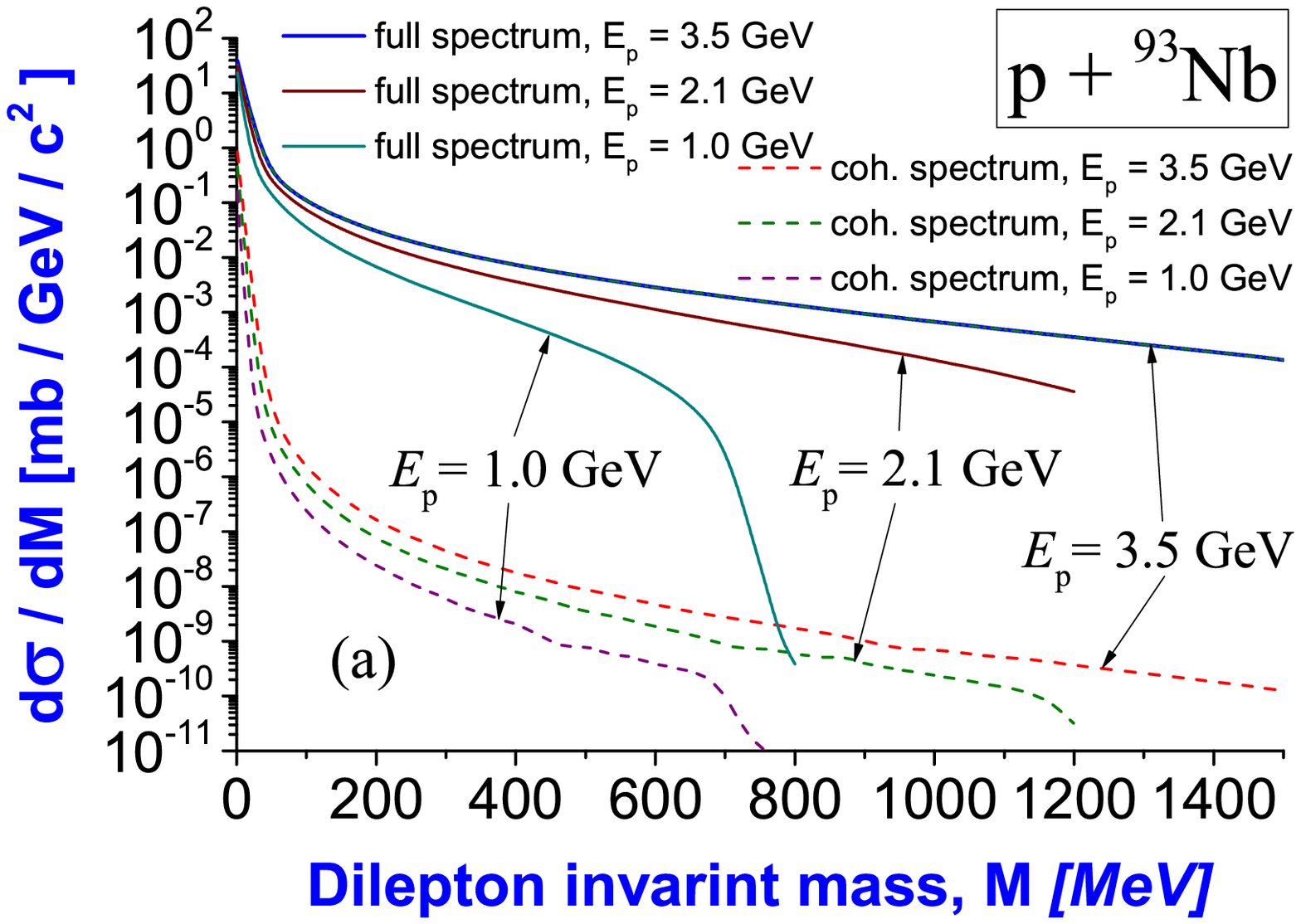}
\hspace{-1mm}\includegraphics[width=90mm]{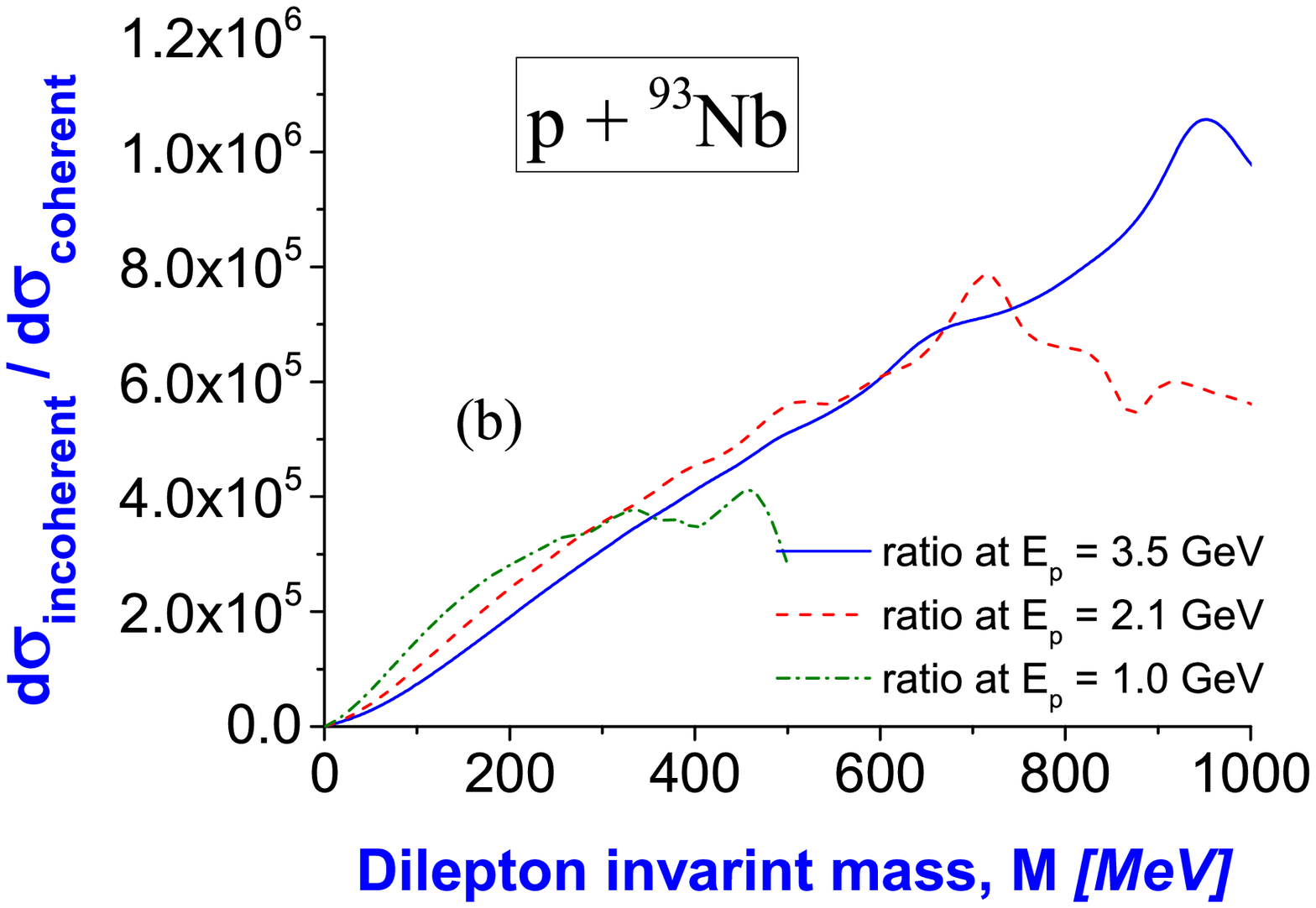}}
\vspace{-4mm}
\caption{\small (Color online)
The calculated cross section of production of leptons pair % (with coherent and incoherent terms)
in the scattering of protons off the \isotope[93]{Nb} nuclei at energies of proton beam of
$E_{\rm p}=1.0$~GeV, $E_{\rm p}=2.1$~GeV and $E_{\rm p}=3.5$~GeV
[% matrix elements are defined in Eqs.~(\ref{eq.resultingformulas.1})--(\ref{eq.resultingformulas.2}),
% cross section is defined in Eq.~(\ref{eq.app.model.crosssection.2.9}),
% function $f^{}$ is defined in Eqs.~(\ref{eq.resultingformula.5}),
% radial integrals are defined in Eqs.~(\ref{eq.resultingformula.6}),
% $Z_{A} (k_{\rm ph})\simeq Z_{A}$ is electric charge of nucleus,
we normalize the full calculated spectrum on one point of experimental data,
obtain factor of normalization,
then renormalize the coherent and incoherent contributions on this one factor,
time of computer calculations is 31 min for 40 points of each calculated spectrum
].
Panel (a):
The coherent contribution and full spectrum at different energies of proton beam.
The incoherent contribution almost coincides with the full spectrum for each energy $E_{\rm p}$.
% Here,
% experimental data given by full rectangles % (Naudet 1989)
% are extracted from Ref.~\cite{Agakishiev.HADEScollab.2012.plb},
% blue solid line is the calculated full cross section (which includes all coherent and incoherent terms),
% red dashed line is coherent contribution defined by $M_{p}^{(E,\, {\rm dip})}$, $M_{p}^{(M,\, {\rm dip})}$,
% green dash-dotted line is incoherent contribution defined by $M_{\Delta M}$ and $M_{k}$
%
% The calculated bremsstrahlung spectra (with coherent and incoherent terms)
% in the scattering of protons off the \isotope[197]{Au} nuclei at energy of proton beam of $E_{\rm p}=190$~MeV
% in comparison with experimental data~\cite{Goethem.2002.PRL}
% [matrix elements are defined in Eqs.~(\ref{eq.resultingformulas.1})--(\ref{eq.resultingformulas.2}),
% $Z_{A} (k_{\rm ph})\simeq Z_{A}$ is electric charge of nucleus].
% blue dashed line is coherent contribution defined by $M_{p}^{(E,\, {\rm dip})}$,
% red solid line is full spectrum with coherent and incoherent contributions defined by $M_{p}^{(E,\, {\rm dip})}$, $M_{p}^{(M,\, {\rm dip})}$, $M_{\Delta M}$ and $M_{k}$.
%
% \vspace{1.5mm}
% \newline
Panel (b):
Ratio between incoherent contribution and coherent contribution for different energies of proton beam.
One can see that role of incoherent contribution for this reaction is essentially larger than
for $p + \isotope[9]{Be}$ at $E_{\rm p}=2.1$~GeV [see Fig.~\ref{fig.6}~(b)].
% The same calculated spectra shown in larger region of mass
% (for more complete picture, and for possible further experimental tests).
% New calculated spectrum of full bremsstrahlung for \isotope[197]{Au} at energy of proton beam of $E_{\rm p}=800$~MeV
%
% Ration between incoherent and coherent contributions in dependence on energy of emitted bremsstrahlung photon
% [we define ratio as $\varepsilon = \sigma_{\rm incoh}/\sigma_{\rm coh}$, in calculations we use factor of incoherence of $f_{\rm incoh}=1.0$]
% related to full spectrum shown by dashed brown line in figure (a)].
% One can see that
% (a) incoherent emission is essentially more intensive than the coherent emission,
% (b) role of incoherent processes is increased at increasing of energy of photon,
% (c) better agreement with experimental data at $f_{\rm incoh}=0.001$ (than at $f_{\rm incoh}=1$) indicates on presence of some unknown effect,
% which highly suppresses the incoherent processes.
%
% factor $f_{\rm incoh}$ which suppresses the intensity of incoherent processes.
% One can see clear changes of the full spectrum in dependence on factor $f_{\rm incoh}$,
% red dash-dotted solid line (at $f=0.001$) corresponds to the best agreement with experimental data.
% This result confirms the important (and not small) role of incoherent emission in bremsstrahlung.
\label{fig.9}}
\end{figure}

In Fig.~\ref{fig.9} we present results of calculations of full cross sections and different contributions
for dilepton production in scattering $p + \isotope[93]{Nb}$ by our model at different energies of proton beam.
% \cite{Maydanyuk_Zhang.2015.PRC,Maydanyuk.2012.PRC,Maydanyuk_Zhang_Zou.2016.PRC,Liu_Maydanyuk_Zhang_Liu.2019.PRC.hypernuclei}
% \cite{Maydanyuk_Zhang.2015.PRC,Maydanyuk.2012.PRC,Maydanyuk_Zhang_Zou.2016.PRC,Liu_Maydanyuk_Zhang_Liu.2019.PRC.hypernuclei,Maydanyuk_Zhang_Zou.2019.PRC.microscopy}
% show that incoherent emission is essentially larger than coherent one.
% By such a reason, we start calculations for \isotope[197]{Au}, where experimental data exist.
% Results of such calculations with inclusion of coherent and incoherent contributions in comparison with experimental data are presented in Fig.~\ref{fig.1}~(a).
%
From these calculations one can conclude that
% \textcolor[rgb]{1.00,0.00,0.00}{\textbf{%
at energies of proton beam from 1~GeV to 3.5~GeV ratio between the incoherent contribution and coherent contribution can be in region $10^{+5}$ -- $10^{+6}$ for \isotope[]{Nb},
i.e. the incoherent processes have leading role in production of dileptons, while the coherent
% \textcolor[rgb]{1.00,0.00,0.00}{\textbf{%
processes are less 1 percent.
% }}
% *******************************************************************************************************************

% *******************************************************************************************************************
\subsection{Role of nuclear part of proton-nucleus potential in production of dilepton
% and experimental data
\label{sec.analysis.potential}}

\begin{figure}[htbp]
\centerline{\includegraphics[width=90mm]{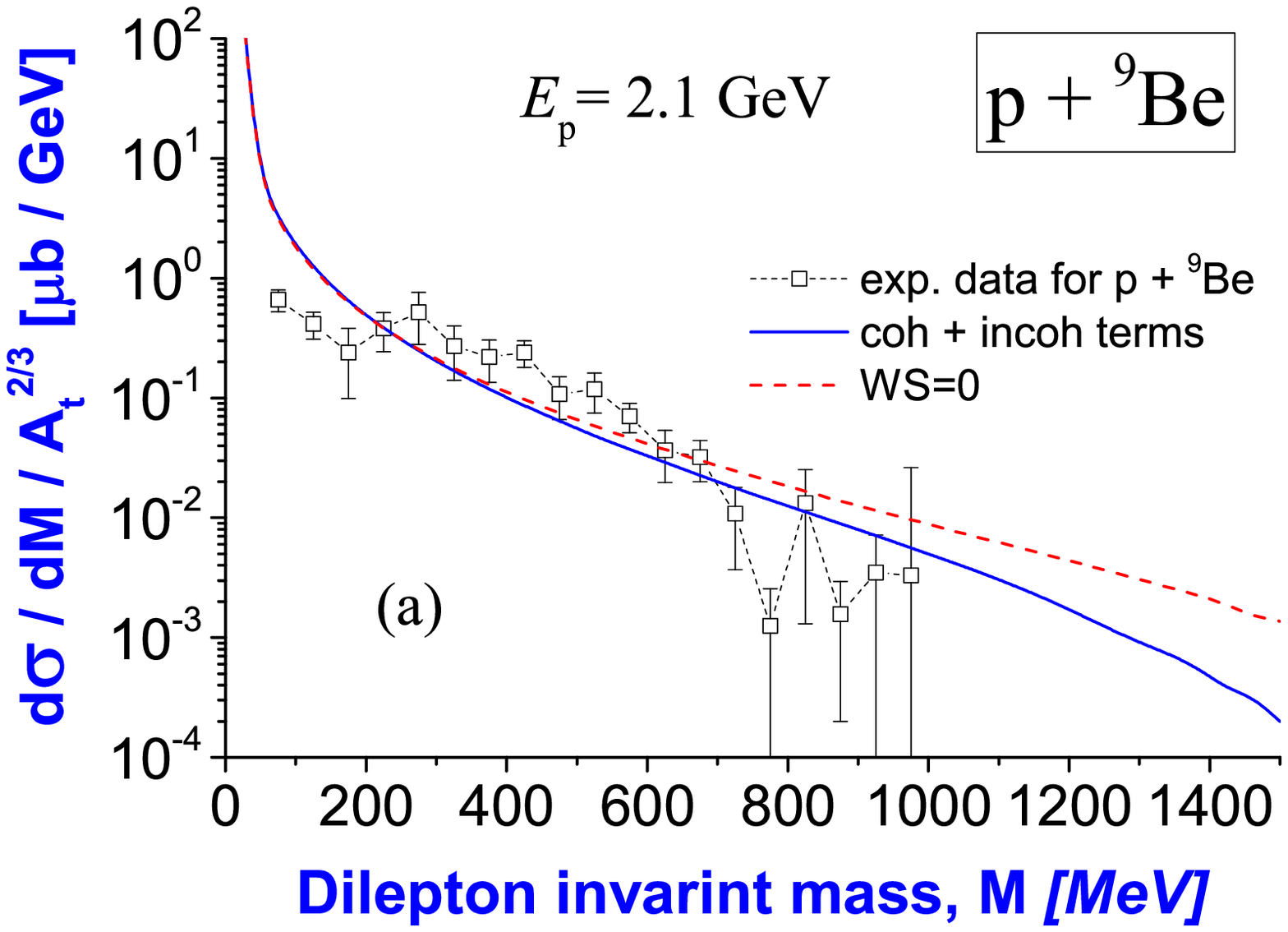}
% \hspace{-1mm}\includegraphics[width=90mm]{Figures/fig.10b_p-93Nb-dileptons_ratio_incoh_coherent.eps}
}
\vspace{-4mm}
\caption{\small (Color online)
The calculated cross sections of production of leptons pair % (with coherent and incoherent terms)
in the scattering of protons off the \isotope[9]{Be} nuclei at energies of proton beam of
$E_{\rm p}=2.1$~GeV
with included nuclear part of potential of proton-nucleus interactions and without this part
[% matrix elements are defined in Eqs.~(\ref{eq.resultingformulas.1})--(\ref{eq.resultingformulas.2}),
parameters of potential are given in App.~\ref{sec.app.4}].
% cross section is defined in Eq.~(\ref{eq.app.model.crosssection.2.9}),
% function $f^{}$ is defined in Eqs.~(\ref{eq.resultingformula.5}),
% radial integrals are defined in Eqs.~(\ref{eq.resultingformula.6}),
% $Z_{A} (k_{\rm ph})\simeq Z_{A}$ is electric charge of nucleus,
% we normalize the full calculated spectrum on one point of experimental data,
% obtain factor of normalization,
% then renormalize the coherent and incoherent contributions on this one factor,
% time of computer calculations is 31 min for 40 points of each calculated spectrum
% ].
Here,
experimental data given by open rectangles (Naudet 1989) are extracted from Ref.~\cite{Naudet.1989.PRL},
blue solid line is calculated cross section with included nuclear part of potential,
red dashed line is calculated cross section without this nuclear part of potential.
\label{fig.10}}
\end{figure}

We will estimate role of nuclear part of potential of interactions between proton and nucleus in calculations of cross section of dilepton production.
Results of such calculations are presented in Fig.~\ref{fig.10} for dilepton production in scattering $p + \isotope[9]{Be}$.
% \cite{Maydanyuk_Zhang.2015.PRC,Maydanyuk.2012.PRC,Maydanyuk_Zhang_Zou.2016.PRC,Liu_Maydanyuk_Zhang_Liu.2019.PRC.hypernuclei}
% \cite{Maydanyuk_Zhang.2015.PRC,Maydanyuk.2012.PRC,Maydanyuk_Zhang_Zou.2016.PRC,Liu_Maydanyuk_Zhang_Liu.2019.PRC.hypernuclei,Maydanyuk_Zhang_Zou.2019.PRC.microscopy}
% show that incoherent emission is essentially larger than coherent one.
% By such a reason, we start calculations for \isotope[197]{Au}, where experimental data exist.
% Results of such calculations with inclusion of coherent and incoherent contributions in comparison with experimental data are presented in Fig.~\ref{fig.1}~(a).
%
From results presented in this figure one can see that difference between two spectra is clearly more visible at higher invariant masses.
Note that it is observed stability in calculations. That provides good basis of further study of role of nuclear potentials in dileptons productions.
This is new result not investigated before.

% *******************************************************************************************************************

% *******************************************************************************************************************
\section{Conclusions and perspective
\label{sec.conclusions}}

In this paper we investigate production of lepton pairs in the scattering of protons off nuclei at intermediate energy region.
For that research we construct a new model.
In description of emission of virtual photons we generalize the formalism of emission of photons in the proton nucleus scattering
(see Refs.~\cite{Maydanyuk_Zhang.2015.PRC,Maydanyuk.2012.PRC,Maydanyuk_Zhang_Zou.2016.PRC,Liu_Maydanyuk_Zhang_Liu.2019.PRC.hypernuclei,Maydanyuk_Zhang_Zou.2019.PRC.microscopy} and reference therein).
% \cite{Maydanyuk_Zhang.2015.PRC,Maydanyuk.2012.PRC,Maydanyuk_Zhang_Zou.2016.PRC,Liu_Maydanyuk_Zhang_Liu.2019.PRC.hypernuclei}
% \cite{Maydanyuk_Zhang.2015.PRC,Maydanyuk.2012.PRC,Maydanyuk_Zhang_Zou.2016.PRC,Liu_Maydanyuk_Zhang_Liu.2019.PRC.hypernuclei,Maydanyuk_Zhang_Zou.2019.PRC.microscopy}
% (see Ref.~\cite{Maydanyuk_Zhang_Zou.2019.PRC.microscopy}, reference therein),
% A focus is directed on question of how much the bremsstrahlung spectrum is changed after transition of one nucleon in nucleus to $\Delta$-resonance.
% We improve previous bremsstrahlung formalism
% (see Refs.~\cite{Maydanyuk_Zhang.2015.PRC,Maydanyuk.2012.PRC,Maydanyuk_Zhang_Zou.2016.PRC,Liu_Maydanyuk_Zhang_Liu.2019.PRC.hypernuclei,Maydanyuk_Zhang_Zou.2019.PRC.microscopy} and reference therein),
% (see Ref.~\cite{Maydanyuk_Zhang_Zou.2019.PRC.microscopy}, reference therein),
% including properties o $\Delta$-resonance in the nucleus-target.
%
% \textcolor[rgb]{1.00,0.00,0.00}{\textbf{%
We do not include here important contribution from Dalitz-decays and direct decays of vector mesons.
On the basis of such a model we obtain the following.

\begin{itemize}
\item
We calculate the cross section of production of lepton pairs in the scattering of protons on the \isotope[9]{B} nuclei at energy of proton beam $E_{\rm p}$ of 2.1~GeV,
where the coherent matrix elements are included to calculations.
This calculated spectrum is in good agreement (with exception of few points)
with experimental data~\cite{Naudet.1989.PRL} obtained by DLS Collaboration
(see Fig.~\ref{fig.2}).
This is good test of the model and algorithms, that is important for further analysis.
\item
We analyze (estimate) cross sections of dilepton production in dependence on different nuclei-targets
in region from low up to large masses at the same energy of proton beam of 2.1~GeV
(see spectra for the \isotope[9]{Be}, \isotope[12]{C}, \isotope[16]{O}, \isotope[24]{Mg}, \isotope[44]{Ca}, \isotope[197]{Au} nuclei
at energy of proton beam of $E_{\rm p}=2.1$~GeV in Fig.~\ref{fig.3}).
We observe that the cross section of dilepton production for the coherent processes is smaller for nuclei-target with larger mass.
%
% I.e. this is a new effect of suppressing of production of dileptons in the proton-nucleus scattering (by nuclear matter of nucleus-target).
%
There is a general tendency of monotonous decreasing of production of dileptons with increasing of mass of the nucleus-target.
Very close similarities in shapes of each calculated spectrum are observed inside the full studied region of masses
that confirms stability of calculations.
% \textcolor[rgb]{1.00,0.00,0.00}{\textbf{%
This is a good test of the developed model and computer algorithms in numerical calculations.
% }}
% *******************************************************************************************************************

% *******************************************************************************************************************
\item
We clarify how much the spectra are sensitive to the choice of different isotopes of nuclei-target with the same fixed charge number.
However, analysis have shown that the spectra are not much sensitive from choice of different isotope with the same charge number,
both for light nuclei and heavy nuclei.
This phenomenon is observed for the proton nucleus scattering
%
% \textcolor[rgb]{1.00,0.00,0.00}{\textbf{%
but we estimate that for nucleus-nucleus scattering (heavy-ion collisions) such dependencies should be much larger).
This is clearly explained by dependencies of effective charge $Z_{\rm eff}^{\rm (mon)} (\vb{k}_{\rm ph})$ on different masses and charges of two participating nuclei
[see Eq.~(\ref{eq.resultingformula.8}), p.~\pageref{eq.resultingformula.8}] in reaction in formula for matrix elements.
% *******************************************************************************************************************

% *******************************************************************************************************************
\item
We analyze
% \textcolor[rgb]{1.00,0.00,0.00}{\textbf{%
the dependence of cross section of production of leptons pair on energy of proton beam for the same fixed nucleus-target
(see Fig.~\ref{fig.4} for calculations for the scattering of protons off the \isotope[9]{Be} nuclei).
%
% [we define ratio as $\varepsilon = \sigma_{\rm incoh}/\sigma_{\rm coh}$, in calculations we use factor of incoherence of $f_{\rm incoh}=1.0$]
% (a) incoherent emission is essentially more intensive than the coherent emission,
% (b) role of incoherent processes is increased at increasing of energy of photon,
% (c) better agreement with experimental data at $f_{\rm incoh}=0.001$ (than at $f_{\rm incoh}=1$) indicates on presence of some unknown effect,
% factor $f_{\rm incoh}$ which suppresses the intensity of incoherent processes.
% One can see clear changes of the full spectrum in dependence on factor $f_{\rm incoh}$,
% red dash-dotted solid line (at $f=0.001$) corresponds to the best agreement with experimental data.
% This result confirms the important (and not small) role of incoherent emission in bremsstrahlung.
%
We observe that production of lepton pairs is more intensive at larger energies of proton beam $E_{\rm p}$.
Each spectrum at increasing of invariant mass $M$ tends to some fixed maximum of this mass, that is explained by kinematic limit
from relation between energies of relative motion of proton-nucleus scattering before emission of virtual photon, after such an emission and
energy (invariant mass) of this emitted photon.

% *******************************************************************************************************************

% *******************************************************************************************************************
% We analyze the coherent and incoherent contributions, electric and magnetic contributions in the full bremsstrahlung for different nuclei and energies of proton beam.
% In the incoherent bremsstrahlung, role of background emission based on $M_{k}$ is a little larger than magnetic contribution based on $M_{\Delta M}$
% ($\sigma_{\rm background}^{\rm (incoh)} / \sigma_{\rm mag}^{\rm (incoh)} = 4.04$ for \isotope[197]{Au} at $E_{\rm p} = 190$~MeV for 10--180~MeV of photons).
% Ratio between incoherent emission and coherent emission is increased at increasing of energy of photon
% (see Fig.~\ref{fig.2}, for \isotope[197]{Au} at $E_{\rm p}=190$~MeV).
%
\item
We analyzed role of incoherent processes in production of pairs of leptons in the proton-nucleus scattering.
For the scattering of $p + \isotope[9]{Be}$ at $E_{\rm p}=2.1$~GeV, we find the following.

% Ration between incoherent and coherent contributions in dependence on energy of emitted bremsstrahlung photon
% [we define ratio as $\varepsilon = \sigma_{\rm incoh}/\sigma_{\rm coh}$, in calculations we use factor of incoherence of $f_{\rm incoh}=1.0$]
% related to full spectrum shown by dashed brown line in figure (a)].
% (a) incoherent emission is essentially more intensive than the coherent emission,
% (b) role of incoherent processes is increased at increasing of energy of photon,
% (c) better agreement with experimental data at $f_{\rm incoh}=0.001$ (than at $f_{\rm incoh}=1$) indicates on presence of some unknown effect,
% which highly suppresses the incoherent processes.
% factor $f_{\rm incoh}$ which suppresses the intensity of incoherent processes.
% One can see clear changes of the full spectrum in dependence on factor $f_{\rm incoh}$,
% red dash-dotted solid line (at $f=0.001$) corresponds to the best agreement with experimental data.
% This result confirms the important (and not small) role of incoherent emission in bremsstrahlung.

\begin{itemize}
\item
Inclusion of incoherent processes to the model and calculations improves agreement with experimental data~\cite{Naudet.1989.PRL} a little
% \textcolor[rgb]{1.00,0.00,0.00}{\textbf{%
(it allows to achieve better description of experimental data at low energies smaller 250~MeV, see Fig.~\ref{fig.5}).
% }}
%
% Our predictions in regions of low energies $M <250$~MeV and high energies $M > 1000$~MeV can be recommended for tests in possible experimental study in future.

\item
The incoherent contribution is essentially larger than the coherent one [see Fig.~\ref{fig.6}~(a)].
By other words, the incoherent contribution has a leading role in production of pairs of leptons for the studied reaction.
This explains improving agreement between calculations and experimental data~\cite{Naudet.1989.PRL} after inclusion of incoherent contribution.
\emph{Ratio between incoherent and coherent contributions is about 10--100 inside main region of invariant mass values}
[see Fig.~\ref{fig.6}~(b)].

\item
\emph{We observe a new phenomenon of suppression of production of lepton pairs at low energy region due to incoherent processes.}
This is explained by that at very low energies coherent contribution is dominant in comparison with incoherent one.
This is confirmed by calculations in Fig.~\ref{fig.6}.

\end{itemize}
% *******************************************************************************************************************

% *******************************************************************************************************************
\item
We also analyzed
% \textcolor[rgb]{1.00,0.00,0.00}{\textbf{%
the role of incoherent processes in production of pairs of leptons
in the scattering of $p + \isotope[93]{Nb}$ inside energy region of proton beam $E_{\rm p}$ from 1~GeV to 3.5~GeV and find the following.

(1) Our model provides the tendencies of the full spectrum at $E_{\rm p}=3.5$~GeV in good agreement with
experimental data obtained by HADES collaboration~\cite{Agakishiev.HADEScollab.2012.plb}
up to 750~MeV
[see Fig.~\ref{fig.8}~(a)].

(2) Role of incoherent processes is really large, while the coherent processes are very small
[see Fig.~\ref{fig.10}].
Incoherent contribution for this reaction is essentially larger than
for $p + \isotope[9]{Be}$ at $E_{\rm p}=2.1$~GeV [see Fig.~\ref{fig.6}~(b)].

(3) Incoherent processes are highly increased at increasing atomic mass of the nucleus-target.

% *******************************************************************************************************************

% *******************************************************************************************************************
\item
We analyzed influence of longitudinal part of virtual photon on the calculated cross sections of production of dilepton pairs.
From analysis of the scattering of $p + \isotope[9]{Be}$ at $E_{\rm p}=2.1$~GeV
(we reduce ourselves by the coherent processes), we find the following
(see Fig.~\ref{fig.7}).
% Results of such calculation of the spectra in comparison with experimental data are presented in Fig.~\ref{fig.7}.
%
% The calculated cross sections of production of leptons pair
% in the scattering of protons off the \isotope[9]{Be} nuclei at energy of proton beam of $E_{\rm p}=2.1$~GeV
% for different virtualities $\chi$ of photon in comparison with experimental data~\cite{Naudet.1989.PRL}.
% (a) incoherent emission is essentially more intensive than the coherent emission,
% (b) role of incoherent processes is increased at increasing of energy of photon,
% (c) better agreement with experimental data at $f_{\rm incoh}=0.001$ (than at $f_{\rm incoh}=1$) indicates on presence of some unknown effect,
% factor $f_{\rm incoh}$ which suppresses the intensity of incoherent processes.
%
The spectra have similar shapes of monotonous type at different virtualities $\chi$.
Increasing of the longitudinal part of virtual photon suppresses the cross section of dilepton production.
However, difference between the spectra is small.

\end{itemize}
%-----------------------------------------------------------------------------------------------------------------------

%-----------------------------------------------------------------------------------------------------------------------
\noindent
% If to compare the spectra for normal nuclei with the spectra for nuclei with the shortly lived $\Delta$-resonance,
% one can see essential difference between these spectra at high energy region of photons.
% This property can be used for proposal for future experiments with measurements of photons,
% as tools to distinguish process of formation of $\Delta$-resonance in the nucleus-target.
%-----------------------------------------------------------------------------------------------------------------------

%-----------------------------------------------------------------------------------------------------------------------
% (note that calculations in Ref.~\cite{Gil_Oset.1998.PLB.v416} were predictions, not tested on the bremsstrahlung experimental data).
% Processes with emission of photons in scattering of protons on the \isotope[12]{C}, \isotope[40]{Ca}, \isotope[208]{Pb} nuclei
% in the $\Delta$-resonance energy region were analyzed~\cite{Gil_Oset.1998.PLB.v416}
% This incoherent bremsstrahlung is highly dependent on magnetic moments of nucleons of the scattered nuclear fragments,
% but those estimations were obtained for magnetic moments of nucleon for vacuum.
%-----------------------------------------------------------------------------------------------------------------------

%-----------------------------------------------------------------------------------------------------------------------
\section*{Acknowledgements
\label{sec.acknowledgements}}

% Author thanks the Wigner Research Centre for Physics in Budapest for warm hospitality and support.
Authors are highly appreciated to
Profs. Pengming Zhang and Liping Zou for useful discussions concerning to elementary processes and emission in nuclear collisions,
Prof.~V.~S.~Vasilevsky for fruitful discussions concerning to different aspects concerning to physics of nuclear scattering,
Prof.~A.~G.~Magner for fruitful discussions concerning to different aspects of compound nuclear system and fusion,
Dr.~S.~A.~Omelchenko for fruitful discussions concerning to aspects of different nuclear models for scattering and decays.
% formalism for energies in the center-of-mass frame an laboratory frame.
% Prof. G.~Wolf for interesting suggestions to apply formalism for study of dilepton productions in collisions.
% Profs.~V.~A.~Plujko, S.~N.~Fedotkin, A.~G.~Magner, F.~A.~Ivanyuk, A.~P.~Ilyin, A.~Ya.~Dzyublik for fruitful and useful discussions.
% S.~P.~M. thanks the Institute of Modern Physics of Chinese Academy of Sciences for warm hospitality and support.
% This work was supported by the National Natural Science Foundation of China (Grant Nos. 11975320 and 11805242).
Authors thank
the support of OTKA grant K138277.
% the support in part by the budget program ``Support for the development of priority   areas of scientific researches'',
% the project of the Academy of Sciences of Ukraine (Code 6541230, No. 0122U000848).

% 11427904 and 11535016).
% the Major State Basic Research Development Program in China (No. 2015CB856903),
% the National Natural Science Foundation of China (Grant Nos. 11575254, 11447105 and 11175215),
% the Chinese Academy of Sciences fellowships for researchers from developing countries (No. 2014FFJA0003).

% \vspace{5mm}
% \textcolor[rgb]{1.00,0.00,0.00}{\hspace{25mm}\LARGE{\textbf{To correct text below!}}}
% \vspace{4mm}
% *******************************************************************************************************************

% *******************************************************************************************************************
\appendix
%-----------------------------------------------------------------------------------------------------------------------

%-----------------------------------------------------------------------------------------------------------------------
\section{Tensor forms for hadronic and leptonic parts of the matrix elements
\label{sec.app.1.DIS_leptonic}}

Let us write down the found matrix elements for interaction between proton (hadron) and leptons
(for example, see Eqs.~(6.8), p.~149 in Ref.~\cite{Halzen.book.1987})
% [see Eqs.~(\ref{eq.3.2.6.4}) (), p.~\pageref{eq.3.2.6.4}]
%
\begin{equation}
\begin{array}{llllllll}
  T_{fi} =
  -i\, (2\pi)^{4}\,
  \delta^{4} (p_{A} + p_{B} - p_{C} - p_{D}) \cdot \mathcal{M},
\end{array}
\label{eq.3.2.1}
\end{equation}
where
\begin{equation}
\begin{array}{llclll}
\Diagram{\vertexlabel^{p'} & & & \vertexlabel^{e^{+}} \\
fdV & & fu\\
& g & \\
\vertexlabel_{p} fuA & & fd \\
  & & & & \vertexlabel^{e^{-}}
}  & \to &

  -\, i\, \mathcal{M} =
  \bar{u}_{C} (p')\: (i e \gamma^{\mu})\: u_{A} (p) \cdot
  \Bigl( -\, \displaystyle\frac{i\, g_{\mu\nu}}{q^{2}} \Bigr) \cdot
  \bar{u}_{D} (k')\: (i e \gamma^{\nu})\: u_{B} (k).
\end{array}
\label{eq.3.2.2}
\end{equation}
For unpolarized scattering, we should take into account all spin states for proton and lepton.
This is done via summation over all possible spin states of proton and leptons
(for example, see Eqs.~(6.10), p.~150 in Ref.~\cite{Halzen.book.1987})
\begin{equation}
\begin{array}{llllllll}
% \textcolor[rgb]{1.00,0.00,0.00}{\mathbf{%
  |\mathcal{M}|^{2} \to
  \overline{| \mathcal{M}|^{2}} \equiv
  \displaystyle\frac{1}{(2s_{A}+1)}
  \displaystyle\sum\limits_{s_{A},\, s_{B}}
    |\mathcal{M}|^{2}.
% }}
\end{array}
\label{eq.3.2.3}
\end{equation}
We separate explicitly summations over spin states for proton and lepton as
[along Eqs.~(6.18) in Ref.~\cite{Halzen.book.1987}, p.~152]
\begin{equation}
\begin{array}{llllllll}
  \overline{| \mathcal{M}|^{2}} =
    \displaystyle\frac{e^{4}}{q^{4}}\,
    L_{\rm p}^{\mu\nu}\, L_{\mu\nu}^{\rm lep},
\end{array}
\label{eq.3.2.4}
\end{equation}
where $L_{\rm p}^{\mu\nu}$ is tensor associated with proton vertex having form
[along Eqs.~(6.19) in Ref.~\cite{Halzen.book.1987}, p.~152]
\begin{equation}
\begin{array}{llllllll}
  L_{\rm p}^{\mu\nu}\, =
  \displaystyle\frac{1}{2}\,
  \displaystyle\sum\limits_{\rm spin\, states\, for\, proton}
    \Bigl[\bar{u}_{C} (p')\, \gamma^{\mu}\, u_{A}(p) \Bigr]\,
    \Bigl[\bar{u}_{C} (p')\, \gamma^{\nu}\, u_{A}(p) \Bigr]^{*}.
\end{array}
\label{eq.3.2.5}
\end{equation}
For tensor $L_{\mu\nu}^{\rm lep}$ associated with lepton vertex we have similar expression.

We calculate summation over spins of proton in the initial and final states,
using technique of calculation of traces of $\gamma$-matrixes.
For proton tensor we obtain
[see (6.20), p.~153, book~\cite{Halzen.book.1987}]
\begin{equation}
\begin{array}{llllllll}
  L_{\rm p}^{\mu\nu}\, =
  \displaystyle\frac{1}{2}\,
  {\rm Tr}\,
    \Bigl[(\hat{p}' + m_{\rm p})\, \gamma^{\mu}\, (\hat{p} + m_{\rm p})\, \gamma^{\nu}\, \Bigr],
\end{array}
\label{eq.3.2.6}
\end{equation}
where $m_{\rm p}$ is mass of proton.
Using trace theorems of $\gamma$-matrixes
(see (6.22), p.~153, book~\cite{Halzen.book.1987})
\begin{equation}
\begin{array}{llllllll}
  {\rm Tr}\, (\hat{a} \, \hat{b}) = 4\, a \cdot b, &

  {\rm Tr}\, (\hat{a}\, \hat{b}\, \hat{c}\, \hat{d}) =
  4\, \Bigl[ (a \cdot b)\, (c \cdot d) - (a \cdot c)\, (b \cdot d) + (a \cdot d)\, (b \cdot c) \Bigr],
\end{array}
\label{eq.3.2.7}
\end{equation}
we calculate tensor $L_{\rm p}^{\mu\nu}$
(see (6.25), (6.26), p.~154, book~\cite{Halzen.book.1987})
\begin{equation}
\begin{array}{llllllll}
% \vspace{1.5mm}
  L_{\rm p}^{\mu\nu}\, & = &
%   \displaystyle\frac{1}{2}\, {\rm Tr}\, \bigl(\hat{p}' \gamma^{\mu}\, \hat{p}\, \gamma^{\nu} \bigr) +
%   \displaystyle\frac{m_{\rm p}^{2}}{2}\, {\rm Tr}\, \bigl(\gamma^{\mu}\, \gamma^{\nu}\, \bigr) = \\

%v  & = &
  2\, \Bigl[ p'^{\mu}\, p^{\nu} + p'^{\nu}\, p^{\mu} - \bigl(p' \cdot p - m_{\rm p}^{2} \bigr)\, g^{\mu\nu} \Bigr].
\end{array}
\label{eq.3.2.8}
\end{equation}
For leptonic tensor we obtain similar formula as
\begin{equation}
\begin{array}{llllllll}
% \vspace{1.5mm}
  L^{\rm lepton}_{\mu\nu}\, & = &
  2\, \Bigl[ k'_{\mu}\, k_{\nu} + k'_{\nu}\, k_{\mu} - \bigl(k' \cdot k - m_{\rm lep}^{2} \bigr)\, g_{\mu\nu} \Bigr],
\end{array}
\label{eq.3.2.9}
\end{equation}
where $m_{\rm lep}$ is mass of lepton.

Multiplying (\ref{eq.3.2.8}) on (\ref{eq.3.2.9}), we obtain complete amplitude from (\ref{eq.3.2.4}) as
(see (6.27), p.~154, book~\cite{Halzen.book.1987})
\begin{equation}
\begin{array}{llllllll}
% \vspace{1.5mm}
  \overline{| \mathcal{M}|^{2}} =
  \displaystyle\frac{8\, e^{4}}{q^{4}}\,
    \Bigl[ (p' \cdot k')\, (p \cdot k) + (p' \cdot k)\, (p \cdot k') -
       m_{\rm p}^{2}\, k' \cdot k -  m_{\rm lep}^{2} p' \cdot p + 2\, m_{\rm p}^{2}\, m_{\rm lep}^{2} \Bigr].
\end{array}
\label{eq.3.2.10}
\end{equation}
%
% In the extreme relativistic limit, we may neglect terms involving masses of proton and leptons and for amplitude we obtain
% (see (6.28), p.~154, book~\cite{Halzen.book.1987})
%
% \textcolor[rgb]{1.00,0.00,0.00}{\textbf{%
% (THIS FORMULA IS USED FOR EXPLANATION OF CONNECTION BETWEEN TWO FORMALISMS IN SECT. II.C)
% }}
%
% \begin{equation}
% \begin{array}{llllllll}
% \textcolor[rgb]{1.00,0.00,0.00}{\mathbf{%
%   \overline{| \mathcal{M}|^{2}} =
%   \displaystyle\frac{8\, e^{4}}{q^{4}}\,
%     \Bigl[ (p' \cdot k')\, (p \cdot k) + (p' \cdot k)\, (p \cdot k') \Bigr].
% }}
% \end{array}
% \label{eq.3.2.11}
% \end{equation}
%
% \textcolor[rgb]{1.00,0.00,0.00}{\textbf{%
This formalism above helps to understand which matrix structure of full matrix element of production of lepton pair should be found (obtained after developments) finally.
% }}
%-----------------------------------------------------------------------------------------------------------------------

%-----------------------------------------------------------------------------------------------------------------------
\section{Matrix element for nucleons of nuclear system in scattering
\label{sec.app.2}}

We define the matrix element of emission, using the wave functions $\Psi_{i}$ and $\Psi_{f}$ of the full nuclear system in states before emission of photons ($i$-state) and after such emission ($f$-state),
as
\begin{equation}
  F = \langle \Psi_{f} |\, \hat{H}_{\gamma} |\, \Psi_{i} \rangle,
% \label{eq.13.1.1}
\label{eq.app.2.1.2.3}
\end{equation}
where we integrate over all independent variables in this matrix element.
% i.e. space variables $\vb{R}$, $\vb{r}$, $\rhobf_{Am}$.
% We should take into account space representation of all used moments $\vu{P}$, $\vu{p}$, $\vb{\tilde{p}}_{A m}$
% (as
% $\vu{P} = -i\hbar\, \vb{d/dR}$,
% $\vu{p} = -i\hbar\, \vb{d/dr}$,
% $\vb{\tilde{p}}_{A m} = -i\hbar\, \vb{d/d} \rhobf_{Am}$).

We define the wave function of the full nuclear system as
\begin{equation}
  \Psi =
  \Phi (\vb{R}) \cdot
  \Phi_{\rm p - nucl} (\vb{r}) \cdot
  \psi_{A} (\beta_{A}) \cdot
  \psi_{B} (\beta_{B}),
\label{eq.app.2.1.2.4}
% \label{eq.app.2.6.1}
\end{equation}
where
\begin{equation}
\begin{array}{lcl}
  \psi_{\rm nucl} (\beta_{A}) =
  \psi_{\rm nucl} (1 \cdots A ) =
  \displaystyle\frac{1}{\sqrt{A!}}
  \displaystyle\sum\limits_{p_{A}}
    (-1)^{\varepsilon_{p_{A}}}
    \psi_{\lambda_{1}}(1)
    \psi_{\lambda_{2}}(2) \ldots
    \psi_{\lambda_{A}}(A),
\end{array}
\label{eq.app.2.1.2.5}
% \label{eq.app.2.6.2}
\end{equation}
following the formalism in Ref.~\cite{Maydanyuk_Zhang.2015.PRC} for the proton-nucleus scattering [see Sect.~II.B, Eqs.~(10)--(13)],
and we add description of many-nucleon structure of the nuclei as in Ref.~\cite{Maydanyuk_Zhang_Zou.2016.PRC}.
Here,
$\beta_{A}$ is the set of numbers $1 \cdots A$ of nucleons of the nucleus labeled by index $A$,
$\Phi (\vb{R})$ is
% \textcolor[rgb]{1.00,0.00,0.00}{\textbf{%
the wave function describing motion of center-of-mass of the full nuclear system in laboratory frame,
$\Phi_{\rm p - nucl} (\vb{r})$ is
% \textcolor[rgb]{1.00,0.00,0.00}{\textbf{%
the wave function describing relative motion between proton and nucleus (without description of internal relative motions of nucleons in nucleus),
% $\psi_{\rm p} (\beta_{p})$ is the wave function of the scattered proton,
$\psi_{\rm nucl} (\beta_{A})$ is
% \textcolor[rgb]{1.00,0.00,0.00}{\textbf{%
the many-nucleon wave function of the nucleus,
defined in Eq.~(12) Ref.~\cite{Maydanyuk_Zhang.2015.PRC} on the basis of one-nucleon wave functions $\psi_{\lambda_{s}}(s)$,
$\beta_{A}$ is the set of numbers $1 \cdots A$ of nucleons of the nucleus.
Summation in Eqs.~(\ref{eq.app.2.1.2.5}) is performed over all $A!$ permutations of coordinates or states of nucleons.
One-nucleon wave functions $\psi_{\lambda_{s}}(s)$ represent the multiplication of space and spin-isospin
functions as $\psi_{\lambda_{s}} (s) = \varphi_{n_{s}} (\vb{r}_{s})\, \bigl|\, \sigma^{(s)} \tau^{(s)} \bigr\rangle$,
where
$\varphi_{n_{s}}$ is the space function of the nucleon with number $s$,
$n_{s}$ is the number of state of the space function of the nucleon with number $s$,
$\bigl|\, \sigma^{(s)} \tau^{(s)} \bigr\rangle$ is the spin-isospin function of the nucleon with number $s$.
%
% \textcolor[rgb]{1.00,0.00,0.00}{\textbf{%
We choose one-nucleon space wave functions in harmonic oscillator basis.
Such a choice allows to describe many nucleon system as unified bound system where only two nucleon interactions are included.
Here, minimum of the full energy of such nuclear system gives binding energy of this studied nucleus and the corresponding oscillator lengths
used in one nucleon space wave functions
\cite{Steshenko.1971.SJNP}
(see also Refs.~\cite{Filippov.1980.SJNP,Filippov.1981.SJNP,Vasilevsky.1990.SJNP,
Filippov.1985.SJPN,Filippov.1986.SJNP,Filippov.1984.NPA,Filippov.1984.SJNP,Vasilevsky.1997.PAN,
Filippov.1994.PPN,Vasilevsky.2001.PRC,Vasilevsky.2012.PRC}
for more detailed approaches in that research line).
By such a way, we obtain definite deformation of the studied nucleus related with the oscillator lengths.
Also we obtain explicit formula for form factors of nucleus in our problem
(see App.~A in Ref.~\cite{Maydanyuk_Zhang_Zou.2016.PRC}, for details of calculations, definitions, explanations, reference therein).
So, our formalism allows to study role of internal structure of nucleus-target in the process of production of dileptons.
For example, one can analyze dependence of cross sections of the dilepton production on variation of the oscillator lengths,
but we omit this in this paper
(some similar calculations were done for bremsstrahlung emission during scattering of light nuclei in cluster approach,
see Fig.~5, Sect.~IV.D in Ref.~\cite{Maydanyuk_Vasilevsky.2023.fold.arXiv}).
% }}
% -----------------------------------------------------------------------------------------------------------------------

% -----------------------------------------------------------------------------------------------------------------------
Operator of emission
% ~(\ref{eq.pauli.5})
is [see Eqs.~(4), (5) in Ref.~\cite{Maydanyuk.2023.PRC.delta}]
\begin{equation}
\begin{array}{llll}
  \hat{H}_{\gamma} & = &
%   \displaystyle\sum_{i=1}^{A_{1}}
  \biggl\{
    - \displaystyle\frac{z_{\rm p} e}{m_{\rm p}c}\; \vu{p}_{\rm p} \cdot \vb{A}_{\rm p} -
    \mu_{N}\, \mu_{\rm p}\, \sigmabf \cdot \vu{H}_{\rm p}
  \biggr\} +

  \displaystyle\sum_{j=1}^{A}
  \biggl\{
    - \displaystyle\frac{z_{j} e}{m_{j}c}\; \vu{p}_{j} \cdot \vb{A}_{j} -
    \mu_{N}\, \mu_{j}\, \sigmabf \cdot \vu{H}_{j}
  \biggr\},
% +
%   \biggl\{
%     - \displaystyle\frac{z_{\Delta} e}{m_{\Delta}c}\; \vu{p}_{\Delta} \cdot \vb{A}_{\Delta} -
%     f_{\Delta} \cdot \mu_{\Delta}\, \sigmabf \cdot \vu{H}_{\Delta}
%   \biggr\}.
\end{array}
\label{eq.app.2.1.2.6}
% \label{eq.2.2.6}
\end{equation}
where
\begin{equation}
  \vu{H} = \vb{rot\: A} = \bigl[ \curl{\vb{A}} \bigr].
\label{eq.app.2.1.2.7}
% \label{eq.2.2.5}
\end{equation}
%
% где $\mu_{\Delta} = f_{\Delta} \cdot \mu_{p\,{\rm or}\, n}$ ($f_{\Delta}=3$).
% Это выражение --- многонуклонное обобщение оператора излучения $\hat{W}$ в~(4) в работе~\cite{Maydanyuk.2012.PRC} с
% включенными аномальными магнитными моментами нуклонов и изобары.
%
In this formalism the following definition for the potential of electromagnetic field is used
[see Eq.~(6) in Ref.~\cite{Maydanyuk.2023.PRC.delta}]:
\begin{equation}
\begin{array}{lcl}
  \vb{A} & = &
  \displaystyle\sum\limits_{\alpha=1,2}
    \sqrt{\displaystyle\frac{2\pi\hbar c^{2}}{w_{\rm ph}}}\; \vb{e}^{(\alpha),\,*}
    e^{-i\, \vb{k_{\rm ph}r}},
\end{array}
\label{eq.app.2.1.2.8}
% \label{eq.2.3.1}
\end{equation}
After calculations, operator of emission obtains form
[see Eqs.~(9) in Ref.~\cite{Maydanyuk.2023.PRC.delta}]
\begin{equation}
\begin{array}{lcl}
  \hat{H}_{\gamma} & = &
  \sqrt{\displaystyle\frac{2\pi\hbar c^{2}}{w_{\rm ph}}}\; \mu_{N}\,
%  \displaystyle\sum_{i=1}^{A_{1}}
  \displaystyle\sum\limits_{\alpha=1,2}
    e^{-i\, \vb{k_{\rm ph}r}_{\rm p}}\,
  \biggl\{
    i\, 2 z_{\rm p}\: \vb{e}^{(\alpha)} \cdot \grad_{\rm p} +
    \mu_{\rm p}\, \sigmabf \cdot \Bigl( i\, \bigl[ \vb{k_{\rm ph}} \times \vb{e}^{(\alpha)} \bigr] - \bigl[ \grad_{\rm p} \times \vb{e}^{(\alpha)} \bigr] \Bigr)
  \biggr\}\; + \\
  % \biggl\{ i\, \mu_{N}\, \displaystyle\frac{2 z_{K} m_{\rm p}}{m_{K}}\: \vb{e}^{(\alpha)} \cdot \grad_{K} \biggr\}\; + \\

  & + &
  \sqrt{\displaystyle\frac{2\pi\hbar c^{2}}{w_{\rm ph}}}\; \mu_{N}\,
  \displaystyle\sum_{j=1}^{A}
  \displaystyle\sum\limits_{\alpha=1,2}
    e^{-i\, \vb{k_{\rm ph}r}_{j}}\;
    \biggl\{
      i\, \displaystyle\frac{2 z_{j} m_{\rm p}}{m_{Aj}}\: \vb{e}^{(\alpha)} \cdot \grad_{j} +
      \mu_{j}\, \sigmabf \cdot \Bigl( i\, \bigl[ \vb{k_{\rm ph}} \times \vb{e}^{(\alpha)} \bigr] - \bigl[ \grad_{j} \times \vb{e}^{(\alpha)} \bigr] \Bigr)
    \biggr\}.
\end{array}
\label{eq.app.2.1.2.9}
% \label{eq.2.3.4}
\end{equation}
%-----------------------------------------------------------------------------------------------------------------------

%-----------------------------------------------------------------------------------------------------------------------
Calculations of matrix element are straightforward, we obtain
% Using formulas (\ref{eq.2.5.2})--(\ref{eq.2.5.7}) for the operator of emission, we calculate
[see Eqs.~(15)--(21) in Ref.~\cite{Maydanyuk.2023.PRC.delta}]
%
% [see Appendix~\ref{sec.app.short.2} for details, Eqs.~(\ref{eq.app.short.2.7.2}), (\ref{eq.app.short.2.12.3})--(\ref{eq.app.short.2.12.7})]
% [see Eqs.~(23), (34) and Appendixes B--D in Ref.~\cite{Maydanyuk_Zhang_Zou.2019.brem_alpha_nucleus.arxiv} adapting calculations for proton-nucleus scattering]:
%
\begin{equation}
\begin{array}{lll}
  \langle \Psi_{f} |\, \hat{H}_{\gamma} |\, \Psi_{i} \rangle \;\; = \;\;
  \sqrt{\displaystyle\frac{2\pi\, c^{2}}{\hbar w_{\rm ph}}}\,
  M_{\rm full}, &
  % \Bigl\{ M_{P} + M_{p}^{(E)} + M_{p}^{(M)} + M_{k} + M_{\Delta E} + M_{\Delta M} \Bigr\}, &

  M_{\rm full} = M_{P} + M_{p}^{(E)} + M_{p}^{(M)} + M_{k} + M_{\Delta E} + M_{\Delta M},
\end{array}
\label{eq.app.2.1.2.10}
% \label{eq.13.1.2}
\end{equation}
% -----------------------------------------------------------------------------------------------------------------------
%
%-----------------------------------------------------------------------------------------------------------------------
where for the proton-nucleus scattering we obtain
\begin{equation}
\begin{array}{lll}
\vspace{-0.2mm}
  M_{p}^{(E)} & = &
  2\,i \hbar\, (2\pi)^{3} \displaystyle\frac{m_{\rm p}}{\mu}\: \mu_{N}
  \displaystyle\sum\limits_{\alpha=1,2}
  \displaystyle\int\limits_{}^{}
    \Phi_{\rm p - nucl, f}^{*} (\vb{r})\;
    e^{-i\, \vb{k}_{\rm ph} \vb{r}} \cdot
    Z_{\rm eff} (\vb{k}_{\rm ph}, \vb{r}) \cdot \vb{e}^{(\alpha)}\, \vb{\displaystyle\frac{d}{dr}} \cdot
    \Phi_{\rm p - nucl, i} (\vb{r})\; \vb{dr}, \\

  M_{p}^{(M)} & = &
  -\, \hbar\, (2\pi)^{3} \displaystyle\frac{m_{\rm p}}{\mu}\: \mu_{N}
  \displaystyle\sum\limits_{\alpha=1,2}
  \displaystyle\int\limits_{}^{}
    \Phi_{\rm p - nucl, f}^{*} (\vb{r})\;
    e^{-i\, \vb{k}_{\rm ph} \vb{r}} \cdot
    \vb{M}_{\rm eff} (\vb{k}_{\rm ph}, \vb{r}) \cdot \Bigl[ \vb{\displaystyle\frac{d}{dr}} \times \vb{e}^{(\alpha)} \Bigr] \cdot
    \Phi_{\rm p - nucl, i} (\vb{r})\; \vb{dr},
\end{array}
\label{eq.app.2.1.2.11}
% \label{eq.13.1.3.p-nucl}
\end{equation}
\begin{equation}
\begin{array}{lllll}
\vspace{-0.1mm}
  M_{P} & = &
  \displaystyle\frac{\hbar\, (2\pi)^{3}}{m_{A} + m_{p}}\, \mu_{N}\,
  \displaystyle\sum\limits_{\alpha=1,2}
  \displaystyle\int\limits_{}^{}
    \Phi_{\rm p - nucl, f}^{*} (\vb{r})\;
  \biggl\{
    2\, m_{\rm p}\;
    \Bigl[
      e^{-i\, c_{A}\, \vb{k_{\rm ph}} \vb{r}} F_{p,\, {\rm el}} + e^{i\, c_{p}\, \vb{k_{\rm ph}} \vb{r}} F_{A,\, {\rm el}}
    \Bigr]\, \vb{e}^{(\alpha)} \cdot \vb{K}_{i}\; + \\

  & + &
    i\: \Bigl[
      e^{-i\, c_{A}\, \vb{k_{\rm ph}} \vb{r}}\, \vb{F}_{p,\, {\rm mag}} + e^{i\, c_{p}\, \vb{k_{\rm ph}} \vb{r}}\, \vb{F}_{A,\, {\rm mag}}
    \Bigr] \cdot
    \bigl[ \vb{K}_{i} \cp \vb{e}^{(\alpha)} \bigr]
  \biggr\} \cdot
  \Phi_{\rm p - nucl, i} (\vb{r})\; \vb{dr},
\end{array}
\label{eq.app.2.1.2.12}
% \label{eq.13.1.4.p-nucl}
\end{equation}
\begin{equation}
\begin{array}{lcl}
% \vspace{-0.1mm}
  M_{k} & = &
  i\, \hbar\, (2\pi)^{3}  \mu_{N}\,
  \displaystyle\sum\limits_{\alpha=1,2}
    \bigl[ \vb{k_{\rm ph}} \cp \vb{e}^{(\alpha)} \bigr]
  \displaystyle\int\limits_{}^{}
    \Phi_{\rm p - nucl, f}^{*} (\vb{r}) \cdot
    \Bigl\{ e^{-i\, c_{A}\, \vb{k_{\rm ph}} \vb{r}}\, \vb{D}_{p,\, {\rm k}} + e^{i\, c_{p}\, \vb{k_{\rm ph}} \vb{r}}\, \vb{D}_{A,\, {\rm k}} \Bigr\} \cdot
    \Phi_{\rm p - nucl, i} (\vb{r})\; \vb{dr},
\end{array}
\label{eq.app.2.1.2.13}
% \label{eq.13.1.5.p-nucl}
\end{equation}
%
% \begin{equation}
% \begin{array}{lll}
% \vspace{-0.1mm}
%   M_{\Delta E} & = &
%   -\, (2\pi)^{3}\, 2\, \mu_{N}
%   \displaystyle\sum\limits_{\alpha=1,2} \vb{e}^{(\alpha)}
%   \displaystyle\int\limits_{}^{}
%     \Phi_{\rm p - nucl, f}^{*} (\vb{r})\;
%   \biggl\{
%     \Bigl[ e^{-i\, c_{A}\, \vb{k_{\rm ph}} \vb{r}}\, \vb{D}_{p 1,\, {\rm el}} + e^{i\, c_{p}\, \vb{k_{\rm ph}} \vb{r}}\, \vb{D}_{A 1,\, {\rm el}} \Bigr]\; - \\
%   &- &
%     \Bigl[ e^{-i\, c_{A}\, \vb{k_{\rm ph}} \vb{r}}\, \vb{D}_{p 2,\, {\rm el}} +
%     \displaystyle\frac{m_{\rm p}}{m_{A}}\, e^{i\, c_{p}\, \vb{k_{\rm ph}} \vb{r}}\, \vb{D}_{A 2,\, {\rm el}} \Bigr]
%   \biggr\} \cdot
%   \Phi_{\rm p - nucl, i} (\vb{r})\; \vb{dr},
% \end{array}
% \label{eq.13.1.6}
% \end{equation}
%
% \begin{equation}
% \begin{array}{lll}
% \vspace{-0.1mm}
%   M_{\Delta M} & = &
%   -\, i\, (2\pi)^{3}\,  \mu_{N}\,
%   \displaystyle\sum\limits_{\alpha=1,2}
%   \displaystyle\int\limits_{}^{}
%     \Phi_{\rm p - nucl, f}^{*} (\vb{r})\;
%   \biggl\{
%     \Bigl[ e^{-i\, c_{A}\, \vb{k_{\rm ph}} \vb{r}}\; D_{p 1,\, {\rm mag}} (\vb{e}^{(\alpha)}) + e^{i\, c_{p}\, \vb{k_{\rm ph}} \vb{r}}\; D_{A 1,\, {\rm mag}} (\vb{e}^{(\alpha)}) \Bigr]\; - \\
%   & - &
%     \Bigl[ e^{-i\, c_{A}\, \vb{k_{\rm ph}} \vb{r}}\; D_{p 2,\, {\rm mag}} (\vb{e}^{(\alpha)}) + e^{i\, c_{p}\, \vb{k_{\rm ph}} \vb{r}}\; D_{A 2,\, {\rm mag}} (\vb{e}^{(\alpha)}) \Bigr]
%   \biggr\} \cdot
%   \Phi_{\rm p - nucl, i} (\vb{r})\; \vb{dr}
% \end{array}
% \label{eq.13.1.7}
% \end{equation}
%
\begin{equation}
\begin{array}{lll}
  M_{\Delta E} & = &
  -\, (2\pi)^{3}\, 2\, \mu_{N}
  \displaystyle\sum\limits_{\alpha=1,2} \vb{e}^{(\alpha)}
  \displaystyle\int\limits_{}^{}
    \Phi_{\rm p - nucl, f}^{*} (\vb{r})\;
  \biggl\{
    e^{i\, c_{p}\, \vb{k_{\rm ph}} \vb{r}}\, \vb{D}_{A 1,\, {\rm el}} -
    \displaystyle\frac{m_{\rm p}}{m_{A}}\, e^{i\, c_{p}\, \vb{k_{\rm ph}} \vb{r}}\, \vb{D}_{A 2,\, {\rm el}}
  \biggr\} \cdot
  \Phi_{\rm p - nucl, i} (\vb{r})\; \vb{dr},
\end{array}
\label{eq.app.2.1.2.14}
% \label{eq.13.1.6.p-nucl}
\end{equation}
\begin{equation}
\begin{array}{lll}
  M_{\Delta M} & = &
  -\, i\, (2\pi)^{3}\,  \mu_{N}\,
  \displaystyle\sum\limits_{\alpha=1,2}
  \displaystyle\int\limits_{}^{}
    \Phi_{\rm p - nucl, f}^{*} (\vb{r})\;
  \biggl\{
    e^{i\, c_{p}\, \vb{k_{\rm ph}} \vb{r}}\; D_{A 1,\, {\rm mag}} (\vb{e}^{(\alpha)}) -
    e^{i\, c_{p}\, \vb{k_{\rm ph}} \vb{r}}\; D_{A 2,\, {\rm mag}} (\vb{e}^{(\alpha)})
  \biggr\} \cdot
  \Phi_{\rm p - nucl, i} (\vb{r})\; \vb{dr},
\end{array}
\label{eq.app.2.1.2.15}
% \label{eq.13.1.7.p-nucl}
\end{equation}
\begin{equation}
\begin{array}{lll}
\vspace{1.0mm}
  Z_{\rm eff} (\vb{k}_{\rm ph}, \vb{r}) =
  e^{i\, \vb{k_{\rm ph}} \vb{r}}\,
  \Bigl[
    e^{-i\, c_{A} \vb{k_{\rm ph}} \vb{r}}\, \displaystyle\frac{m_{A}}{m_{p} + m_{A}}\, F_{p,\, {\rm el}} -
    e^{i\, c_{p} \vb{k_{\rm ph}} \vb{r}}\, \displaystyle\frac{m_{p}}{m_{p} + m_{A}}\, F_{A,\, {\rm el}}
  \Bigr], \\

  \vb{M}_{\rm eff} (\vb{k}_{\rm ph}, \vb{r}) =
  e^{i\, \vb{k_{\rm ph}} \vb{r}}\,
  \Bigl[
    e^{-i\, c_{A} \vb{k_{\rm ph}} \vb{r}}\,  \displaystyle\frac{m_{A}}{m_{p} + m_{A}}\, \vb{F}_{p,\, {\rm mag}} -
    e^{i\, c_{p} \vb{k_{\rm ph}} \vb{r}}\,  \displaystyle\frac{m_{p}}{m_{p} + m_{A}}\, \vb{F}_{A,\, {\rm mag}}
  \Bigr].
\end{array}
\label{eq.app.2.1.2.16}
% \label{eq.13.1.8.p-nucl}
% \label{eq.2.13.1.3}
\end{equation}
Here,
$F_{{\rm p},\, {\rm el}}$,
$F_{A,\, {\rm el}}$,
$\vb{F}_{{\rm p},\, {\rm mag}}$,
$\vb{F}_{A,\, {\rm mag}}$,
$\vb{D}_{A 1,\, {\rm el}}$,
$\vb{D}_{A 2,\, {\rm el}}$,
$D_{A 1,\, {\rm mag}}$,
$D_{A 2,\, {\rm mag}}$,
$\vb{D}_{{\rm p},\, {\rm k}}$,
$\vb{D}_{A,\, {\rm k}}$,
$D_{{\rm p}, P\, {\rm el}}$,
$D_{A,P\, {\rm el}}$,
$\vb{D}_{{\rm p}, P\, {\rm mag}}$,
$\vb{D}_{A,P\, {\rm mag}}$
are electric and magnetic form factors defined in
Sect.~I in Supplemental Material in Ref.~\cite{Maydanyuk.2023.PRC.delta} [see Eqs. (16), (21), (23), (25) there].
%
% Appendix~\ref{sec.app.short.2.9}
% [see Eqs.~(\ref{eq.app.short.2.9.7}), (\ref{eq.app.short.2.9.b.4}), (\ref{eq.app.short.2.9.b.4}), (\ref{eq.app.short.2.9.c.2}), (\ref{eq.app.short.2.9.d.2})].
% -----------------------------------------------------------------------------------------------------------------------

%-----------------------------------------------------------------------------------------------------------------------
\subsection{Monopole approximation of effective electric charge and magnetic moment of nuclear system
\label{sec.app.2.resultingformulas}}

We obtain the matrix elements for the coherent bremsstrahlung
[see Eqs.~(22) in Ref.~\cite{Maydanyuk.2023.PRC.delta},
little modification due to new Eq.~(\ref{eq.app.2.resultingformulas.6})]
% [see Appendix~\ref{sec.result.formulas}, Eqs.~(\ref{eq.result.formulas.3})]
%
\begin{equation}
\begin{array}{lll}
\vspace{0.2mm}
  M_{p}^{(E,\, {\rm mon},0)} & = &
  i \hbar\, (2\pi)^{3} \displaystyle\frac{2\, \mu_{N}\,  m_{\rm p}}{\mu}\;
  Z_{\rm eff}^{\rm (mon,\, 0)}\;
  \displaystyle\sum\limits_{\alpha=1,2} \vb{e}^{(\alpha)} \cdot \vb{I}_{1}, \\

% \end{equation}
% \begin{array}{lll}
  M_{p}^{(M,\, {\rm mon},\, 0)} & = &
  \hbar\, (2\pi)^{3}\, \mu_{N}\, \alpha\;
  (\vb{e}_{\rm x} + \vb{e}_{\rm z})\,
  \displaystyle\sum\limits_{\alpha=1,2} \Bigl[ \vb{I}_{1} \times \vb{e}^{(\alpha)} \Bigr],
\end{array}
\label{eq.app.2.resultingformulas.1}
% \label{eq.result.formulas.3}
\end{equation}
and for incoherent bremsstrahlung
[see Eqs.~(23) in Ref.~\cite{Maydanyuk.2023.PRC.delta}]
% [Appendixes~\ref{sec.result.formulas}, Eqs.~(\ref{eq.result.formulas.6}), p.~\pageref{eq.result.formulas.6}]
%
\begin{equation}
\begin{array}{lll}
\vspace{1.4mm}
  M_{\Delta E} = 0, \\

\vspace{1.4mm}
  M_{\Delta M} =
  i\, \hbar\, (2\pi)^{3}\, \mu_{N}\, k_{\rm ph} \cdot
%   \Bigl[
    f_{A} \cdot Z_{\rm A} (\vb{k}_{\rm ph}) \cdot
    I_{2} (- c_{\rm p}), \\
%     + f_{B1} \cdot Z_{\rm B} (\vb{k}_{\rm ph}) \cdot I_{2} (+c_{A})
%   \Bigr], \\

  M_{k} =
    -\, i\, \hbar\, (2\pi)^{3}\, \mu_{N} \cdot k_{\rm ph}\, z_{\rm p}\: \mu_{\rm p} \cdot I_{2} (+ c_{A}) -
    \displaystyle\frac{\bar{\mu}_{\rm pn}}{f_{1}} \cdot M_{\Delta M}, \\

  M_{\Delta M} + M_{k} =
    -\, i\, \hbar\, (2\pi)^{3}\, \mu_{N}\, k_{\rm ph}\,
    \Bigl\{
      \displaystyle\frac{A+1}{2A}\, \bar{\mu}_{\rm pn}^{(A)}\, Z_{\rm A} (\vb{k}_{\rm ph}) \cdot I_{2} (-c_{\rm p}) +
      \mu_{\rm p}\, z_{\rm p} \cdot I_{2} (+ c_{A})

      % z_{\rm p}\: \mu_{\rm p} \cdot I_{3} - \displaystyle\frac{\bar{\mu}_{\rm pn}}{f_{1}} \cdot M_{\Delta M}
    \Bigr\}

\end{array}
\label{eq.app.2.resultingformulas.2}
% \label{eq.resultingformulas.2}
\end{equation}
and
[see Eqs.~(26) in Ref.~\cite{Maydanyuk.2023.PRC.delta}]
% [see Eq.~(\ref{eq.result.formulas.7}), p.~\pageref{eq.result.formulas.7}]
%
\begin{equation}
\begin{array}{lll}
  f_{A} = \displaystyle\frac{A-1}{2A}\: \bar{\mu}_{\rm pn}^{\rm (A)}, &
  \bar{\mu}_{\rm pn}^{\rm (A)} - f_{A} = \displaystyle\frac{A+1}{2A}\: \bar{\mu}_{\rm pn}^{\rm (A)}.
%   f_{B1} = \displaystyle\frac{B-1}{2B}\: \bar{\mu}_{\rm pn}^{\rm (B)}.
\end{array}
\label{eq.app.2.resultingformulas.3}
% \label{eq.resultingformulas.3}
% \label{eq.result.formulas.7}
\end{equation}
Here,
% Здесь
$\bar{\mu}_{\rm pn}^{\rm (A)} = \mu_{\rm p}^{\rm (an)} + \kappa_{A}\,\mu_{\rm n}^{\rm (an)}$,
% $\bar{\mu}_{\rm pn}^{\rm (B)} = \mu_{\rm p}^{\rm (an)} + \kappa_{B}\,\mu_{\rm n}^{\rm (an)}$,
$\mu_{\rm p}^{\rm (an)} = 2.79284734462$ is anomalous magnetic moment for proton,
$\mu_{\rm n}^{\rm (an)} = -1.91304273$ is anomalous magnetic moment for neutron
% $\mu_{j}$ are magnetic moments of protons or neutrons of nucleus
(measured in units of nuclear magneton $\mu_{N}$, see Ref.~\cite{RewPartPhys_PDG.2018}),
$\kappa_{A} = (A-N_{A})/N_{A}$,
% $\kappa_{B} = (B-N_{B})/N_{B}$,
$A$ and $N_{A}$ are numbers of nucleons and neutrons in nucleus with index $A$.
% $A$ and $N_{A}$ (or $B$ and $N_{B}$) are numbers of nucleons and neutrons in nnucleus with index $A$ (or numbers of nucleons and neutrons in nucleus with index $B$).
% $A$ и $N_{A}$ (или $B$ и $N_{B}$) --- числа нуклонов и нейтронов в ядре с номером $A$ (или числа нуклонов и нейтронов в ядре с номером $B$).
% -----------------------------------------------------------------------------------------------------------------------
%
% -----------------------------------------------------------------------------------------------------------------------
Integrals are
[see Eqs.~(24) in Ref.~\cite{Maydanyuk.2023.PRC.delta}]
% [see Eqs.~(\ref{eq.result.formulas.8}), p.~\pageref{eq.result.formulas.8}]
% Интегралы определены так:
%
\begin{equation}
\begin{array}{lll}
\vspace{1.0mm}
  \vb{I}_{1} = \biggl\langle\: \Phi_{\rm p - nucl, f} (\vb{r})\; \biggl|\, e^{-i\, \vb{k}_{\rm ph} \vb{r}}\; \vb{\displaystyle\frac{d}{dr}} \biggr|\: \Phi_{\rm p - nucl, i} (\vb{r})\: \biggr\rangle_\mathbf{r}, \\
% \vspace{0.5mm}
  I_{2} (\pm c) =
  \Bigl\langle \Phi_{\rm p - nucl, f} (\vb{r})\; \Bigl|\, e^{\pm i\, c\, \vb{k_{\rm ph}} \vb{r}}\, \Bigr|\, \Phi_{\rm p - nucl, i} (\vb{r})\: \Bigr\rangle_\mathbf{r}.
%   I_{2} (\pm c) =
%   \Bigl\langle \Phi_{\rm p - nucl, f} (\vb{r})\; \Bigl|\, e^{\pm i\, c_{p}\, \vb{k_{\rm ph}} \vb{r}}\, \Bigr|\, \Phi_{\rm p - nucl, i} (\vb{r})\: \Bigr\rangle_\mathbf{r}, \\
%   I_{3} = \Bigl\langle \Phi_{\rm p - nucl, f} (\vb{r})\; \Bigl|\, e^{-i\, c_{A}\, \vb{k_{\rm ph}} \vb{r}}\, \Bigr|\, \Phi_{\rm p - nucl, i} (\vb{r})\: \Bigr\rangle_\mathbf{r}.
\end{array}
\label{eq.app.2.resultingformulas.4}
% \label{eq.resultingformulas.4}
% \label{eq.result.formulas.8}
\end{equation}

The effective electric charge and magnetic moment are
[see Eqs.~(25) in Ref.~\cite{Maydanyuk.2023.PRC.delta}]
% [see Eqs.~(\ref{eq.app.2.13.2.1}) in p.~\pageref{eq.app.2.13.2.1}, (\ref{eq.app.2.13.5.1}) in p.~\pageref{eq.app.2.13.5.1}]
%
\begin{equation}
\begin{array}{llll}
  Z_{\rm eff}^{\rm (mon)} (\vb{k}_{\rm ph}) = \displaystyle\frac{m_{A}\, z_{\rm p} - m_{\rm p}\, Z_{\rm A}(\vb{k}_{\rm ph})}{m_{\rm p} + m_{A}}, \\

  \vb{M}_{\rm eff}^{\rm (mon)} (\vb{k}_{\rm ph}) =
  \Bigl[
    z_{\rm p}\, m_{A}\, \mu_{\rm p} -
    Z_{\rm A} (\vb{k}_{\rm ph})\: m_{p}\, \bar{\mu}_{\rm pn}
  \Bigr] \cdot
  \displaystyle\frac{m_{p}}{m_{p} + m_{A}}\; (\vb{e}_{\rm x} + \vb{e}_{\rm z})
\end{array}
\label{eq.app.2.resultingformulas.5}
% \label{eq.resultingformulas.5}
% \label{eq.app.2.13.2.1}
\end{equation}
and we obtain
[$\alpha = \alpha_{M} \cdot m_{\rm p} / \mu^{2}$ , $\alpha_{M}$ is defined in Eqs.~(26) in Ref.~\cite{Maydanyuk.2023.PRC.delta},
new parameter $\alpha$ is more useful for nucleus-nucleus formalism]
% [see Eq.~(\ref{eq.app.matr_el.coh_mag.9}), p.~\pageref{eq.app.matr_el.coh_mag.9}]
%
\begin{equation}
\begin{array}{lll}
  \alpha & = &
  \displaystyle\frac{m_{\rm p}}{\mu}\;
  \Bigl[
    \displaystyle\frac{m_{\rm p}}{m_{A}}\, Z_{\rm A} (\vb{k}_{\rm ph})\, \bar{\mu}_{\rm pn}^{\rm (A)} -
    % \displaystyle\frac{m_{\rm p}}{m_{p}}\,
    z_{\rm p}\, \mu_{\rm p}
  \Bigr] =

  \displaystyle\frac{m_{\rm p}^{2}}{\mu^{2}}\;
  \Bigl[
    \displaystyle\frac{m_{p}\, Z_{\rm A} (\vb{k}_{\rm ph})\, \bar{\mu}_{\rm pn}^{\rm (A)} - m_{A}\, z_{\rm p}\, \mu_{\rm p}}{m_{A} + m_{p}}
  \Bigr] \ne

  \displaystyle\frac{m_{\rm p}^{2}}{\mu^{2}}\,
  Z_{\rm eff}^{\rm (mon)} (\vb{k}_{\rm ph})\,
  % \Bigl[ \displaystyle\frac{m_{B}\, Z_{\rm A} (\vb{k}_{\rm ph}) - m_{A}\, Z_{\rm B} (\vb{k}_{\rm ph})}{m_{A} + m_{B}} \Bigr]\;
  \bar{\mu}_{\rm pn}^{\rm (A)}.
\end{array}
\label{eq.app.2.resultingformulas.6}
% \label{eq.resultingformulas.6}
% \label{eq.model.incoh.1.5}
\end{equation}

% \begin{equation}
% \begin{array}{lll}
%   \alpha_{M} =
%   \Bigl[ Z_{\rm A} (\vb{k}_{\rm ph})\: m_{p}\, \bar{\mu}_{\rm pn} - z_{\rm p}\, m_{A}\, \mu_{\rm p} \Bigr]\, \displaystyle\frac{m_{p}}{m_{p} + m_{A}}.
%
%    \alpha & = &
%    \displaystyle\frac{m_{\rm p}^{2}}{\mu^{2}}\;
%    \Bigl[
%      \displaystyle\frac{m_{B}\, Z_{\rm A} (\vb{k}_{\rm ph})\, \bar{\mu}_{\rm pn}^{\rm (A)} - m_{A}\, Z_{\rm B} (\vb{k}_{\rm ph})\, \bar{\mu}_{\rm pn}^{\rm (A)}}{m_{A} + m_{B}}
%    \Bigr].
% \end{array}
% \label{eq.resultingformulas.6}
% \end{equation}
% -----------------------------------------------------------------------------------------------------------------------

%-----------------------------------------------------------------------------------------------------------------------
\subsection{Multiple expansion of matrix elements with ``explicit'' integrations
\label{sec.app.2.multiple}}

Calculation of integrals (\ref{eq.app.2.resultingformulas.4}) is straightforward.
% In result, we obtain:
% We rewrite the found solutions [see Eqs.~(\ref{eq.simplestcase.4}), (\ref{eq.simplestcase.7})]:
But in this paper we present more accurate solution for the matrix elements than in Ref.~\cite{Maydanyuk.2023.PRC.delta}  [see Eqs.~(27) in that paper]
%-----------------------------------------------------------------------------------------------------------------------
%
%-----------------------------------------------------------------------------------------------------------------------
% Using solutions (\ref{eq.result.formulas.8}) for integrals, resulting formulas are
% [see Eqs.~(\ref{eq.app.simplestcase.10}), (\ref{eq.app.simplestcase.12}), (\ref{eq.app.simplestcase.incoh.3}), (\ref{eq.app.simplestcase.incoh.4}),
% p.~\pageref{eq.app.simplestcase.10}, \pageref{eq.app.simplestcase.12}]
%
\begin{equation}
\begin{array}{lll}
\vspace{1.0mm}
  M_{p}^{(E,\, {\rm mon},\, 0)} =
  -\, \hbar\, (2\pi)^{3}\,
  \displaystyle\frac{\mu_{N}}{\sqrt{3}}\,
  \displaystyle\frac{m_{\rm p}\, Z_{\rm eff}^{\rm (mon,\, 0)}}{\mu}\,
  \Bigl(
    J_{1}(0,1,0) -
    \displaystyle\frac{47}{40} \sqrt{\displaystyle\frac{1}{2}}\, J_{1}(0,1,2)
  \Bigr), \\

\vspace{1.0mm}
  M_{p}^{(E,\, {\rm mon},\, 0)} + M_{p}^{(M,\, {\rm mon},\, 0)} =
  M_{p}^{(E,\, {\rm mon},\, 0)}\,
  \Bigl( 1 + i\: \displaystyle\frac{\alpha_{M}} {2\, m_{\rm p}\, Z_{\rm eff}^{\rm (mon,\, 0)}} \Bigr) =

  M_{p}^{(E,\, {\rm mon},\, 0)}\,
  \Bigl( 1 + i\: \displaystyle\frac{\mu^{2}} {2\, m_{\rm p}^{2}\, Z_{\rm eff}^{\rm (mon,\, 0)}}\; \alpha \Bigr), \\

\vspace{1.0mm}
  M_{\Delta M} =
  \hbar\, (2\pi)^{3}\,
  \displaystyle\frac{\sqrt{3}}{2}\,
  \mu_{N}\, k_{\rm ph}\,
%   \Bigl\{
  f_{A} \cdot Z_{\rm A} (\vb{k}_{\rm ph}) \cdot \tilde{J}\, (- c_{\rm p}, 0,1,1), \\
%   + f_{B1} \cdot Z_{\rm B} (\vb{k}_{\rm ph}) \cdot \tilde{J}\, (+c_{A}, 0,1,1) \Bigr\}, \\

  M_{\Delta M} + M_{k} =
  - \hbar\, (2\pi)^{3}\,
  \displaystyle\frac{\sqrt{3}}{2}\,
  \mu_{N}\, k_{\rm ph}\,
  \Bigl\{
    \displaystyle\frac{A+1}{2A}\, \bar{\mu}_{\rm pn}^{(A)} \cdot Z_{\rm A} (\vb{k}_{\rm ph})\, \tilde{J}\, (- c_{\rm p}, 0,1,1) +
    \mu_{\rm p}\, z_{\rm p}\, \tilde{J}\, (c_{A}, 0,1,1)
    % \displaystyle\frac{B+1}{2B}\, \bar{\mu}_{\rm pn}^{(B)} \cdot Z_{\rm B} (\vb{k}_{\rm ph}) \cdot \tilde{J}\, (+c_{A}, 0,1,1)
  \Bigr\}.
\end{array}
\label{eq.app.2.multiple.1}
\end{equation}
%
% where
% [see Eq.~(\ref{eq.app.matr_el.coh_mag.9})]
%
% \begin{equation}
% \begin{array}{lll}
%   \alpha_{M} & = &
%   \displaystyle\frac{m_{\rm p}}{\mu}\;
%   \Bigl[
%     \displaystyle\frac{m_{\rm p}}{m_{A}}\, Z_{\rm A} (\vb{k}_{\rm ph})\, \bar{\mu}_{\rm pn}^{\rm (A)} -
%     \displaystyle\frac{m_{\rm p}}{m_{B}}\, Z_{\rm B} (\vb{k}_{\rm ph})\, \bar{\mu}_{\rm pn}^{\rm (B)}
%   \Bigr] =
%
%   \displaystyle\frac{m_{\rm p}^{2}}{\mu^{2}}\;
%   \Bigl[
%     \displaystyle\frac{m_{B}\, Z_{\rm A} (\vb{k}_{\rm ph})\, \bar{\mu}_{\rm pn}^{\rm (A)} - m_{A}\, Z_{\rm B} (\vb{k}_{\rm ph})\, \bar{\mu}_{\rm pn}^{\rm (A)}}{m_{A} + m_{B}}
%   \Bigr] \ne
%
%   \displaystyle\frac{m_{\rm p}^{2}}{\mu^{2}}\,
%   Z_{\rm eff}^{\rm (mon)} (\vb{k}_{\rm ph})\,
%   % \Bigl[ \displaystyle\frac{m_{B}\, Z_{\rm A} (\vb{k}_{\rm ph}) - m_{A}\, Z_{\rm B} (\vb{k}_{\rm ph})}{m_{A} + m_{B}} \Bigr]\;
%   \bar{\mu}_{\rm pn}^{\rm (A)}.
% \end{array}
% \label{eq.multimple.2}
% \end{equation}

% \vspace{3.0mm}
% \noindent
At  $\bar{\mu}_{\rm pn}^{(A)} \approx \mu_{\rm p}$ we have
% \textcolor[rgb]{0.00,0.00,1.00}{\textbf{\large{\underline{Approximation $\bar{\mu}_{\rm pn}^{(A)} \approx \bar{\mu}_{\rm pn}^{(B)}$:}}}}
%
\begin{equation}
\begin{array}{lll}
\vspace{1.0mm}
  M_{p}^{(E,\, {\rm mon},\, 0)} + M_{p}^{(M,\, {\rm mon},\, 0)} =
  M_{p}^{(E,\, {\rm mon},\, 0)}\,
  \Bigl( 1 - i\, \displaystyle\frac{\bar{\mu}_{\rm pn}^{(A)} }{2} \Bigr), &

  \alpha \approx
  -\, \displaystyle\frac{m_{\rm p}^{2}}{\mu^{2}}\,
  Z_{\rm eff}^{\rm (mon)} (\vb{k}_{\rm ph})\,
  \bar{\mu}_{\rm pn}^{\rm (A)}.
\end{array}
\label{eq.app.2.multiple.2}
% \label{eq.multimple.3}
% \label{eq.result.formulas.11}
% \label{eq.app.simplestcase.incoh.4}
\end{equation}
%-----------------------------------------------------------------------------------------------------------------------
%
%-----------------------------------------------------------------------------------------------------------------------
Radial integrals are
\begin{equation}
\begin{array}{llllll}
  J_{1}(l_{i},l_{f},n) & = & \displaystyle\int\limits^{+\infty}_{0} \displaystyle\frac{dR_{i}(r, l_{i})}{dr}\: R^{*}_{f}(l_{f},r)\, j_{n}(k_{\rm ph}r)\; r^{2} dr, &
  \tilde{J}\,(c, l_{i},l_{f},n) & = & \displaystyle\int\limits^{+\infty}_{0} R_{i}(l_{i}, r)\, R^{*}_{f}(l_{f},r)\, j_{n}(c\, k_{\rm ph}r)\; r^{2} dr.
%  \breve{J}\,(c_{A}, l_{i}, l_{f},n) & = & \displaystyle\int\limits^{+\infty}_{0} R_{i}(r)\, R^{*}_{l,f}(r)\, V(\mathbf{r})\, j_{n}(c_{A}\,kr)\; r^{2} dr.
\end{array}
\label{eq.app.2.multiple.3}
% \label{eq.multimple.4}
% \label{eq.resultingformulas.7}
\end{equation}
Here, $R_{i,f}$ is radial part of wave function $\Phi_{p - nucl} (\vb{r})$ in $i$-state or $f$-state,
$j_{n}(k_{\rm ph}r)$ is spherical Bessel function of order $n$.
% *******************************************************************************************************************
%
% *******************************************************************************************************************
Previously, we defined cross-sections of the emitted bremsstrahlung photons on the basis of the full matrix element $p_{fi}$ % (\ref{eq.2.5.6.2})
in frameworks of formalism given in Refs.~\cite{Maydanyuk_Zhang_Zou.2016.PRC,Maydanyuk.2012.PRC,Maydanyuk_Zhang.2015.PRC}
(see Eq.~(22) in Ref.~\cite{Maydanyuk_Zhang_Zou.2016.PRC}, reference therein) as
%
% \footnote{We obtain the formula (\ref{eq.2.6.1}) in dependence on mass of proton $m_{\rm p}$ while in Ref.~\cite{Maydanyuk.2012.PRC} we had the bremsstrahlung probability (49) in dependence on the reduced mass $\mu$.
% Such a difference is explained by that in the current paper we develop formalism on the basis of the emission operator of the many-nucleon system (\ref{eq.2.2.3})
% while in Ref.~\cite{Maydanyuk.2012.PRC} we started the formalism on the basis of the operator of emission (4) of the proton-nucleus system defined via the reduced mass of proton and nucleus.}
%
\begin{equation}
\begin{array}{llll}
  \displaystyle\frac{d \sigma^{\rm (br)}}{dw_{\rm ph}} =
    \displaystyle\frac{e^{2}}{2\pi\,c^{5}}\: \displaystyle\frac{w_{\rm ph}\,E_{i}}{m_{\rm p}^{2}\,k_{i}}\:
    \bigl| p_{fi} \bigr|^{2}, &

%   \displaystyle\frac{d^{2} \sigma}{dw_{\rm ph}\, d \cos \theta} =
%     \displaystyle\frac{e^{2}}{2\pi\,c^{5}}\: \displaystyle\frac{w_{\rm ph}\,E_{i}}{m_{\rm p}^{2}\,k_{i}}\:
%     \bigl\{ p_{fi} \displaystyle\frac{d\,p_{fi}^{*}}{d\, \cos \theta}  + c.\,c. \bigr\}, &
  M_{\rm full} = - \displaystyle\frac{e}{m_{\rm p}}\, p_{fi}.
\end{array}
\label{eq.app.2.bremprobability.1}
\end{equation}
%
% where $p_{fi}$ is proportional to the electrical component $p_{\rm el}$ in Eqs.~(10) in \cite{Maydanyuk.2012.PRC}
% [with the additional factor of $2\, e^{-\, (a^{2} k_{x}^{2} + b^{2} k_{y}^{2} + c^{2} k_{z}^{2})\,/4}$ and the included effective charge $\tilde{Z}_{\rm eff}^{\rm (dip)}$] and
% $d\,p_{fi} (\theta_{f})\, / d\,\cos{\theta_{f}}$ is defined by the same way as $d\,p\, (k_{i}, k_{f}, \theta_{f})\, / d\,\cos{\theta_{f}}$ in Ref.~\cite{Maydanyuk.2012.PRC}.
% where c.\,c. is complex conjugation.
% We calculated the different contributions of the emitted photons to the full bremsstrahlung spectrum.
% For estimation of the interesting contribution, we use the corresponding matrix element of emission.
% *******************************************************************************************************************

% *******************************************************************************************************************
% \newpage
\section{Inclusion of longitudinal polarization of virtual photon
\label{sec.virtual}}

% \subsection{Explicit separation of tensor associated with production of leptons pair in square of $S$-matrix element
% \label{sec.7.1}}

In description of emission of virtual photon
we should take into account its non-zero longitudinal polarization.
To understand that, we follow to the formalism developed in Sect.~\ref{sec.4}.
We need to calculate the following matrix element
\begin{equation}
\begin{array}{lllll}
% \vspace{0.5mm}
  \vb{I}_{1,\,\mu} =
  \biggl\langle\: \Phi_{f} (\vb{r})\; \biggl|\, e^{-i\, \vb{k}_{\rm ph} \vb{r}}\; \vb{\displaystyle\frac{d}{dr^{\mu}}} \biggr|\: \Phi_{i} (\vb{r})
  \biggr\rangle,
% \vspace{0.5mm}
%   I_{2} = \Bigl\langle \Phi_{f} (\vb{r})\; \Bigl|\, e^{i\, c_{p}\, \vb{k_{\rm ph}} \vb{r}}\, \Bigr|\, \Phi_{i} (\vb{r})\: \Bigr\rangle_\mathbf{r}, \\
% \vspace{0.5mm}
%   I_{3} = \Bigl\langle \Phi_{f} (\vb{r})\; \Bigl|\, e^{-i\, c_{A}\, \vb{k_{\rm ph}} \vb{r}}\, \Bigr|\, \Phi_{i} (\vb{r})\: \Bigr\rangle_\mathbf{r}, \\
%   I_{4} = \Bigl\langle \Phi_{f} (\vb{r})\; \Bigl|\, e^{- i\, c_{A}\, \vb{k_{\rm ph}} \vb{r}}\, V(\vb{r})\, \Bigr|\, \Phi_{i} (\vb{r})\: \Bigr\rangle_\mathbf{r}.
\end{array}
\label{eq.virtual.0.1}
% \label{eq.app.2.1.7}
\end{equation}
where we have to take into account the longitudinal part of virtual photon.
%-----------------------------------------------------------------------------------------------------------------------

%-----------------------------------------------------------------------------------------------------------------------
% \textcolor[rgb]{1.00,0.00,0.00}{\textbf{%
% \subsection{Separation of the wave vector of (virtual) photon on the transverse and longitudinal parts
% \label{sec.virtual.1}}
% }}

In QED,  two independent gauges are imposed on wave function of the emitted real photon.
These are Lorentz gauge like $\partial^{\mu} A_{\mu}\, (\vb{r},t) = 0$ and Coulomb gauge like $\vb{\displaystyle\frac{d}{dr}\, A} = 0$ (or alternative one),
where $A_{\mu}^{\lambda}(x)$ can be defined in Eq.~(\ref{eq.4.3.1}).
In result, we obtain two independent relations between vectors of polarization $\vb{e}_{\mu}^{\lambda}$ and wave vector of photon $\vb{k_{\rm ph}}$.
In particular, Coulomb gauge gives requirement that space vectors of polarization are perpendicular to wave vector of photon,
i.e. $\vb{e}^{(1,2)} \cdot \vb{k_{\rm ph}} = 0$.

In the case of virtual photon,
Coulomb gauge
% \textcolor[rgb]{1.00,0.00,0.00}{\textbf{%
is not imposed, and only one relation still applied on vectors of polarization and wave vector of photon.
So, in a general case, we do not have requirement that space vectors of polarization are perpendicular to wave vector of photon,
i.e. we obtain $\vb{e}^{(1,2)} \cdot \vb{k_{\rm ph}} \ne 0$.
By such a reason, we write down the full wave vector of virtual photon as
\begin{equation}
\begin{array}{lcl}
  \vb{k}_{\rm ph} =
  \vb{k}_{\rm ph}^{\parallel} + \vb{k}_{\rm ph}^{\perp},
\end{array}
\label{eq.virtual.1.1}
\end{equation}
where $\vb{k}_{\rm ph}^{\perp}$ is transverse part of wave vector which is perpendicular to the space vectors of polarization $\vb{e}^{(1,2)}$ (like for real photon),
$\vb{k}_{\rm ph}^{\parallel}$ is rest, i.e longitudinal part of wave vector of the virtual photon.
%-----------------------------------------------------------------------------------------------------------------------
%
%-----------------------------------------------------------------------------------------------------------------------
% To take the longitudinal part of wave vector of virtual photon into account, we separate the full wave vector of this photon on
So, we write down
\begin{equation}
\begin{array}{lllll}
\vspace{1.5mm}
  \vb{k}_{\rm ph} \vb{r} =
  \vb{k}_{\rm ph}^{\parallel} \vb{r} + \vb{k}_{\rm ph}^{\perp} \vb{r}, &

% \vspace{1.5mm}
  e^{\pm i\, \vb{k}_{\rm ph} \vb{r}} =
  e^{\pm i\, \vb{k}_{\rm ph}^{\parallel} \vb{r}} \cdot e^{\pm i\, \vb{k}_{\rm ph}^{\perp} \vb{r}}.
\end{array}
\label{eq.virtual.1.2}
% \label{eq.app.2.1.1}
\end{equation}
% ***************************************************************************

% ***************************************************************************
% \subsection{Spherical waves expansion
% \label{sec.virtual.2}}

For
% \textcolor[rgb]{1.00,0.00,0.00}{\textbf{%
numerical calculations of the integral (\ref{eq.virtual.0.1}),
let's apply the expansion of the plane wave over the spherical waves
[for example, see Ref.~\cite{Landau.v3.1989}, p.~144, (34.1)]:
\begin{equation}
  e^{ikz} =
    \displaystyle\sum\limits_{l=0}^{+\infty}
    (-i)^{l} (2l+1) P_{l} (\cos{\beta})
    \biggl(\displaystyle\frac{r}{k} \biggr)^{l}
    \biggl(\displaystyle\frac{1}{r} \displaystyle\frac{d}{dr} \biggr)^{l}
    \displaystyle\frac{\sin {kr}}{kr},
\label{eq.virtual.2.1}
\end{equation}
where $z=r \cos{\beta}$
(axis $z$ is along vector $\vb{k}$, $k = |\vb{k}|$).
Introducing the spherical Bessel functions (see book~\cite{Landau.v3.1989},
p.~139, (33.9), (33.10) and (33.11)):
\begin{equation}
  j_{l}(kr) =
    (-1)^{l} \biggl(\displaystyle\frac{r}{k} \biggr)^{l}
    \biggl(\displaystyle\frac{1}{r} \displaystyle\frac{d}{dr} \biggr)^{l}
    \displaystyle\frac{\sin {kr}}{kr},
\label{eq.virtual.2.2}
% \label{eq.4.2}
\end{equation}
we obtain:
\begin{equation}
  e^{-ikr \cos{\beta}} =
  \biggl(e^{ikr \cos{\beta}} \biggr)^{*} =
    \displaystyle\sum\limits_{l=0}^{+\infty}
    i^{l} (-1)^{l} (2l+1) P_{l} (\cos{\beta}) j_{l}(kr).
\label{eq.virtual.2.3}
% \label{eq.4.3}
\end{equation}
So, for the longitudinal wave vector we write down
\begin{equation}
  e^{-i\, \vb{k}_{\rm ph}^{\parallel} \vb{r}} =
  e^{-ik^{\parallel} r \cos{\beta}} =
  \biggl(e^{ik^{\parallel} r \cos{\beta}} \biggr)^{*} =
    \displaystyle\sum\limits_{l_{\parallel} =0}^{+\infty}
    i^{l_{\parallel}} (-1)^{l_{\parallel}} (2l_{\parallel} +1) P_{l_{\parallel}} (\cos{\beta}) j_{l}(k^{\parallel} r),
\label{eq.virtual.2.4}
% \label{eq.4.3}
\end{equation}
and $\beta$ is angle between vectors $\vb{r}$ and $\vb{k}^{\parallel}$.
Now we write down the matrix element (\ref{eq.virtual.0.1}) as
\begin{equation}
\begin{array}{lllll}
% \vspace{0.5mm}
  \vb{I}_{1,\,\mu} =
  \biggl\langle\: \Phi_{f} (\vb{r})\; \biggl|\,
    e^{-i\, \vb{k}_{\rm ph} \vb{r}}\; \vb{\displaystyle\frac{d}{dr^{\mu}}} \biggr|\: \Phi_{i} (\vb{r})\: \biggr\rangle_\mathbf{r} =

  \biggl\langle\: \Phi_{f} (\vb{r})\; \biggl|\,
    e^{- i\, \vb{k}_{\rm ph}^{\parallel} \vb{r}} \cdot e^{- i\, \vb{k}_{\rm ph}^{\perp} \vb{r}}\;
    % e^{-i\, \vb{k}_{\rm ph} \vb{r}}\;
    \vb{\displaystyle\frac{d}{dr^{\mu}}} \biggr|\: \Phi_{i} (\vb{r})\: \biggr\rangle_\mathbf{r}.
\end{array}
\label{eq.virtual.2.5}
% \label{eq.app.2.1.7}
\end{equation}
Substituting Eq.~(\ref{eq.virtual.2.4}) to this formula, we obtain:
\begin{equation}
\begin{array}{lllll}
% \vspace{0.5mm}
  \vb{I}_{1,\,\mu} & = &

  \displaystyle\sum\limits_{l_{\parallel} =0}^{+\infty}
    i^{l_{\parallel}} (-1)^{l_{\parallel}} (2l_{\parallel}+1)\,
  \biggl\langle\: \Phi_{f} (\vb{r})\; \biggl|\,
    P_{l_{\parallel}} (\cos{\beta}) j_{l_{\parallel}}(k^{\parallel} r) \cdot
    e^{- i\, \vb{k}_{\rm ph}^{\perp} \vb{r}}\;
    \vb{\displaystyle\frac{d}{dr^{\mu}}} \biggr|\: \Phi_{i} (\vb{r})\: \biggr\rangle_\mathbf{r} .
\end{array}
\label{eq.virtual.2.6}
% \label{eq.app.2.1.7}
\end{equation}
Assuming approximately applicability of the Coulomb gauge, we rewrite matrix element as
\begin{equation}
\begin{array}{lllll}
% \vspace{0.5mm}
  \vb{I}_{1,\,\mu} & = &
  \displaystyle\sum\limits_{l_{\parallel} =0}^{+\infty}
    i^{l_{\parallel}} (-1)^{l_{\parallel}} (2l_{\parallel} +1)\,
  \biggl\langle\: \Phi_{f} (\vb{r})\; P_{l_{\parallel} } (\cos{\beta})\;
    j_{l_{\parallel}}(k^{\parallel} r)\; \biggl|\,
    % e^{- i\, \vb{k}_{\rm ph}^{\parallel} \vb{r}} \cdot
    e^{- i\, \vb{k}_{\rm ph}^{\perp} \vb{r}}\;
    % e^{-i\, \vb{k}_{\rm ph} \vb{r}}\;
    \vb{\displaystyle\frac{d}{dr^{\mu}}} \biggr|\:
    \Phi_{i} (\vb{r})\: \biggr\rangle_\mathbf{r}.
\end{array}
\label{eq.virtual.2.7}
% \label{eq.app.2.1.7}
\end{equation}
From these formulas one can see that this matrix element can be calculated on the basis of the
formalism of multipolar expansion developed before (for emission of real photons).
% in this new case after inclusion of the longitudinal polarization of virtual photon.
One can change the old wave functions
% \textcolor[rgb]{1.00,0.00,0.00}{\textbf{%
to new ones as
\begin{equation}
\begin{array}{lllll}
% \vspace{0.5mm}
  \Phi_{f} (\vb{r}) \to
  \Phi_{f} (\vb{r}) \; P_{l_{\parallel}} (\cos{\beta})\;
    j_{l_{\parallel}}(k^{\parallel} r) &
    {\rm or} &

\left\{
\begin{array}{lllll}
\vspace{1.5mm}
  {\rm change\: of\: radial\: function:}  & R_{l} (r) & \to & R_{l} (r) \cdot j_{l_{\parallel}}(k^{\parallel} r), \\
  {\rm change\: of\: angular\: function:} & Y_{lm}(\theta, \phi) & \to & Y_{lm}(\theta, \phi) \cdot P_{l_{\parallel}} (\cos{\beta}).
\end{array}
\right.
\end{array}
\label{eq.virtual.2.8}
\end{equation}
In result, we obtain similar formulas for all nuclear matrix elements
[see Eqs.~(\ref{eq.model.incoh.1.4}), for details]
%-----------------------------------------------------------------------------------------------------------------------
%
%-----------------------------------------------------------------------------------------------------------------------
\begin{equation}
\begin{array}{lll}
\vspace{1.0mm}
  M_{p}^{(E,\, {\rm mon},\, 0)} =
  -\, \hbar\, (2\pi)^{3}\,
  \displaystyle\frac{\mu_{N}}{\sqrt{3}}\,
  \displaystyle\frac{m_{\rm p}\, Z_{\rm eff}^{\rm (mon,\, 0)}}{\mu}\,
  \Bigl(
    J_{1}^{vir}(0,1,0, l_{\parallel}) -
    \displaystyle\frac{47}{40} \sqrt{\displaystyle\frac{1}{2}}\, J_{1}^{vir}(0,1,2, l_{\parallel})
  \Bigr), \\

\vspace{1.0mm}
  M_{p}^{(E,\, {\rm mon},\, 0)} + M_{p}^{(M,\, {\rm mon},\, 0)} =
  M_{p}^{(E,\, {\rm mon},\, 0)}\,
  \Bigl( 1 + i\: \displaystyle\frac{\alpha_{M}} {2\, m_{\rm p}\, Z_{\rm eff}^{\rm (mon,\, 0)}} \Bigr) =

  M_{p}^{(E,\, {\rm mon},\, 0)}\,
  \Bigl( 1 + i\: \displaystyle\frac{\mu^{2}} {2\, m_{\rm p}^{2}\, Z_{\rm eff}^{\rm (mon,\, 0)}}\; \alpha \Bigr), \\

\vspace{1.0mm}
  M_{\Delta M} =
  \hbar\, (2\pi)^{3}\,
  \displaystyle\frac{\sqrt{3}}{2}\,
  \mu_{N}\, k_{\rm ph}\,
  f_{A} \cdot Z_{\rm A} (\vb{k}_{\rm ph}) \cdot \tilde{J}^{vir}\, (- c_{\rm p}, 0,1,1, l_{\parallel}), \\

  M_{\Delta M} + M_{k} =
  - \hbar\, (2\pi)^{3}\,
  \displaystyle\frac{\sqrt{3}}{2}\,
  \mu_{N}\, k_{\rm ph}\,
  \Bigl\{
    \displaystyle\frac{A+1}{2A}\, \bar{\mu}_{\rm pn}^{(A)} \cdot Z_{\rm A} (\vb{k}_{\rm ph})\, \tilde{J}^{vir}\, (- c_{\rm p}, 0,1,1, l_{\parallel}) +
    \mu_{\rm p}\, z_{\rm p}\, \tilde{J}^{vir}\, (c_{A}, 0,1,1, l_{\parallel})
  \Bigr\},
\end{array}
\label{eq.virtual.2.9}
\end{equation}
where parameters are defined and calculated in Sec.~\ref{sec.app.2.multiple}.
% [see Eq.~(\ref{eq.app.matr_el.coh_mag.9}), p.~\pageref{eq.app.matr_el.coh_mag.9}]
%-----------------------------------------------------------------------------------------------------------------------
%
%-----------------------------------------------------------------------------------------------------------------------
The new integrals should be used [instead of integrals (\ref{eq.app.2.multiple.1})]
\begin{equation}
\begin{array}{llllll}
  J_{1}^{vir}(l_{i},l_{f},n,\; l_{\parallel}) & = &
  \displaystyle\int\limits^{+\infty}_{0}
    \displaystyle\frac{dR_{i}(r, l_{i})}{dr}\:
    R^{*}_{f}(l_{f},r)\, j_{n}(k_{\rm ph}^{\bot}r)\;
    j_{l_{\parallel}}(k^{\parallel} r)\; r^{2} dr, \\

  \tilde{J}^{vir}\,(c, l_{i},l_{f},n;\; l_{\parallel}) & = &
  \displaystyle\int\limits^{+\infty}_{0} R_{i}(l_{i}, r)\, R^{*}_{f}(l_{f},r)\, j_{n}(c\, k_{\rm ph}^{\bot}r)\;
  j_{l_{\parallel}}(k^{\parallel} r)\; r^{2} dr.

%  \breve{J}\,(c_{A}, l_{i}, l_{f},n) & = & \displaystyle\int\limits^{+\infty}_{0} R_{i}(r)\, R^{*}_{l,f}(r)\, V(\mathbf{r})\, j_{n}(c_{A}\,kr)\; r^{2} dr.
\end{array}
\label{eq.virtual.2.10}
% \label{eq.multimple.4}
% \label{eq.resultingformulas.7}
\end{equation}
Here, $R_{i,f}$ is radial part of wave function $\Phi_{\rm p - nucl} (\vb{r})$ in $i$-state or $f$-state,
$j_{n}(k_{\rm ph}r)$ is spherical Bessel function of order $n$.
In particular, at $l_{\parallel} = 0$ formulas (\ref{eq.virtual.2.9})--(\ref{eq.virtual.2.10}) are obtained exactly, without approximation.
Influence of longitudinal part of virtual photon can be studied as the first correction,
as $j_{l_{\parallel}=0}(k^{\parallel} r) \ne 1$.
So, we can start calculations of matrix elements with including longitudinal part of virtual photon from such an approximation
$l_{\parallel} = 0$.

% *******************************************************************************************************************

% *******************************************************************************************************************
\section{The parameters of the proton--nucleus potential
\label{sec.app.4}}

We calculate the radial wave functions $R_{l}(r)$ for proton nucleus scattering numerically concerning the chosen potential
of interaction between the proton and the spherically symmetric core.
For description of proton-nucleus interaction we use the potential as
$V (r) = v_{c} (r) + v_{N} (r) + v_{\rm so} (r) + v_{l} (r)$,
where $v_{c} (r)$, $v_{N} (r)$, $v_{\rm so} (r)$ and $v_{l} (r)$ are Coulomb,
nuclear, spin-orbital and centrifugal components
having the form~\cite{Becchetti.1969.PR}:
\begin{equation}
\begin{array}{ccc}
  \vspace{1mm}
  v_{N} (r) = - \displaystyle\frac{V_{R}} {1 + \exp{\displaystyle\frac{r-R_{R}} {a_{R}}}},
  \hspace{2mm}
  v_{l} (r) = \displaystyle\frac{l\,(l+1)} {2mr^{2}}, % \\
  \hspace{2mm}
  v_{\rm so} (r) =
    V_{\rm so}\,
    {\mathbf {q \cdot l}}\,
    \displaystyle\frac{\lambda_{\pi}^{2}}{r}\,
    \displaystyle\frac{d}{dr}\, \Bigl[1
    + \exp\Bigl(\displaystyle\frac{r-R_{\rm so}} {a_{\rm so}} \Bigr)\Bigr]^{-1}, \\
  v_{c} (r) =
  \left\{
  \begin{array}{ll}
    \displaystyle\frac{Z e^{2}} {r}, &
      \mbox{at  } r \ge R_{c}, \\
    \displaystyle\frac{Z e^{2}} {2 R_{c}}\;
      \biggl\{ 3 -  \displaystyle\frac{r^{2}}{R_{c}^{2}} \biggr\}, &
      \mbox{at  } r < R_{c}.
  \end{array}
  \right.
\end{array}
\label{eq.app.4.1}
\end{equation}
We use the parametrization proposed by Becchetti and Greenlees in~\cite{Becchetti.1969.PR}
which has been tested in numerous research papers:
\begin{equation}
\begin{array}{ll}
\begin{array}{llll}
  V_{R} = 54.0 - 0.32\,E + 0.4\,Z/A^{1/3} + 24.0\,I, &
  V_{\rm so} =6.2, \\
\end{array} \\
\begin{array}{llll}
  R_{R} = r_{R}\, A^{1/3}, &
  R_{c} = r_{c}\, A^{1/3}, &
  R_{\rm so} = r_{\rm so}\, A^{1/3}, \\
  r_{\rm so} = 1.01\;{\rm fm}, &
  a_{R} = 0.75\; {\rm fm}, &
  a_{\rm so} = 0.75\; {\rm fm}.
\end{array}
\end{array}
\label{eq.app.4.2}
\end{equation}
Here,
$I = (N-Z)/A$, $A$ and $Z$ are mass and proton numbers of the daughter nucleus,
$E$ is incident lab energy,
$V_{R}$ and $V_{\rm so}$ are strength of nuclear and spin-orbital
components defined in MeV,
$R_{c}$ and $R_{R}$ are Coulomb and nuclear radiuses of nucleus, $a_{R}$
and $a_{\rm so}$ are diffusion parameters.
In the calculations we restrict ourselves by approximation $r_{c} = r_{R}$.
Also we neglect by term $V_{\rm so}$
(previously, we estimated role of term $V_{\rm so}$ in study of bremsstrahlung in proton-nucleus scattering at low energies, which was small
\cite{Maydanyuk_Zhang.2015.PRC}).

\end{document}